\documentclass[aps,prd,nofootinbib,twocolumn]{revtex4-1}
\usepackage{amsmath}
\usepackage[scientific-notation=true]{siunitx}
\usepackage{graphicx}
\usepackage{placeins}
\usepackage{bm}
\usepackage[export]{adjustbox}
\usepackage{enumitem}

\usepackage{scalerel}
\usepackage{tikz}
\usetikzlibrary{svg.path}

\definecolor{orcidlogocol}{HTML}{A6CE39}
\tikzset{
	orcidlogo/.pic={
		\fill[orcidlogocol] svg{M256,128c0,70.7-57.3,128-128,128C57.3,256,0,198.7,0,128C0,57.3,57.3,0,128,0C198.7,0,256,57.3,256,128z};
		\fill[white] svg{M86.3,186.2H70.9V79.1h15.4v48.4V186.2z}
		svg{M108.9,79.1h41.6c39.6,0,57,28.3,57,53.6c0,27.5-21.5,53.6-56.8,53.6h-41.8V79.1z M124.3,172.4h24.5c34.9,0,42.9-26.5,42.9-39.7c0-21.5-13.7-39.7-43.7-39.7h-23.7V172.4z}
		svg{M88.7,56.8c0,5.5-4.5,10.1-10.1,10.1c-5.6,0-10.1-4.6-10.1-10.1c0-5.6,4.5-10.1,10.1-10.1C84.2,46.7,88.7,51.3,88.7,56.8z};
	}
}

\newcommand\orcidicon[1]{\href{https://orcid.org/#1}{\mbox{\scalerel*{
				\begin{tikzpicture}[yscale=-1,transform shape]
				\pic{orcidlogo};
				\end{tikzpicture}
			}{1}}}}

\usepackage{hyperref} 

\let\vec\bm

\newcommand{\approptoinn}[2]{\mathrel{\vcenter{
			\offinterlineskip\halign{\hfil$##$\cr
				#1\propto\cr\noalign{\kern2pt}#1\sim\cr\noalign{\kern-2pt}}}}}

\begin{document}
\title{Evolution of dark matter microhalos through stellar encounters}
\author{M. Sten Delos \orcidicon{0000-0003-3808-5321}}
\email{delos@unc.edu}
\affiliation{Department of Physics and Astronomy, University of North Carolina at Chapel Hill, Phillips Hall CB3255, Chapel Hill, North Carolina 27599, USA}

\begin{abstract}
In the cold dark matter scenario, the smallest dark matter halos may be earth mass or smaller.  These microhalos would be the densest dark matter objects in the Universe, making their accurate characterization important for astrophysical dark matter detection efforts.  Moreover, their properties are closely linked to the nature of dark matter and the physics of the early universe, making them valuable cosmological probes.  Dark matter microhalos survive as subhalos within larger galactic halos, but due to their small size, they are susceptible to encounters with individual stars.  We use a large number of $N$-body simulations to develop a framework that can predict the evolution of a microhalo's density profile due to stellar encounters.  We find that there is a universal density profile for microhalos subjected to stellar encounters, and each encounter alters a microhalo's scale parameters in a way that is simply related to the energy the encounter injects.  Our framework can rapidly and accurately characterize the impact of stellar encounters on whole ensembles of microhalos, making it a promising tool for understanding the populations of microhalos within galactic halos.
\end{abstract}

                              
\maketitle

\section{Introduction}

Dark matter structure grows through hierarchical assembly; the smallest halos form first, and larger halos grow through subsequent mergers and accretion of smaller halos.  The latter are not destroyed by this assembly process but instead broadly survive as subhalos of their new hosts, and cosmological simulations reveal copious substructure within each dark matter halo (e.g., Refs.~\cite{ghigna1998dark,tormen1998survival,gao2004subhalo,kravtsov2004dark,giocoli2008population,giocoli2010substructure}).  In the cold dark matter scenario, the smallest halos may be earth mass or smaller \cite{hofmann2001damping,green2004power,diemand2005earth}, and many of these microhalos are expected to persist today within galactic halos \cite{diemand2005earth,goerdt2007survival,schneider2010impact,ishiyama2010gamma}.

The present-day abundance and structure of dark matter microhalos are important to astrophysical dark matter detection efforts.  Thanks to their early formation, these halos would possess the highest characteristic density of any dark matter objects, making them important contributors to prospective signals from dark matter annihilation \cite{ishiyama2010gamma,anderhalden2013density,*anderhalden2013erratum,ishiyama2014hierarchical,ishiyama2019abundance,berezinsky2003small,diemand2005earth,pieri2008dark,springel2008prospects,berezinsky2008remnants,gao2011will,belotsky2014gamma,sanchez2014flattening,bartels2015boosting,anderson2016search,stref2017modeling,hiroshima2018modeling,stref2019remnants} (see Ref.~\cite{ando2019halo} for a review).  Their high density also makes them promising targets for gravitational searches, whether through microlensing (e.g., Refs.~ \cite{boden1998astrometric,chen2010gravitational,erickcek2011astrometric,van2018halometry}), timing delays (e.g., Refs.~\cite{siegel2007probing,baghram2011prospects,kashiyama2018detectability}), or the dynamics of stars or other astrophysical systems (e.g., Refs.~\cite{gonzalez2013constraining,erkal2015forensics,erkal2016number,buschmann2018stellar,penarrubia2017fluctuations}).  Conversely, any observational constraints on the structure and abundance of these microhalos serve as cosmological probes.  The mass scale of the smallest halos directly reflects the free-streaming scale of the dark matter particle.  Meanwhile, the abundance and internal structures of microhalos are closely linked to the statistics of the primordial density fluctuations from which they formed \cite{delos2019predicting} (see also Refs.~\cite{diemer2015universal,okoli2015concentration,ludlow2016mass,diemer2019accurate}), and these fluctuations are sensitive to the details of both inflation (e.g., Refs.~\cite{silk1987double,salopek1989designing,starobinsky1992,ivanov1994inflation,randall1996supernatural,stewart1997flattening,adams1997multiple,starobinsky1998beyond,covi1999running,martin2000nonvacuum,chung2000probing,martin2001trans,joy2008new,barnaby2009particle,barnaby2010features,ben2010cosmic,gong2011waterfall,lyth2011contribution,bugaev2011curvature,barnaby2011large,achucarro2011features,cespedes2012importance,barnaby2012gauge}) and the postinflationary cosmic history (e.g., Refs.~\cite{erickcek2011reheating,barenboim2014structure,fan2014nonthermal,erickcek2015dark,redmond2018growth,blanco2019annihilation}).

However, microhalos are subjected to complex dynamical processes after accretion onto a host halo, and the impact of these processes must be understood in order to accurately predict microhalo populations.  These microhalos experience gradual disruption by means of tidal forces from the host, encounters with other substructures, and dynamical friction.  Subhalo survival prospects resulting from these processes have been widely studied (e.g., Refs.~\cite{taylor2001dynamics,hayashi2003structural,penarrubia2005effects,van2005mass,zentner2005physics,kampakoglou2006tidal,goerdt2007survival,berezinsky2008remnants,gan2010improved,penarrubia2010impact,pullen2014nonlinear,jiang2016statistics,van2017dissecting,van2017disruption,van2018dark,ogiya2019dash,errani2019can,delos2019tidal}).  Additionally, within galaxies microhalos are susceptible to encounters with individual stars.  The objective of this work is to develop a general framework, applicable to a wide variety of systems, that can predict the evolution of a microhalo's density profile through stellar encounters.

Previous works have explored the impact of stellar encounters on dark matter microhalos.  A common strategy (e.g., Refs.~\cite{berezinsky2006destruction,berezinsky2014small,goerdt2007survival,green2007mini,schneider2010impact}) is to compare the total energy injected by a stellar encounter to the total binding energy of the microhalo.  However, as Ref.~\cite{van2017disruption} has noted, this comparison is not directly connected to the question of halo survival because such energy injections are not distributed efficiently; the least bound particles, at large radii, receive the most energy.  Prior works have also employed semianalytic models \cite{zhao2007tidal} and numerical simulations \cite{green2007mini,schneider2010impact,angus2007cold,ishiyama2010gamma} to study the impact of stellar encounters.  However, the scopes of these investigations have been limited; they typically aim to understand the microhalo's energy injection or mass loss or focus on the survival of microhalos near the solar neighborhood.  Our work is much more general.  We study the evolution of a microhalo's full density profile due to an arbitrary series of stellar encounters.

The smallest microhalos are expected to form with $\rho\propto r^{-3/2}$ inner density profiles \cite{ishiyama2010gamma,anderhalden2013density,*anderhalden2013erratum,ishiyama2014hierarchical,polisensky2015fingerprints,ogiya2017sets,delos2018ultracompact,delos2018density,angulo2017earth,delos2019predicting,ishiyama2019abundance}, but successive mergers tend to shallow their inner cusps toward $\rho\propto r^{-1}$ \cite{ogiya2016dynamical,angulo2017earth,delos2019predicting}.  Accordingly, we study microhalos that initially possess the Navarro-Frenk-White (NFW) density profile \cite{navarro1996structure,navarro1997universal},
\begin{equation}\label{NFW}
\rho(r) = \frac{\rho_s}{(r/r_s)(1+r/r_s)^{2}},
\end{equation}
which has scale parameters $r_s$ and $\rho_s$ and a $\rho\propto r^{-1}$ cusp.  We use high-resolution $N$-body simulations of 96 stellar encounters to explore the parameter space of encounters with these microhalos, and we consider both first and successive encounters.  For validation we also simulate randomized realistic sequences of encounters.  The framework we develop predicts the density profile of a microhalo after arbitrarily many stellar encounters predominantly as a function of the energy injected by each encounter.  We also find that the density profile after stellar encounters is nearly universal, which simplifies the problem considerably.

This article is organized as follows.  In Sec.~\ref{sec:sims}, we detail how our simulations are carried out.  Section~\ref{sec:param} develops a parametrization of stellar encounters and shows how our simulated encounters are distributed.  In Sec.~\ref{sec:model}, we use our simulation results to develop a model for the impact of distant encounters, where the impact parameter is much larger than the halo, and we consider both initial and successive encounters.  Section~\ref{sec:b} discusses the impact of closer encounters.  In Sec.~\ref{sec:fields}, we explore the application of our model to microhalos passing through fields of stars.  Section~\ref{sec:conclusion} concludes.  Finally, Appendix~\ref{sec:impulse} explores the range of validity of the impulse approximation that we employ throughout this work, while Appendix~\ref{sec:multiple} explores the impact of encounters occurring in close succession.

\section{Simulations}\label{sec:sims}

We follow Refs.~\cite{green2007mini,schneider2010impact,angus2007cold,ishiyama2010gamma} in using $N$-body simulations to study a microhalo's response to a stellar encounter.  We prepare the microhalo with an NFW density profile using the same procedure as Ref.~\cite{delos2019tidal}.  We draw particles from an isotropic distribution function computed using the fitting formula in Ref.~\cite{widrow2000distribution}.  Additionally, to better resolve the microhalo's central region, we sample particles whose orbital pericenters are below $r_s/3$, where $r_s$ is the microhalo's scale radius, with $64$ times the number density and $1/64$ the mass of the other particles.\footnote{Mixing particles of different masses can induce a discreteness artifact wherein two-body interactions transfer energy from heavy to light particles \cite{binney2002two}.  However, this effect is suppressed in our halo construction by the high particle count and the initial radial segregation between the particle types.  Reference~\cite{delos2019tidal} verified that there is no tendency for the heavier particles to sink to lower radii within our halos even after hundreds of dynamical time intervals.}  We cut off the density profile at $r = 500 r_s$; subhalo particles this far out are completely stripped by even a glancing encounter, so as long as the cutoff radius is much larger than $r_s$, the precise choice makes no difference.  We represent the subhalo using a total of $8\times 10^6$ particles, and roughly 70\% of them, carrying roughly 4\% of the total mass, are high-resolution particles.

To simulate a stellar encounter, we follow Refs.~\cite{angus2007cold,ishiyama2010gamma} in perturbing the velocities of microhalo particles using the impulse approximation, in which these particles are treated as stationary while the passing star exerts a tidal acceleration on each particle.  Consider an encounter with impact parameter $b$ and relative velocity $V$, and center a coordinate system on the microhalo such that the star is at position $(-V t,b,0)$ at time $t$.  Since a microhalo is typically much larger than a solar system (e.g., Sec.~\ref{sec:fields}), the star can be accurately treated as a point mass.  By integrating the tidal acceleration it is straightforward to show that a particle at position $(x,y,z)$ experiences velocity injection
\begin{equation}\label{dv}
\Delta \vec v = -\frac{2G M_*}{V}\frac{1}{(y-b)^2+z^2}\left(0,\frac{(y-b)y+z^2}{b},z\right),
\end{equation}
where $M_*$ is the mass of the star.

Application of the impulsive velocity injections given by Eq.~(\ref{dv}) is computationally faster and more numerically stable than adding a new particle to the simulation to represent the star, owing to the star's immense velocity and mass compared to those of microhalo particles.  Moreover, we expect the impulse approximation to be valid in most scenarios because due to the size difference, a microhalo's internal velocity dispersion is typically much smaller than that of its host.  A passing star's relative velocity is thus expected to be much greater than that of microhalo particles.  However, to clarify the conditions under which the impulse approximation is valid, we explicitly simulate a passing star in Appendix~\ref{sec:impulse}; we find that the approximation yields accurate results as long as
\begin{equation}\label{impulse}
t_\mathrm{dyn} \gtrsim 5 b/V,
\end{equation}
where $t_\mathrm{dyn}$ is the initial microhalo's dynamical timescale given by \cite{binney1987galactic}
\begin{equation}\label{tdyn}
t_\mathrm{dyn}\equiv\sqrt{3\pi/(16G\rho_s)}.
\end{equation}
Here, $\rho_s$ is the microhalo's scale density.

After perturbing the velocities of microhalo particles, we use the $N$-body simulation code \textsc{Gadget-2} to simulate the microhalo's response.  The force-softening length is set at $\epsilon=0.003 r_s$, a small value that allows radii as small as $2.8\epsilon\simeq 0.01 r_s$ to be resolved.  When studying microhalo density profiles, we use the procedure in Ref.~\cite{delos2019tidal} to set the resolution limit as the largest of the limits set by the softening length, Poisson noise, and artificial relaxation.

Each microhalo may experience a large number of stellar encounters, so it is necessary to also study a microhalo's response to successive encounters.  To prepare a stellar encounter after the first, we extract the self-bound remnant of the microhalo at the end of a prior simulation using an iterative procedure detailed in Ref.~\cite{delos2019tidal}.  We compute the potential energy of each particle due to other bound particles and use that energy to determine whether each particle is bound; this process is iterated until the number of bound particles converges.  The resulting bound remnant is then subjected to a stellar encounter using Eq.~(\ref{dv}).

\begin{figure}[t]
	\includegraphics[width=.49\columnwidth]{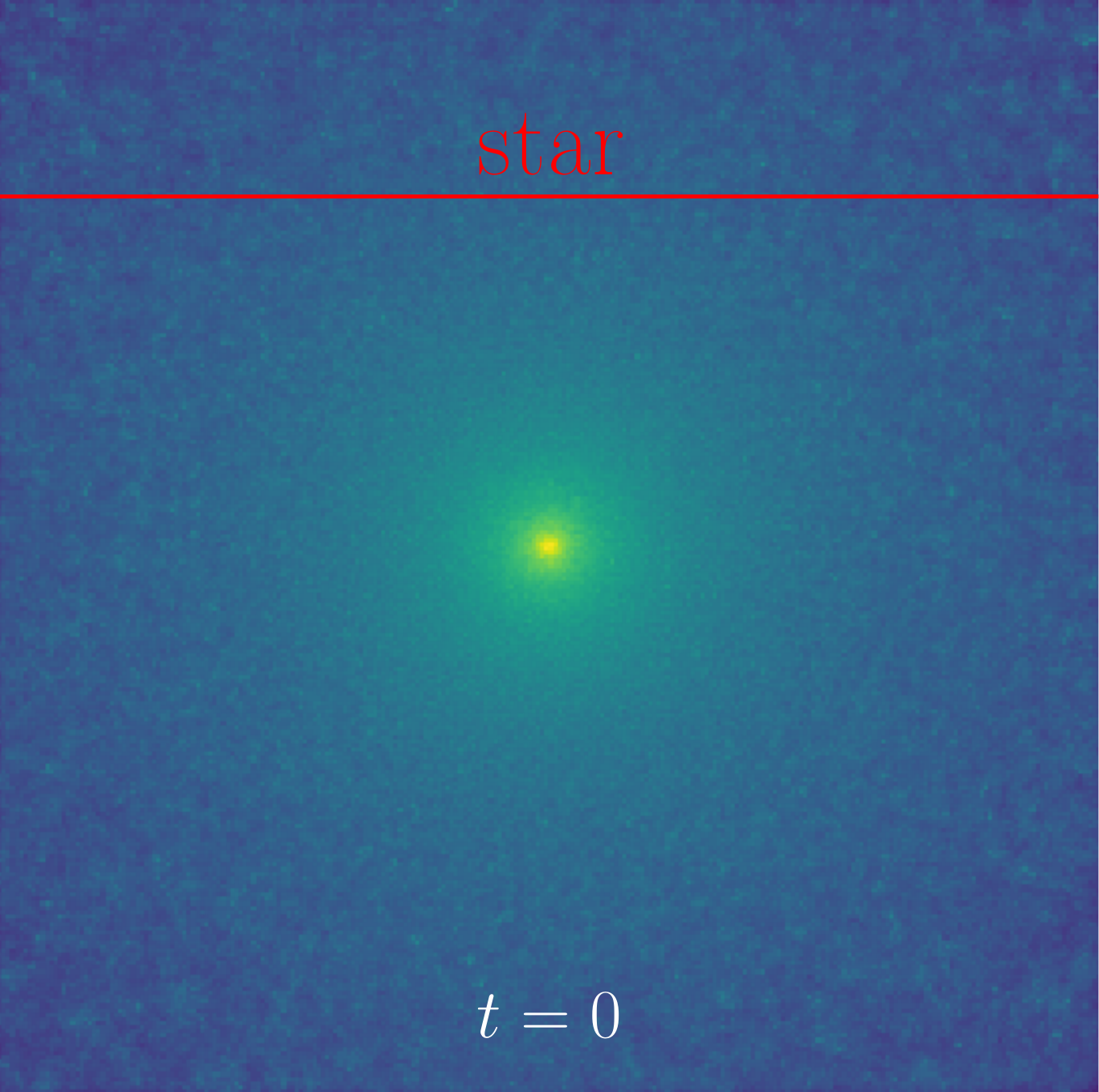} \hspace{-2mm}
	\includegraphics[width=.49\columnwidth]{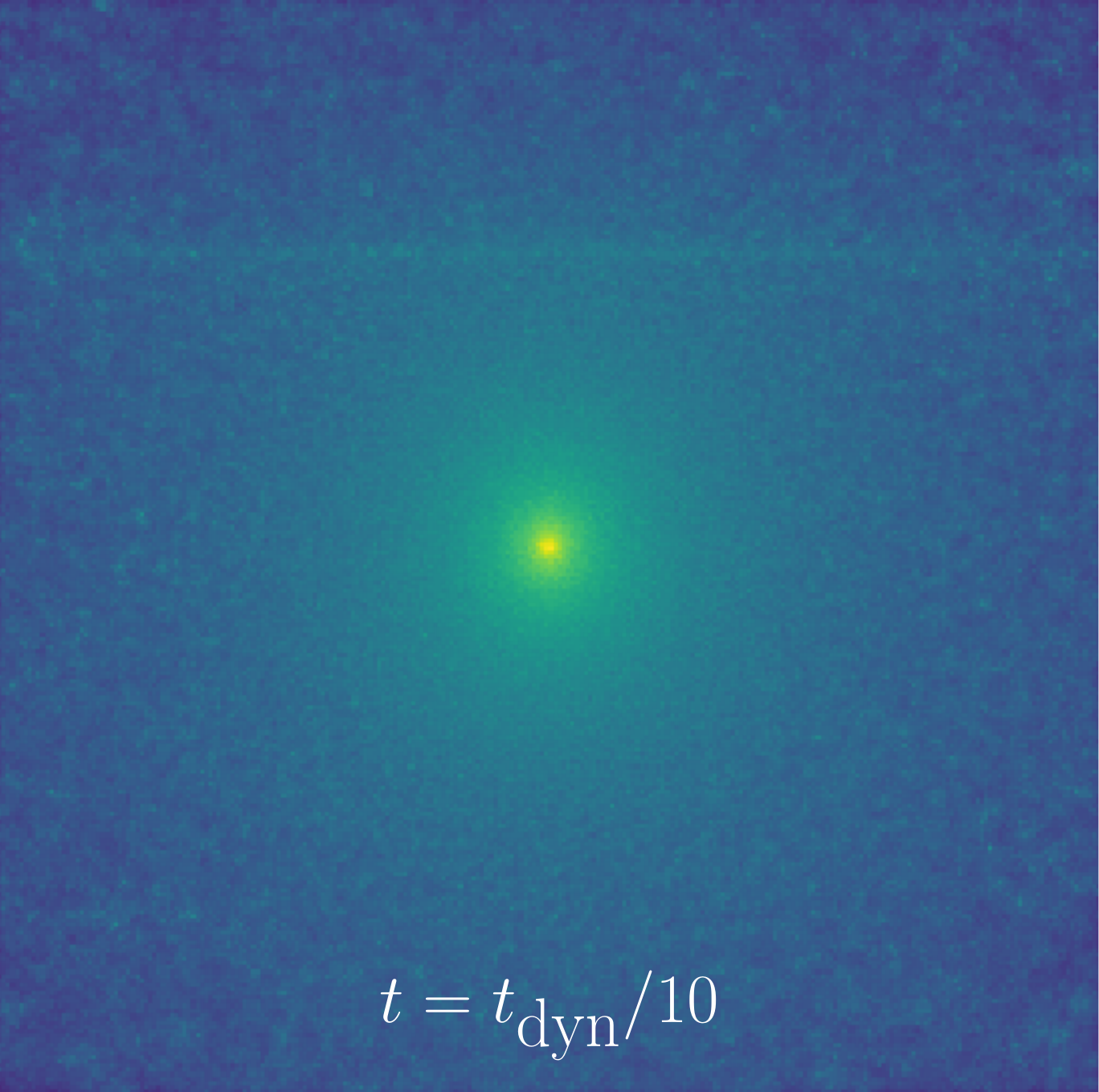} \\
	\vspace{-0mm}
	\includegraphics[width=.49\columnwidth]{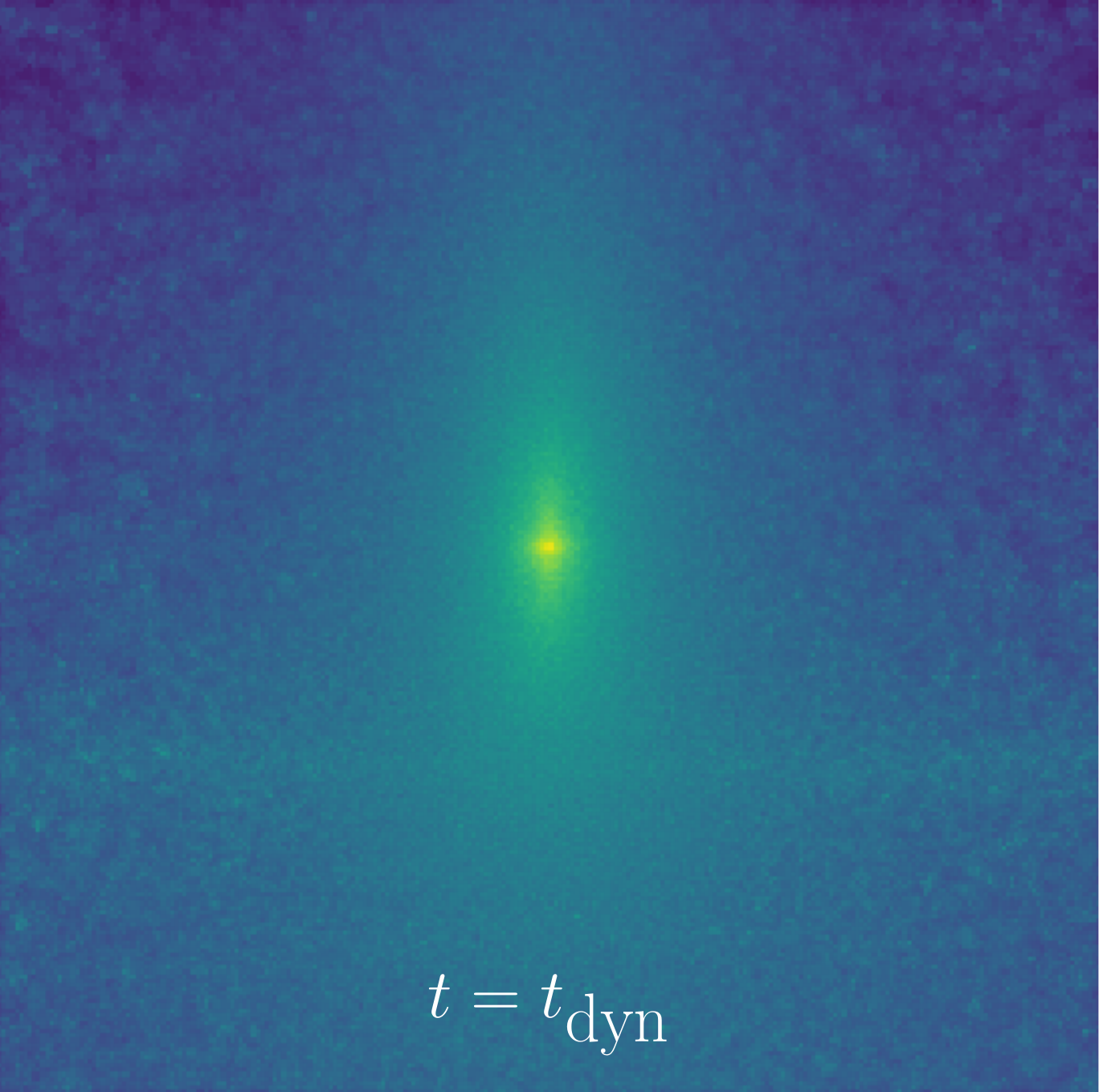} \hspace{-2mm}
	\includegraphics[width=.49\columnwidth]{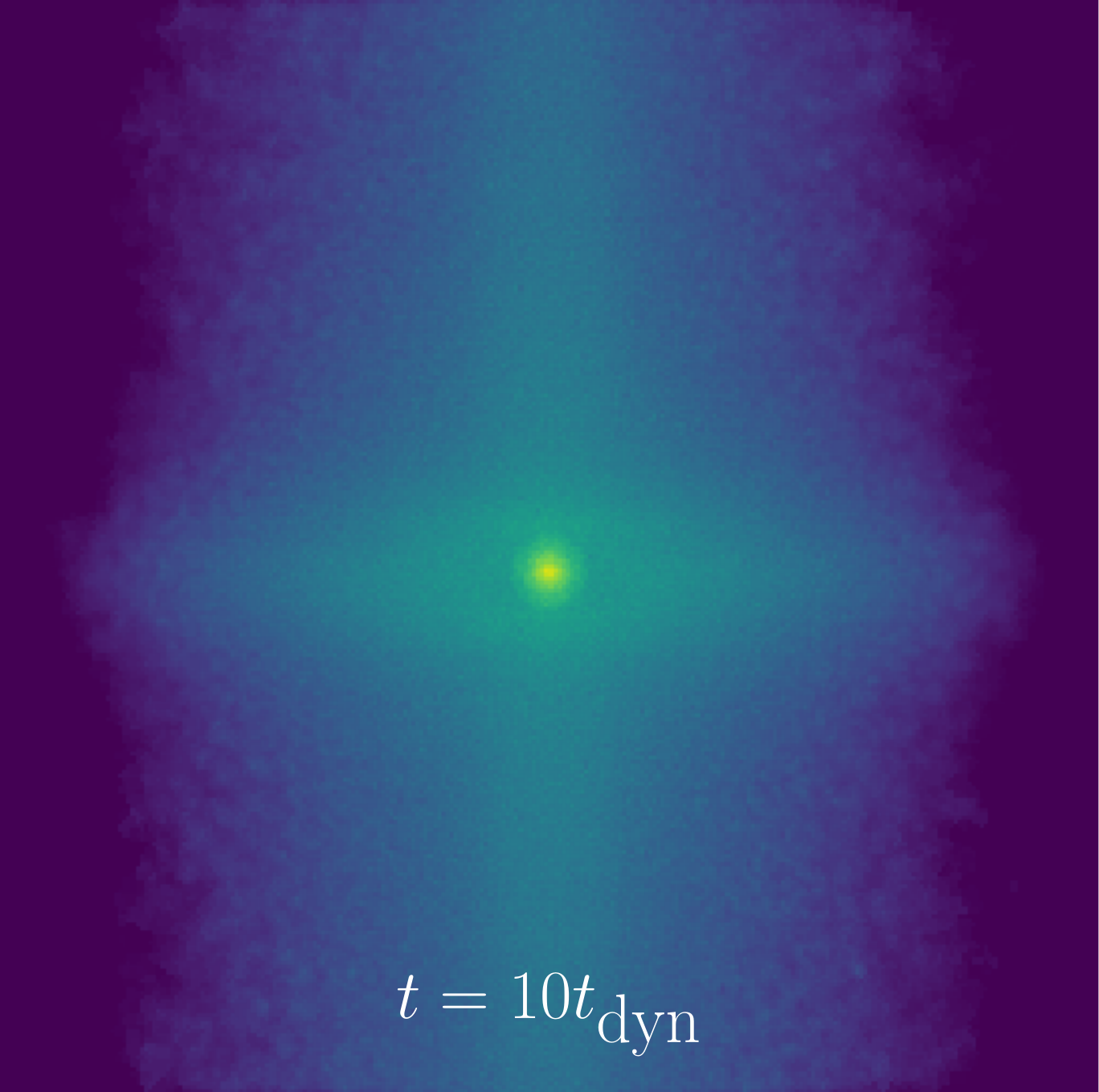}
	\caption{\label{fig:field} The projected density field of a microhalo subjected to a stellar encounter.  Each panel has width $50r_s$, where $r_s$ is the microhalo scale radius.  The density is computed using a $k$-nearest-neighbor density estimate with $k=50$ and is plotted with a logarithmic color scale (lighter is denser).}
\end{figure}

\begin{figure}[t]
	\includegraphics[width=\columnwidth]{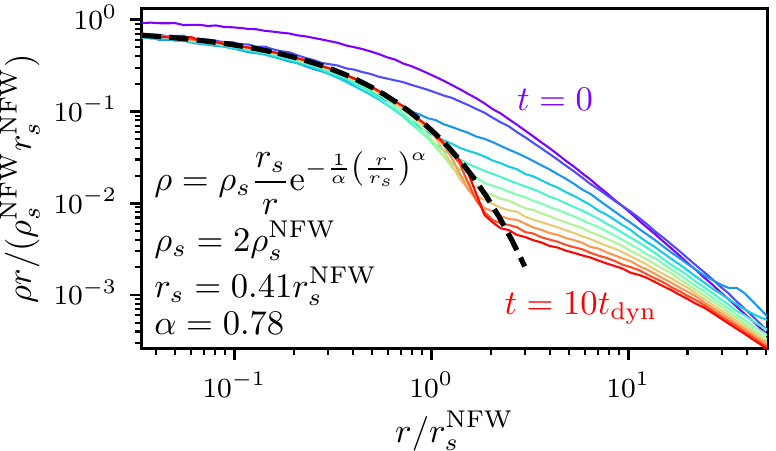}
	\caption{\label{fig:profile} The density profile of the microhalo depicted in Fig.~\ref{fig:field}.  Each curve indicates the passage of a single dynamical time $t_\mathrm{dyn}$ given by Eq.~(\ref{tdyn}).  This profile, initially an NFW profile with parameters $\rho_s^\mathrm{NFW}$ and $r_s^\mathrm{NFW}$, stabilizes into the form given by Eq.~(\ref{profile}) with fitting parameters shown.  All particles are plotted, not only particles bound to the halo, and beyond the stable part of the density profile the mean radial motion is typically outward.}
\end{figure}

Figure~\ref{fig:field} illustrates a stellar encounter with an NFW microhalo simulated in this way.  96\% of the initial mass of the microhalo is freed by the encounter, but a highly dense remnant remains.  While the final system is significantly nonspherical, the dense remnant itself is highly spherical, and Fig.~\ref{fig:profile} shows the evolution of the microhalo's spherically averaged density profile.  The central profile stabilizes over the course of a single dynamical time interval $t_\mathrm{dyn}$, and the stable region grows over time.  The final density profile is well fit by
\begin{equation}\label{profile}
\rho=\rho_s^\prime \frac{r_s^\prime}{r} \exp\left[-\frac{1}{\alpha}\left(\frac{r}{r_s^\prime}\right)^\alpha\right]
\end{equation}
with different scale parameters $r_s^\prime$ and $\rho_s^\prime$ and an additional parameter $\alpha= 0.78$.  For this profile, $\rho_s^\prime$ and $r_s^\prime$ are defined analogously to the parameters of the NFW profile: $r_s^\prime$ is the radius at which $\mathrm{d}\ln \rho/\mathrm{d}\ln r=-2$, and at small radii, the profile asymptotes to $\rho(r)=\rho_s^\prime r_s^\prime/r$.  However, the profile in Eq.~(\ref{profile}) decays much more rapidly than the NFW profile at large radii.  In this respect it is similar to density profiles proposed to result from tidal evolution within host halos (e.g., Refs.~\cite{hayashi2003structural,penarrubia2010impact,green2019tidal}), although its functional form is more closely inspired by the Einasto profile \cite{einasto2965}, which differs only in lacking the $r_s^\prime/r$ factor.

We simulated a variety of stellar encounter scenarios, including up to five successive encounters with a single microhalo.\footnote{Successive encounters are cosmologically relevant.  For instance, we find in Sec.~\ref{sec:fields} that microhalos in the solar vicinity are expected to have experienced potentially thousands of stellar encounters, although the precise number depends on the threshold impact parameter for inclusion.}  In each case, we simulated the microhalo for the same duration of $10t_\mathrm{dyn}^\mathrm{NFW}$, where $t_\mathrm{dyn}^\mathrm{NFW}$ is the dynamical timescale of the initial NFW halo.\footnote{Stellar encounters raise the characteristic density of the subhalo remnant by stripping its less-dense outskirts, thereby reducing its dynamical timescale.  Hence, our simulations always last at least 10 times the dynamical timescale of the initial microhalo, even when the initial microhalo has already experienced an encounter.}  We find that the density profile of a microhalo subjected to any number of stellar encounters almost universally follows Eq.~(\ref{profile}) with the same $\alpha=0.78$.  In subsequent sections, we will demonstrate the near universality of this profile and discuss when it fails to apply.

\section{Parametrization of encounters}\label{sec:param}

Our goal is to understand the impact of a stellar encounter with mass $M_*$, relative velocity $V$, and impact parameter $b$ on a microhalo with scale density $\rho_s$ and scale radius $r_s$.  The stellar encounter is characterized by these five parameters, but they exhibit substantial degeneracy.  Equation~(\ref{dv}) implies that for a halo of characteristic size $r_s$, velocity injections are proportional to a characteristic velocity injection
\begin{equation}\label{dvc}
\Delta v \equiv \frac{G M_*}{V b^2} r_s f\!\left(\frac{r_s}{b}\right),
\end{equation}
where $f(x)=1+\mathcal{O}(x)$ is a nonlinear function.  Meanwhile, particle velocities within a halo of characteristic scale $r_s$ and density $\rho_s$ are proportional to a characteristic velocity $v\equiv r_s\sqrt{G\rho_s}$, so
\begin{equation}\label{dvv}
\frac{\Delta v}{v} = \sqrt{\frac{G}{\rho_s}} \frac{M_*}{V b^2} f\!\left(\frac{r_s}{b}\right)
\end{equation}
is the characteristic relative velocity injection.  The dynamical impact of a stellar encounter is thus a function of just two parameters: $\sqrt{G/\rho_s} M_*/(V b^2)$ and $r_s/b$.

We are free to choose a parametrization that is any set of independent functions of these two parameters, and we do so in the following way.  In the $b\gg r_s$ limit, Eq.~(\ref{dv}) implies that the energy $(1/2)|\Delta \vec v|^2$ (per mass) injected into a particle at cylindrical radius $r_s$ (in the $y$-$z$ plane) is
\begin{equation}\label{dE_}
\Delta E = \frac{2 G^2 M_*^2 r_s^2}{V^2 b^4}.
\end{equation}
The $b^{-4}$ divergence at small $b$ owes to the choice of reference frame: Eq.~(\ref{dv}) specifies the velocity change relative to that of the initial halo's center.  However, to understand the dynamics of the halo remnant, the more relevant energy injection is that relative to the halo remnant's center of mass, whose trajectory may differ from that of the halo's center if $b\lesssim r_s$.  Reference~\cite{moore1993upper} found that the energy injected by an impulsive encounter is well fit by a form proportional to $1/[1+(b/r_s)^4]$.  With this scaling, Eq.~(\ref{dE_}) becomes
\begin{equation}\label{dE}
\Delta E = \frac{2 G^2 M_*^2 r_s^2}{V^2 \left(b^4+r_s^4\right)}.
\end{equation}
While Ref.~\cite{moore1993upper} studied star clusters with King density profiles \cite{king1966structure} instead of dark matter halos, we will see later that this scaling works well for our simulated dark matter halos.

The characteristic binding energy (per mass) of a particle within a microhalo with scale radius $r_s$ and scale density $\rho_s$ is
\begin{equation}\label{Eb}
E_b = -4\pi G \rho_s r_s^2.
\end{equation}
We define one parameter as the encounter's relative energy injection $q\equiv \Delta E/|E_b|$, which implies that
\begin{equation}\label{x}
q=\frac{G}{2\pi}\frac{M_*^2}{\rho_s V^2\left(b^4+r_s^4\right)}.
\end{equation}
We define our second parameter as the relative distance of the encounter,
\begin{equation}\label{y}
p\equiv b/r_s.
\end{equation}

\begin{figure}[t]
	\includegraphics[width=\columnwidth]{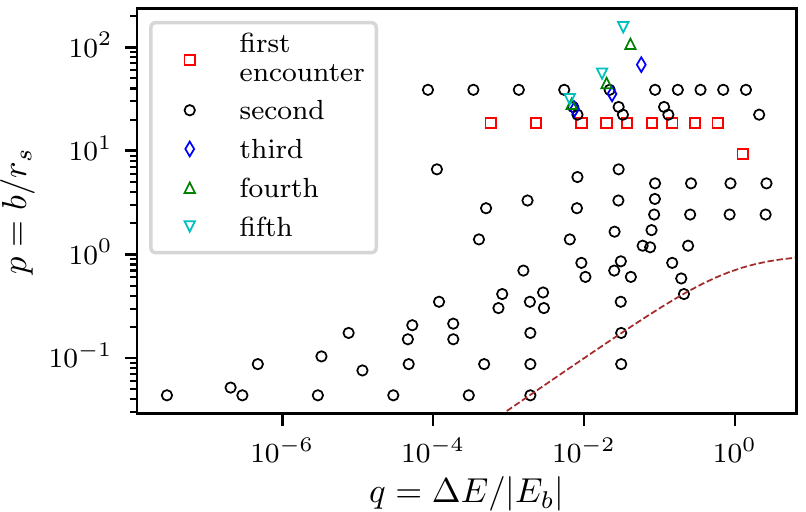}
	\caption{\label{fig:sims} The distribution of our 96 stellar encounter simulations.  For first encounters, the initial halo has an NFW profile.  Higher-order encounters begin with the remnant of a previous encounter.  The dashed curve corresponds to $(1+q^{-1})^{1/2}p=1$; as we discuss in Sec.~\ref{sec:b}, this outlines the region wherein nonlinearities in the impulsive velocity injections become dominant, typically resulting in disruption of the halo's central cusp.  For first encounters, $q$ and $p$ are defined using the results of Sec.~\ref{sec:nfw}.}
\end{figure}

To probe the impact of stellar encounters, we now explore the $q$-$p$ parameter space.  We carried out 96 simulations using the procedure in Sec.~\ref{sec:sims}, and the distribution of these simulations is depicted in Fig.~\ref{fig:sims}.  The choices of encounters will be motivated in subsequent sections.

\section{Impact of distant encounters}\label{sec:model}

\begin{figure}[t]
	\includegraphics[width=\columnwidth]{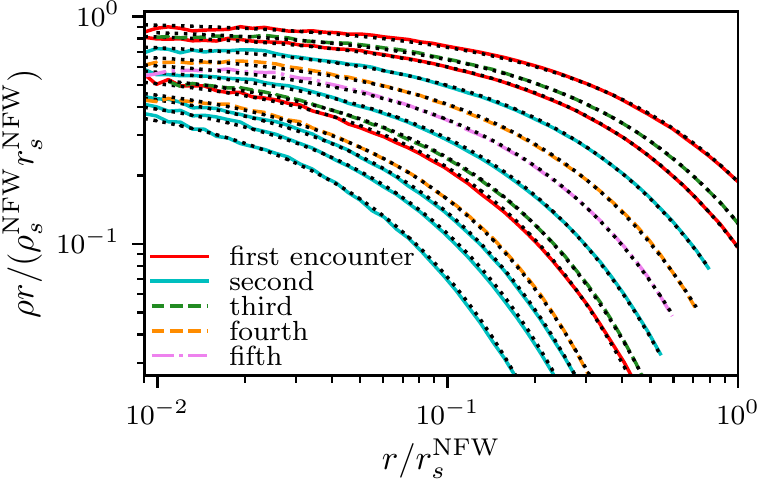}
	\caption{\label{fig:profiles} A demonstration of the universality of the density profile given by Eq.~(\ref{profile}) with $\alpha=0.78$.  This figure shows the density profiles (solid, dashed, and dot-dashed lines) resulting from 13 of our 36 simulations with $b\gg r_s$; this sample spans $5\times 10^{-3}<q<2$.  The dotted lines indicate the fits to each profile using Eq.~(\ref{profile}) with $\alpha=0.78$ enforced.  Density and radius are expressed in units of the parameters of the original NFW microhalo.}
\end{figure}

We begin by studying distant encounters with $b\gg r_s$, and we employ the 36 simulations depicted in Fig.~\ref{fig:sims} for which $p=b/r_s>8$.  In this regime, the nonlinearities (with respect to spatial coordinates) in the velocity injections given by Eq.~(\ref{dv}) are negligible, and the encounter is solely described by the parameter $q=\Delta E/|E_b|$ given by Eq.~(\ref{x}).  Additionally, the density profile that results from stellar encounters in this regime is universal.  Figure~\ref{fig:profiles} demonstrates that Eq.~(\ref{profile}) with $\alpha=0.78$ can accurately fit the outcome of any succession of stellar encounters.  Consequently, it is not necessary to track the full succession of stellar encounters.  Instead, we need only to consider two cases:
\begin{enumerate}[label={(\arabic*)}]
	\item encounters with microhalos whose density profiles are given by Eq.~(\ref{profile}) with $\alpha=0.78$;
	\item encounters with NFW microhalos.
\end{enumerate}
The second case represents a microhalo's first stellar encounter, but it is the more complicated of the two scenarios because it changes the shape of the halo's density profile.  We begin by instead studying the first case, which corresponds to successive stellar encounters.  This scenario is simple because the microhalo's density profile after an encounter is purely rescaled from its profile beforehand.

\subsection{Successive encounters}\label{sec:successive}

We consider a stellar encounter with a microhalo that already experienced an encounter; that is, a microhalo with a density profile given by Eq.~(\ref{profile}) with $\alpha=0.78$.  Let $r_s$ and $\rho_s$ be the scale parameters of the microhalo prior to the encounter and $r_s^\prime$ and $\rho_s^\prime$ be its parameters afterward; the encounter parameter $q$ is defined by Eq.~(\ref{x}) using $r_s$ and $\rho_s$.  Our objective is to find $r_s^\prime$ and $\rho_s^\prime$ as functions of $r_s$, $\rho_s$, and $q$.

\begin{figure}[t]
	\includegraphics[width=\columnwidth]{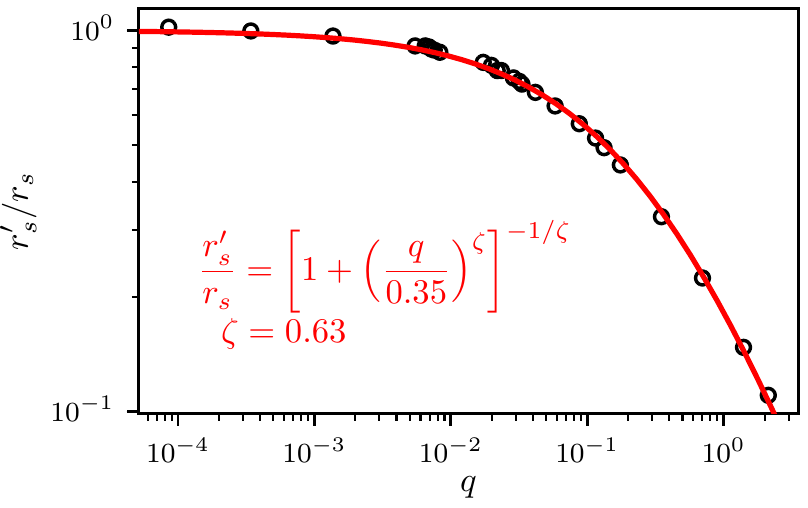}
	\caption{\label{fig:x_re} The change $r_s^\prime/r_s$ in a microhalo's scale radius in response to an encounter with parameter $q$.  We plot our simulations as circles, and the relationship between $r_s^\prime/r_s$ and $q$ is well fit (solid line) by the equation shown.}
\end{figure}

\begin{figure}[t]
	\includegraphics[width=\columnwidth]{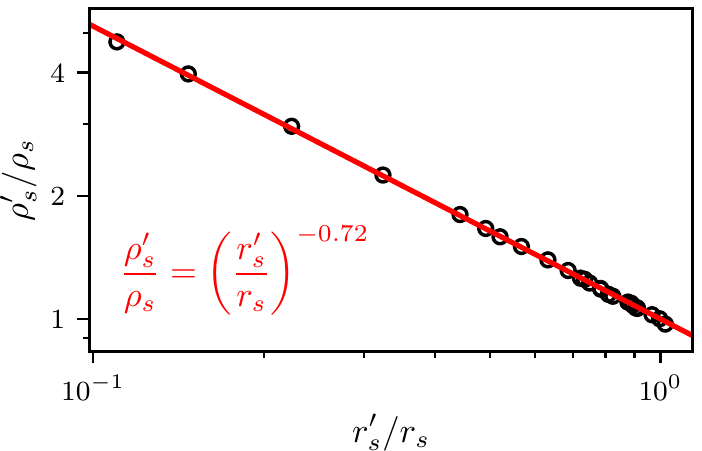}
	\caption{\label{fig:rhovsr_re} The change $\rho_s^\prime/\rho_s$ in a microhalo's scale density in response to an impulsive encounter, as a function of the change $r_s^\prime/r_s$ in its radius.  This relationship is evidently a power law (solid line).  Each circle represents a simulation.}
\end{figure}

We employ the 26 of our 36 $b\gg r_s$ simulations for which the initial halo is a remnant from a previous encounter.  We obtain the scale parameters of the initial and final halos by fitting Eq.~(\ref{profile}), with $\alpha=0.78$ enforced, to the stabilized part of the density profile (see Fig.~\ref{fig:profile}).  In Fig.~\ref{fig:x_re}, we plot the radius change $r_s^\prime/r_s$ as a function of the relative energy injection $q$.  This relationship is well fit by the remarkably simple form
\begin{equation}\label{rs}
\frac{r_s^\prime}{r_s}=\left[1+\left(\frac{q}{q_0}\right)^{\zeta}\right]^{-1/\zeta}
\end{equation}
with just the two fitting parameters $q_0=0.35$ and ${\zeta=0.63}$.  Meanwhile, Fig.~\ref{fig:rhovsr_re} shows that the density change $\rho_s^\prime/\rho_s$ is a power law in $r_s^\prime/r_s$,
\begin{equation}\label{rhos}
\frac{\rho_s^\prime}{\rho_s}=\left(\frac{r_s^\prime}{r_s}\right)^{-\eta}
\end{equation}
with $\eta=0.72$.


There is a theoretical reason to expect the behavior in Eq.~(\ref{rs}) wherein $r_s^\prime\propto q^{-1}$ for $q\gg 1$.  In the $q\gg 1$ regime, all material above the halo's scale radius $r_s$ is fully stripped, and only the halo's inner $\rho\propto r^{-1}$ density profile is relevant to dynamics.  For this density profile circular velocities scale as $r^{1/2}$, so the characteristic energies of particles at radius $r$ scale as $E\propto r$.  Meanwhile, the energy injection $\Delta E$ on a particle at radius $r$ is proportional to $q r^2$.  If the radius $r_s^\prime$ of the halo remnant is proportional to the radius $r$ at which $\Delta E = E$, then $r_s^\prime \propto q^{-1}$.

At $q\ll 1$, Eq.~(\ref{rs}) has another useful interpretation.  In this regime, 
\begin{equation}\label{rsx}
\frac{r_s^\prime}{r_s}=\exp\left[-\frac{1}{\zeta}\left(\frac{q}{q_0}\right)^{\zeta}\right],
\ \text{if}\ q\ll 1.
\end{equation}
Consider two successive encounters with the same $q$, each producing a change $r_s^\prime/r_s$ in the target halo's scale radius.  Since the density profile does not change significantly with each encounter, the net change in $r_s$ is $(r_s^\prime/r_s)^2$.  Now suppose instead that the two encounters happen simultaneously.  While the geometry of the two encounters sets precisely how their velocity injections add together, on average, we expect the energy injection $\Delta E$ to double relative to a single encounter.\footnote{Additivity of energy injections follows from the theory of random walks.  If velocity injections are in random directions, then their squared magnitudes are additive, on average.}  But $q\propto \Delta E$, so Eq.~(\ref{rsx}) implies that the net change in $r_s$ is $(r_s^\prime/r_s)^{2^\zeta}$ in this scenario.  Since $2^\zeta\simeq 1.55 < 2$, this calculation tells us precisely how much less efficiently a halo is altered by simultaneous stellar encounters than by successive ones.
A halo's postencounter relaxation makes it more susceptible to subsequent encounters.  In Appendix~\ref{sec:multiple} we show that two encounters can be treated as simultaneous if they occur within a few dynamical time intervals defined by Eq.~(\ref{tdyn}).  Section~\ref{sec:fields} will explore how to quantify this behavior more precisely.

Despite the evident precision of the relationships depicted in Figs. \ref{fig:x_re} and~\ref{fig:rhovsr_re}, it is unclear whence the values $\zeta=0.63$ and $\eta=0.72$ in Eqs. (\ref{rs}) and~(\ref{rhos}) arise.  The change in a microhalo's density profile is set by a complicated combination of the initial energy injection and subsequent relaxation.  In Appendix~\ref{sec:phase} we explore how the distribution of particle energies is altered by a stellar encounter, but we find no simple interpretation that describes it.

\subsection{First encounter}\label{sec:nfw}

Next, we explore how an NFW microhalo with parameters $\rho_s^\mathrm{NFW}$ and $r_s^\mathrm{NFW}$ evolves through a stellar encounter into a microhalo whose profile is given by Eq.~(\ref{profile}) with $\alpha=0.78$ and scale parameters $\rho_s^\prime$ and $r_s^\prime$.  We define the encounter parameter $q^\mathrm{NFW}$ as the parameter obtained through Eq.~(\ref{x}) using $\rho_s^\mathrm{NFW}$ and $r_s^\mathrm{NFW}$.  From the results of the previous section, we anticipate that $r_s^\prime/r_s^\mathrm{NFW}$ and $\rho_s^\prime/\rho_s^\mathrm{NFW}$ should follow similar functional forms to Eqs. (\ref{rs}) and~(\ref{rhos}).  To make the connection explicit, we make the ansatz that the following two scenarios yield a microhalo with exactly the same profile parameters $\rho_s^\prime$ and $r_s^\prime$:
\begin{enumerate}[label={(\arabic*)}]
	\item an NFW microhalo with parameters $r_s^\mathrm{NFW}$ and $\rho_s^\mathrm{NFW}$ experiencing an encounter with mass $M_*$, velocity $V$, and impact parameter $b$;
	\item a microhalo whose density profile is given by Eq.~(\ref{profile}) with $\alpha=0.78$ and scale parameters $r_s=Ar_s^\mathrm{NFW}$ and $\rho_s=B\rho_s^\mathrm{NFW}$ experiencing the same encounter, where $A$ and $B$ are universal.
\end{enumerate}
This ansatz implies $q=B^{-1}q^\mathrm{NFW}$ in the $b\gg r_s$ regime.  We additionally assume that the two halos have the same asymptote, $\rho_s^\mathrm{NFW} r_s^\mathrm{NFW}=\rho_s r_s$, which implies $B=A^{-1}$.

\begin{figure}[t]
	\includegraphics[width=\columnwidth,right]{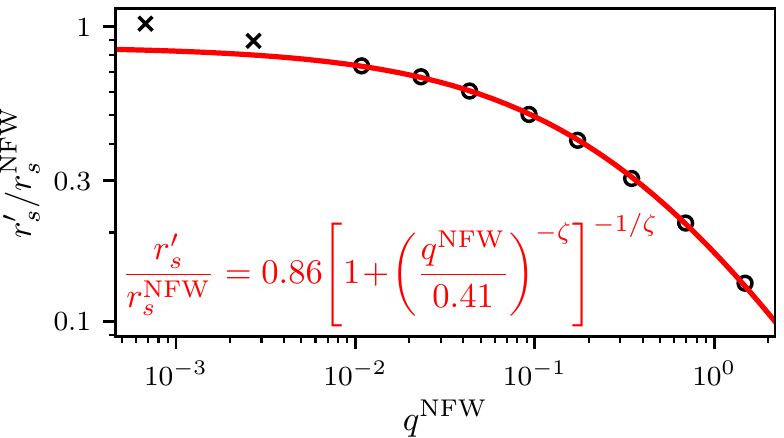}
	\includegraphics[width=.971\columnwidth,right]{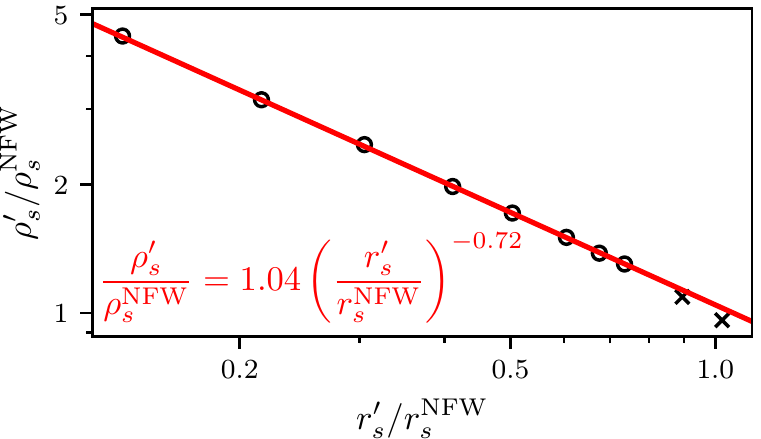}
	\caption{\label{fig:init} Evolution of NFW microhalos through stellar encounters.  Top: The sensitivity of the scale radius $r_s^\prime$ of the final halo to properties of the initial halo and the encounter.  Bottom: The dependence of the change in scale density on the change in scale radius.  For encounters with $q^\mathrm{NFW}\gtrsim 10^{-2}$ (circles), we fit Eqs. (\ref{rs_NFW}) and~(\ref{rhos_NFW}) with only one free parameter, as discussed in the text; the fits are shown as solid lines.  Encounters with $q^\mathrm{NFW}\lesssim 10^{-2}$ (crosses) do not obey the same relationships, an outcome we attribute to the density profile not fully evolving from NFW to the form given by Eq.~(\ref{profile}).}
\end{figure}

Using this ansatz, Eqs. (\ref{rs}) and~(\ref{rhos}) imply that
\begin{equation}\label{rs_NFW}
\frac{r_s^\prime}{r_s^\mathrm{NFW}}=A\left[1+\left(\frac{q^\mathrm{NFW}}{Bq_0}\right)^{\zeta}\right]^{-1/\zeta}
\end{equation}
and
\begin{equation}\label{rhos_NFW}
\frac{\rho_s^\prime}{\rho_s^\mathrm{NFW}}=BA^{\eta}\left(\frac{r_s^\prime}{r_s^\mathrm{NFW}}\right)^{-\eta}
\end{equation}
with $B=A^{-1}$ and the same $q_0=0.35$, $\zeta=0.63$, and $\eta=0.72$.  We now use our ten simulations of stellar encounters with NFW microhalos to test these relationships.  Figure~\ref{fig:init} shows how the final density profile depends on $r_s^\mathrm{NFW}$, $\rho_s^\mathrm{NFW}$, and $q^\mathrm{NFW}$.  We find that as long as $q^\mathrm{NFW}\gtrsim 10^{-2}$ the anticipated relationship is borne out, and by fitting Eqs. (\ref{rs_NFW}) and~(\ref{rhos_NFW}) to this regime, we obtain $A=0.86$ and $B=A^{-1}=1.17$.  For $q^\mathrm{NFW}\lesssim 10^{-2}$, on the other hand, the final density profiles deviate from these relationships.  An interpretation of this outcome is that too little energy is injected to fully convert the halo from the NFW profile to the form given by Eq.~(\ref{profile}).

\begin{figure}[t]
	\includegraphics[width=\columnwidth]{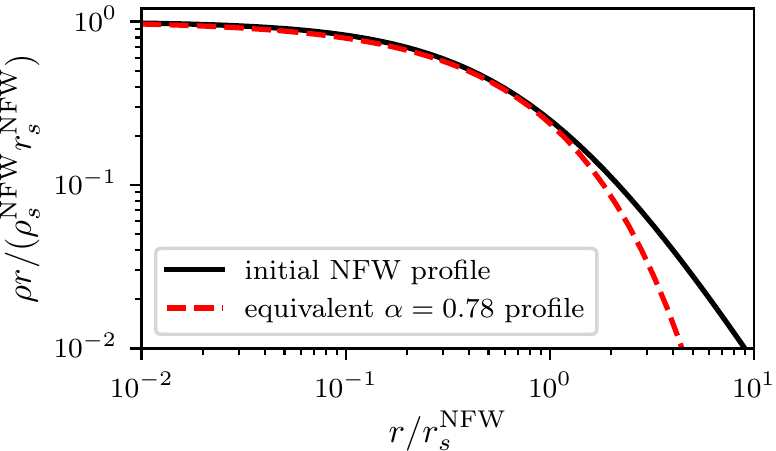}
	\caption{\label{fig:initial} A comparison between an NFW profile and the ``equivalent'' profile (see the text) given by Eq.~(\ref{profile}) with $\alpha=0.78$ using the relations in Eq.~(\ref{initial}).  The two profiles are nearly identical except at large radii.}
\end{figure}

In conclusion, as long as a microhalo's particles experience minimal energy injections of order $1/100$ their binding energy, an NFW microhalo with scale parameters $r_s^\mathrm{NFW}$ and $\rho_s^\mathrm{NFW}$ can be treated as having the density profile given by Eq.~(\ref{profile}) with $\alpha=0.78$ and scale parameters
\begin{equation}\label{initial}
r_s = 0.86 r_s^\mathrm{NFW}
\ \ \text{and}\ \ 
\rho_s = 1.17 \rho_s^\mathrm{NFW}.
\end{equation}
In a realistic scenario in which a microhalo experiences tidal forces not only from stars but also from other substructures and the galactic host, we expect that this condition will usually be satisfied.  Figure~\ref{fig:initial} shows a comparison between these two density profiles.  The equivalent Eq.~(\ref{profile}) density profile drops off more quickly than the NFW profile at large radii, but the two profiles are otherwise nearly identical.

\section{Penetrative encounters}\label{sec:b}

We now turn to penetrative encounters with $b\lesssim r_s$.  In this regime, two new effects become important.
\begin{enumerate}[label={(\arabic*)}]
	\item Nonlinear terms in the velocity injections given by Eq.~(\ref{dv}) become significant.
	\item Equation~(\ref{dv}) no longer accurately describes the velocity injection relative to the halo remnant's center of mass because it describes the velocity injection relative to that on the halo's initial center, whose motion may differ.\footnote{Equation~(\ref{dv}) is still suitable for initializing simulations, with the caveat that it can induce a bulk velocity on the halo remnant.}
\end{enumerate}
We anticipate that our definition of the parameter $q$ in Sec.~\ref{sec:param} will account for the second effect.  However, it is not clear how to account for the first.

\begin{figure}[t]
	\includegraphics[width=\columnwidth]{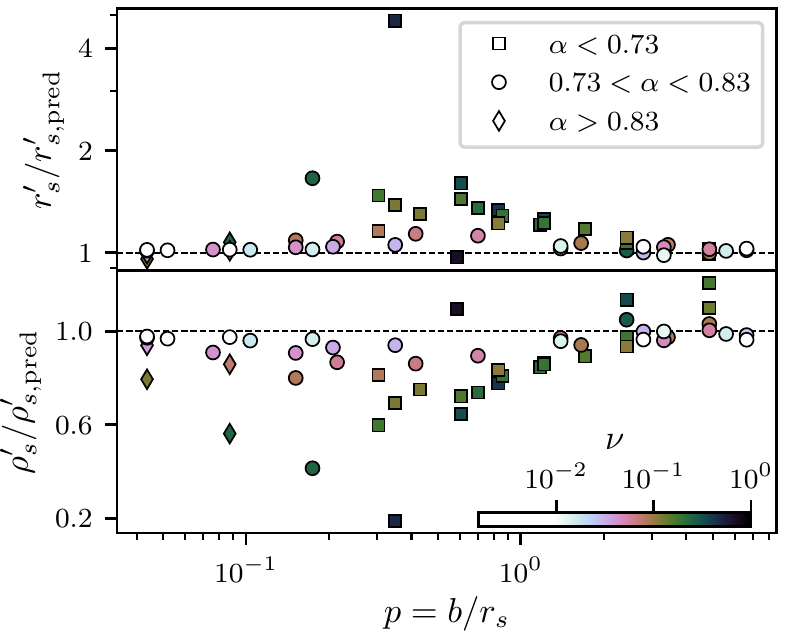}
	\caption{\label{fig:b} A test of the predictions from Eqs. (\ref{rs}) and~(\ref{rhos}) for encounters with $b\lesssim r_s$.  These predictions work well for a broad range of parameters, but they fail when $\nu=p^{-1}(1+q^{-1})^{-1/2}\gtrsim 0.1$ (color scale).  The relative contributions of nonlinear terms in the velocity injections are of order $\nu$, so we find that our predictions are accurate as long as these nonlinear terms can be neglected.}
\end{figure}

We simulated 60 encounters with $p=b/r_s<8$, as shown in Fig.~\ref{fig:sims}.  After each encounter, we fit Eq.~(\ref{profile}) to the microhalo's density profile to obtain its scale parameters $r_s$ and $\rho_s$.  Additionally, we find that it is now necessary to allow the parameter $\alpha$ to vary.  Figure~\ref{fig:b} compares the scale parameters obtained in these simulations to those predicted from Eqs. (\ref{rs}) and~(\ref{rhos}).  Evidently, the model developed for the $b\gg r_s$ regime accurately describes evolution by stellar encounters for a large portion of the $b\lesssim r_s$ regime, including cases where $b\ll r_s$.  However, there are some encounters for which it predicts wildly inaccurate results.

To understand where our predictions fail, we investigate when nonlinear terms in the velocity injections become important.  For a halo with characteristic internal velocities $v$, Eq.~(\ref{dvv}) implies that the characteristic velocity injection is $\Delta v\propto q^{1/2}[1+\mathcal{O}(p^{-1})]v$, which we can separate into linear part $\Delta v_\mathrm{lin}\propto q^{1/2}v$ and nonlinear part $\Delta v_\mathrm{nl}\propto q^{1/2}p^{-1}v$.  If the nonlinear terms can be neglected, then the characteristic final velocity is $v^\prime_\mathrm{lin}= \sqrt{v^2+\Delta v_\mathrm{lin}^2}$ (where we assume the direction of the particle's velocity injection is random relative to its velocity).  Hence, the relative contribution of nonlinear terms is of order
\begin{equation}\label{nl}
\Delta v_\mathrm{nl}/v^\prime_\mathrm{lin}\sim \nu\equiv p^{-1}(1+q^{-1})^{-1/2}.
\end{equation}

\begin{figure}[t]
	\includegraphics[width=\columnwidth]{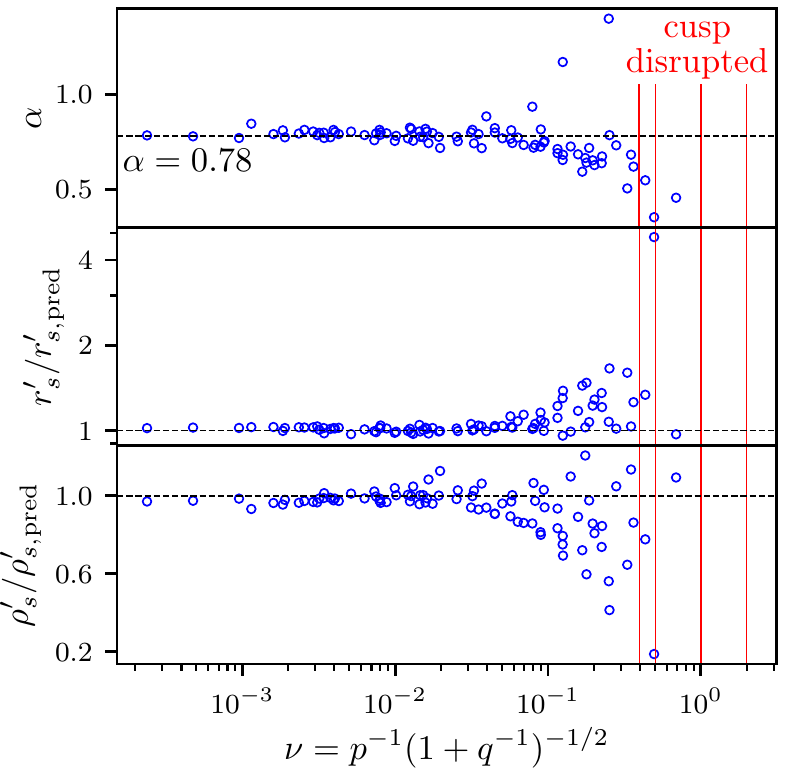}
	\caption{\label{fig:alpha} The impact of nonlinear terms in the velocity injections, which have relative magnitude of order $\nu$.  Top: The parameter $\alpha$ of the best-fitting density profile given by Eq.~(\ref{profile}).  Middle and bottom: Deviations from the predicted values of $r_s$ and $\rho_s$ using Eqs. (\ref{rs}) and~(\ref{rhos}).  When $\nu\gtrsim 0.1$, deviations start to appear in all three parameters.  Additionally, when $\nu\gtrsim 1$, the central $\rho\propto r^{-1}$ cusp can be disrupted (red lines; see Fig.~\ref{fig:disrupt}).  There is no fit for these cases.}
\end{figure}

\begin{figure}[t]
	\includegraphics[width=\columnwidth]{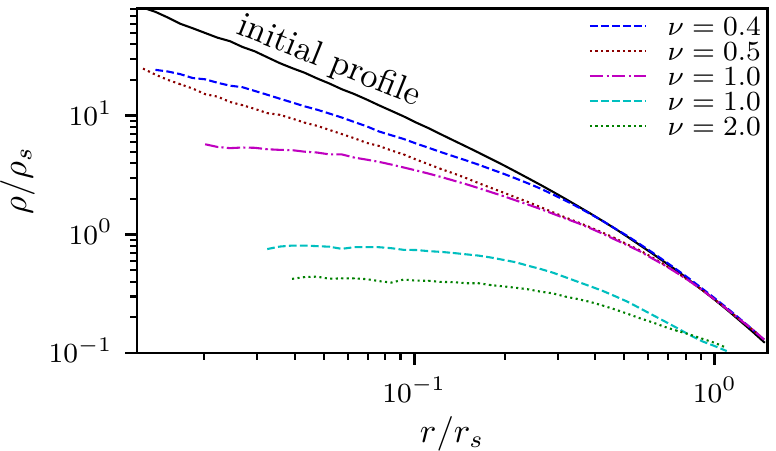}
	\caption{\label{fig:disrupt} The density profiles of microhalos whose central cusps are disrupted by stellar encounters with $\nu=p^{-1}(1+q^{-1})^{-1/2}\sim 1$.  The logarithmic slopes $\gamma$ of their $\rho\propto r^{-\gamma}$ inner profiles become smaller than 1, and in some cases, a uniform-density core develops ($\gamma=0$).  One disruption scenario, with $\nu=1$, is excluded from the plot because the resulting core density is too small.  The axes units are the scale density and radius of the halo prior to the encounter.}
\end{figure}

We indicate the value of $\nu$ in Fig.~\ref{fig:b} with a color scale, and Fig.~\ref{fig:alpha} shows the impacts of the relative nonlinearity $\nu$ more directly.  If $\nu\ll 1$, then nonlinear effects are unimportant, and surely enough, we find that deviations from the predictions of Eqs. (\ref{rs}) and~(\ref{rhos}) are minimal when $\nu\lesssim 0.1$.  On the other hand, when $\nu\gtrsim 0.1$, these deviations can become large.  Additionally, the density profile is no longer universal when $\nu\lesssim 0.1$; the best-fitting value of $\alpha$ can deviate significantly from $\alpha=0.78$.  When $\nu\gtrsim 1$, the halo's central $\rho\propto r^{-1}$ density cusp can even be disrupted so that the logarithmic slope $\gamma$ of its $\rho\propto r^{-\gamma}$ inner profile becomes smaller than 1.  In some cases, a uniform-density core ($\gamma=0$) results.  These disruption scenarios are indicated in Fig.~\ref{fig:alpha} in red, and Fig.~\ref{fig:disrupt} shows the resulting density profiles.

The precise sensitivity of the density profile to $q$ and $p$ after an encounter with $\nu\gtrsim 0.1$ is complicated.  However, as an approximate treatment, the predictions of Eqs.~(\ref{rs}) and~(\ref{rhos}) are reasonably accurate for $\nu\lesssim 1/3$, and when $\nu\gtrsim 1/3$, the resulting halo is close to disruption.  We will see in the next section that the $\nu\gtrsim 0.1$ regime is relatively unimportant to typical microhalo scenarios.

\section{Stellar fields}\label{sec:fields}

We now explore the implications of the model given by Eqs. (\ref{rs}) and~(\ref{rhos}) for microhalos passing through fields of stars.  As a representative example, we study microhalos traversing the solar neighborhood.  We take the microhalos to have NFW scale parameters $\rho_s^\mathrm{NFW}=1.7\times 10^9\ M_\odot/\text{kpc}^{3}$ and $r_s^\mathrm{NFW}=5\times 10^{-6}\ \text{kpc}$.  These parameters correspond to halos with mass $m_\mathrm{vir}=10^{-6}\ M_\odot$ and concentration $r_\mathrm{vir}/r_s=2$ at redshift $z=32$, which are typical parameters for the smallest halos in a cold dark matter scenario (e.g., Ref.~\cite{ishiyama2019abundance}).  Meanwhile, the stellar mass density of the Galactic disk at the sun's altitude is roughly $\rho_*=4\times 10^7$ $M_\odot/\text{kpc}^3$ \cite{binney1987galactic}.  We assume the microhalos have velocity $V_\mathrm{halo}=200\ \text{km}/\text{s}$ relative to the disk while the stars have mass $M_*=0.5M_\odot$ and velocity dispersion $\sigma_*=50\ \text{km}/\text{s}$ within the disk.  We consider a total duration of $t=160\ \text{Myr}$, which is roughly the amount of time microhalos whose orbits cross the solar vicinity are expected to spend inside the disk over the Galactic age \cite{schneider2010impact}.\footnote{For clarity, we note that the vast majority of microhalos within the Galactic halo are not expected to encounter a star due to the small relative volume occupied by the disk.  In this section we specifically treat microhalos that cross the solar vicinity.}

We sample stellar encounter positions uniformly within the cylinder of radius $b_\mathrm{max}=80 r_s$ and length $V_\mathrm{halo}t$.  For our scenario this choice of $b_\mathrm{max}$ implies that only encounters with relative energy injection $q\lesssim 10^{-7}$ are excluded.  A total of 1305 encounters are expected within this volume, and we sample the encounter count from the corresponding Poisson distribution.  Each encounter velocity $V$ is the vector sum of $V_\mathrm{halo}$ and a stellar velocity $V_*$ randomly sampled using the stellar velocity dispersion.

\begin{figure}[t]
	\includegraphics[width=\columnwidth]{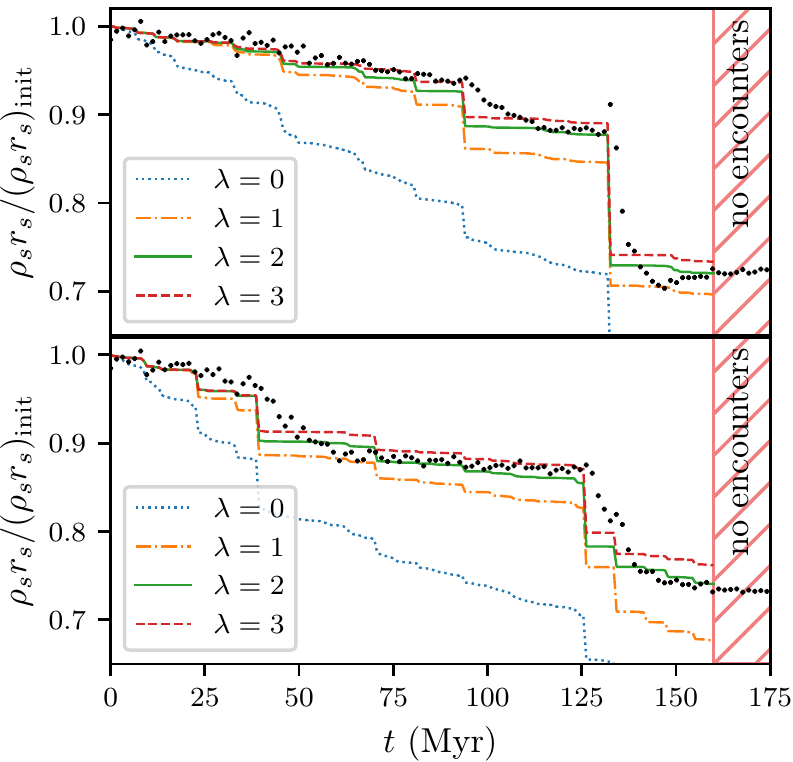}
	\caption{\label{fig:sf} Comparing model predictions to simulations for a microhalo traversing a field of stars.  We plot the evolution of the density profile asymptote $\rho_s r_s$ for a microhalo crossing two random stellar distributions (upper and lower panels) representative of the solar neighborhood.  Predictions using Eqs. (\ref{rs+}) and~(\ref{rhos}) are shown as solid, dashed, dot-dashed, and dotted lines for various values of the parameter $\lambda$ (see the text).  For comparison, the black circles represent simulation results, where the asymptote is determined by fitting Eq.~(\ref{profile}) with $\alpha=0.78$.  These points are averaged over ten simulations, each with different randomized encounter orientations, and the simulations continue after the last encounters (hatched region) to allow the halo to relax.  The predictions for $\lambda=2$ match the simulations reasonably well outside of relaxation periods occurring after major encounters.  We express the asymptote in units of the initial asymptote $(\rho_s r_s)_\mathrm{init}$.}
\end{figure}

The model developed in the previous sections may now be applied to this scenario, but there is a complication.  The dynamical timescale of the initial halo is about 8 Myr, so a large number of stellar encounters are expected to occur within each dynamical time interval.  Meanwhile, as we found in Appendix~\ref{sec:multiple}, encounters should be treated as simultaneous if they occur within a few dynamical time intervals $t_\mathrm{dyn}$ defined by Eq.~(\ref{tdyn}).  To accommodate this requirement we adopt the following procedure.  For an encounter $i$ occurring at time $t_i$, we consider all $n$ encounters (including the encounter $i$) within the previous time interval $t_i-\Delta t<t<t_i$, where
\begin{equation}\label{lambda}
\Delta t = \lambda t_\mathrm{dyn}
\end{equation}
for some number $\lambda$.  Using these $n$ encounters we compute two effective encounter parameters:
\begin{equation}\label{xeff}
q_\mathrm{eff}^{+}=\!\!\sum_{j=i-n+1}^i\!\!\!\! q_j
\ \ \text{and}\ \ 
q_\mathrm{eff}^{-}=q_\mathrm{eff}^{+}- q_i.
\end{equation}
$q_\mathrm{eff}^{+}$ is the combined relative energy injection from all encounters within the last time interval $\Delta t$ including the $i$th encounter, while $q_\mathrm{eff}^{-}$ excludes the $i$th encounter.  Rather than apply the scaling prescribed by Eq.~(\ref{rs}) using the $i$th encounter's parameter $q_i$, we apply this scaling using $q_\mathrm{eff}^{+}$ and the reciprocal scaling using $q_\mathrm{eff}^{-}$.  In other words, we take the $i$th encounter to change the microhalo's scale radius $r_s$ by the factor
\begin{equation}\label{rs+}
\frac{r_s^\prime}{r_s}=\left[\frac{1+(q_\mathrm{eff}^{+}/q_0)^{\zeta}}{1+(q_\mathrm{eff}^{-}/q_0)^{\zeta}}\right]^{-1/\zeta}.
\end{equation}
This procedure treats encounters occurring within the time interval $\lambda t_\mathrm{dyn}$ as simultaneous in a self-consistent way.\footnote{It is easy to see that this procedure yields the desired results in two limiting cases.  If a cluster of encounters occurs within $\lambda t_\mathrm{dyn}$, then each encounter cancels the effect of the previous one, and the final encounter applies the scaling given by Eq.~(\ref{rs}) using the summed energy injection.  If an encounter is separated from all others by intervals longer than $\lambda t_\mathrm{dyn}$, then $q_\mathrm{eff}^{-}=0$, and the scaling given by Eq.~(\ref{rs}) is applied using that encounter's energy injection alone.}  We use Eq.~(\ref{initial}) to initially rescale the microhalo parameters, and for two random stellar encounter distributions, Fig.~\ref{fig:sf} shows the predicted microhalo evolution using this procedure for several values of $\lambda$.

To test this procedure and tune the parameter $\lambda$, we instructed the \textsc{Gadget-2} simulation code to apply velocity injections given by Eq.~(\ref{dv}) according to a preset list of stellar encounters.  These velocity injections are computed taking the point of least potential as the origin, and each encounter's spatial orientation is randomized.  With this arrangement, we carried out numerical simulations of the two stellar field scenarios depicted in Fig.~\ref{fig:sf}.  In order to facilitate direct comparison, we began the simulations with a microhalo that was a remnant from a previous stellar encounter; this remnant was rescaled to have the correct scale parameters predicted by Eq.~(\ref{initial}).  We simulated ten instances of each scenario with different encounter orientations, and Fig.~\ref{fig:sf} shows the orientation-averaged evolution of the microhalo's inner density asymptote $\rho_s r_s$ in these simulations as computed by fitting Eq.~(\ref{profile}) with $\alpha=0.78$ to the density profile.\footnote{Specifically, we use the density profile of the instantaneous bound remnant computed using the procedure in Sec.~\ref{sec:sims}, and we only fit out to the radius at which $\rho r = \rho_s r_s/3$.  We picked the asymptote $\rho_s r_s$ because it is the quantity least sensitive to details of the fit.}  Outside of relaxation periods after major encounters, Fig.~\ref{fig:sf} shows that our predictions with $\lambda=2$ match the simulation results well.

\begin{figure}[t]
	\includegraphics[width=\columnwidth]{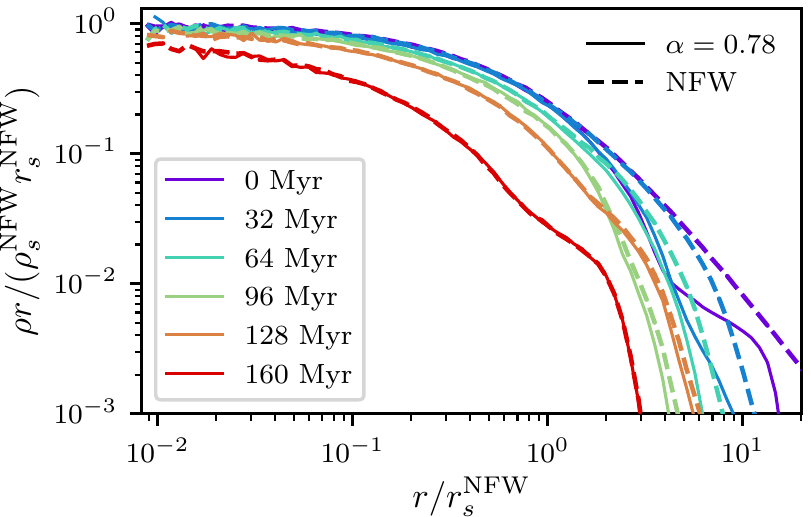}
	\caption{\label{fig:sfnfw} Equivalence of the density profile given by Eq.~(\ref{profile}) with $\alpha=0.78$ to the NFW density profile, for the purpose of stellar encounters, if the two profiles are related by Eq.~(\ref{initial}).  We subjected microhalos with these two density profiles to the same stellar field scenario (corresponding to the upper panel of Fig.~\ref{fig:sf}), and the density profiles that result are identical except at large radii.  The profiles plotted here include only particles bound to the halo as computed using the procedure in Sec.~\ref{sec:sims}.}
\end{figure}

We also subjected a microhalo with an NFW profile to one of the same series of stellar encounters, and we compare this halo's evolution to that of the stellar encounter remnant with the profile given by Eq.~(\ref{profile}) with $\alpha=0.78$.  Figure~\ref{fig:sfnfw} shows the evolution of the two density profiles.  Except at large radii, the resulting halos develop identical density profiles, which further confirms the accuracy of the scaling given by Eq.~(\ref{initial}).

\begin{figure}[t]
	\includegraphics[width=\columnwidth]{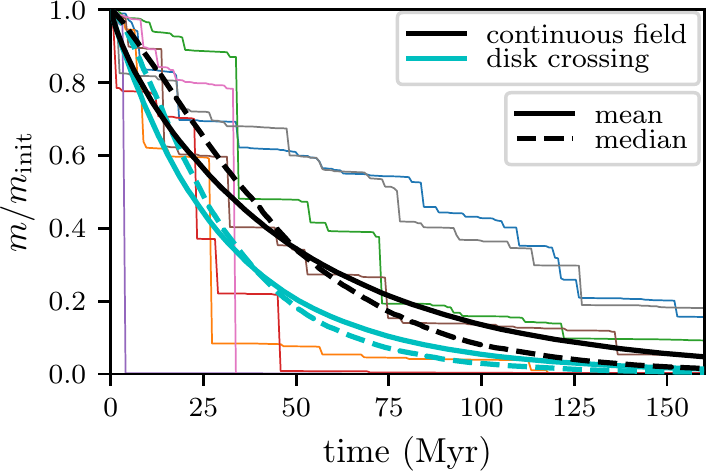}
	\caption{\label{fig:solar} Mass evolution for microhalos crossing the solar neighborhood as predicted using Eqs. (\ref{rs+}) and (\ref{rhos}).  The random distribution of stars induces variance in the mass evolution; we plot eight example mass trajectories along with the mean and median of 10\,000 trajectories (black curves).  The cyan curves represent the trajectories if the halo is allowed to fully relax every 2 Myr, which corresponds to a typical disk-crossing time interval.  We express the mass in units of $m_\mathrm{init}$, which we take to be the mass of the halo after the initial rescaling given by Eq.~(\ref{initial}).}
\end{figure}

Finally, to illustrate the power of our predictive framework, we sample 10\,000 random sequences of stellar encounters for microhalos traversing the solar neighborhood.  Figure~\ref{fig:solar} shows the predicted trajectory of the mass
\begin{equation}\label{mass}
m=4\pi \alpha^{2/\alpha-1}\Gamma(2/\alpha)\rho_s r_s^3\simeq 11.9\rho_s r_s^3
\end{equation}
(for $\alpha=0.78$; $\Gamma$ is the gamma function) of microhalos subjected to different sequences.  We plot the median and mean mass trajectories along with eight examples.  There is substantial variation between halos, but in general, the microhalos lose almost all of their original mass by $t=160\ \text{Myr}$.  These results are similar to those of Ref.~\cite{schneider2010impact}, which simulated a similar scenario.  We note, however, that in the true solar-neighborhood scenario halos should be allowed to fully relax between disk crossings.  In Fig.~\ref{fig:solar}, we also plot the mean and median mass trajectories if the halo is allowed to fully relax every 2~Myr; that is, the summation in Eq.~(\ref{rs+}) is made to exclude any encounters prior to the 2~Myr interval.  The trajectories change significantly in this scenario, implying it is important to properly account for the halo's relaxation.

Our predictions assume that encounters are
\begin{enumerate}[label={(\arabic*)}]
	\item in the impulsive regime with $t_\mathrm{dyn}\gtrsim 5 b/V$; and
	\item in the linear regime with $\nu= p^{-1}(1+q^{-1})^{-1/2}\lesssim 0.1$.
\end{enumerate}
In the solar neighborhood scenario, all encounters are impulsive with $t_\mathrm{dyn}\sim 10^3 b/V$, as expected from the discussion in Sec.~\ref{sec:sims}.  Roughly half of the microhalo instances experience encounters with $\nu>0.1$, but their impact turns out to be minimal.  We tested an alternative procedure where any encounter with $\nu>0.1$ was assumed to destroy the halo, and the mean and median trajectories in Fig.~\ref{fig:solar} did not move appreciably.  Evidently, the halos that underwent these encounters were already effectively destroyed by them.  Thus, neither requirement significantly hinders the model's applicability to this scenario.

\section{Conclusion}\label{sec:conclusion}

In this work, we developed a framework that can predict the evolution of microhalo density profiles as a result of successive stellar encounters.  We found that the density profiles of microhalos subjected to stellar encounters follow an almost universal form given by
\begin{equation}
\rho=\rho_s \frac{r_s}{r} \exp\left[-\frac{1}{\alpha}\left(\frac{r}{r_s}\right)^\alpha\right]
\end{equation}
with $\alpha=0.78$, and Eq.~(\ref{initial}) describes how this form is related to the initial NFW profile.  If each stellar encounter is parametrized by the energy it injects using Eq.~(\ref{x}), then Eqs. (\ref{rs}) and~(\ref{rhos}) describe the microhalo's response to that energy injection.  Successive encounters occurring within roughly $\lambda=2$ dynamical time intervals should be treated as simultaneous and their energy injections added; Sec.~\ref{sec:fields} discusses how to implement this effect.  This framework is accurate assuming that encounters are impulsive [see Eq.~(\ref{impulse})] and the resulting velocity injections are in the linear regime defined by Eq.~(\ref{nl}).  However, these conditions do not significantly hinder its applicability, as we discuss in Sec.~\ref{sec:fields}.

Through Monte Carlo methods, this model can rapidly characterize the impact of stellar encounters on whole ensembles of microhalos.  For instance, we were able to generate in minutes 10\,000 randomized realizations of a stellar field scenario similar to the single realization simulated in Ref.~\cite{schneider2010impact}.  In Ref.~\cite{delos2019gamma} (forthcoming), we use this model to aid in characterizing the microhalo-dominated dark matter annihilation signals that are expected to arise from certain early-universe scenarios.  In the process, we develop a method to combine the impact of stellar encounters with the subhalo tidal evolution modeled by Ref.~\cite{delos2019tidal}.

This framework is limited to microhalos initially possessing NFW density profiles.  While merger events drive halos' inner density profiles toward the $\rho\propto r^{-1}$ of the NFW profile \cite{ogiya2016dynamical,angulo2017earth,delos2019predicting}, it is possible that the smallest halos might retain steeper density cusps today.  An exploration of different density profiles is beyond this work's scope, but we anticipate that because of its simplicity, our model will extend readily, albeit with a possibly different universal density profile.  Additionally, this model describes microhalos subjected to stellar encounters alone.  Further work is needed to precisely understand the combined impact of stellar encounters, galactic tides, and other disruptive processes; Ref.~\cite{delos2019gamma} represents a start.  Nevertheless, the model presented here will enable more accurate characterizations of the microhalo population within galactic halos.

\begin{acknowledgments}
	The simulations for this work were carried out on the Dogwood computing cluster at the University of North Carolina at Chapel Hill.  This work was supported by a fellowship from the North Carolina Space Grant Consortium.  The author thanks Adrienne Erickcek and Tim Linden for helpful discussions.  Figure~\ref{fig:b} employs the cube-helix color scheme developed by Ref.~\cite{green2011colour}.
\end{acknowledgments}

\appendix

\section{Accuracy of the impulse approximation}\label{sec:impulse}

In this appendix, we explicitly simulate microhalo-stellar encounters in order to test the validity of the impulse approximation employed in Sec.~\ref{sec:sims}.  Reference~\cite{angus2007cold} used such simulations to demonstrate that the impulse approximation is accurate in the case where the stellar encounter timescale is of order $10^{-3}$ the microhalo dynamical timescale.  However, we aim to explore how far the impulse approximation can be taken.

\begin{figure}[t]
	\includegraphics[width=\columnwidth]{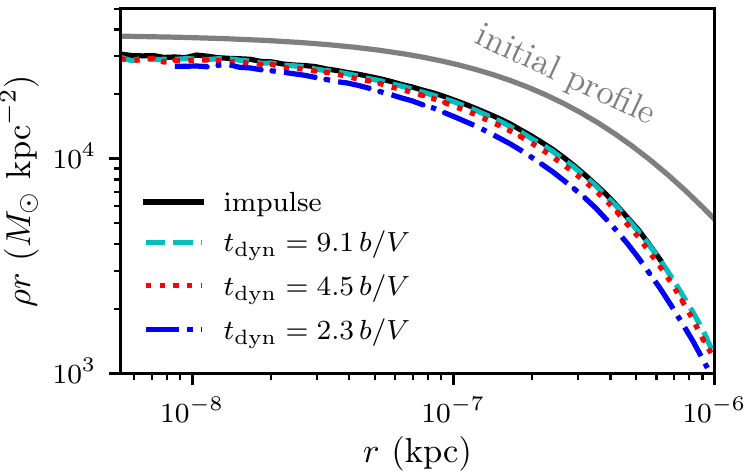}
	\caption{\label{fig:impulse} A test of the impulse approximation.  This figure shows the microhalo density profile resulting from several different stellar encounters; $b$ and $M_*/V$ are held fixed, but we vary the velocity $V$.  The impulse approximation (solid line) corresponds to $V\to\infty$, and we find that as long as the microhalo's internal dynamical timescale $t_\mathrm{dyn}\gtrsim 5 b/V$, the final density profile is identical to that resulting from the impulse approximation.}
\end{figure}

For these simulations, we set the encounter parameters  $b=16r_s$ and $M_*/(V b^2)=0.76\sqrt{\rho_s/G}$.  As discussed in Sec.~\ref{sec:param}, these parameters suffice to fully describe the encounter, at least in the impulsive regime, and they correspond to the parameters $q=0.093$ and $p=16$ as defined in that section.  With these parameters, we prepare an $N$-body microhalo as in Sec.~\ref{sec:sims}, but instead of perturbing the particle velocities using the impulse approximation, we insert the star as a point mass in the simulation.  The star is placed at position $(800r_s,b,0)$ with velocity $(-V,0,0)$, and the simulation is carried out for a duration of $1600 r_s/V$, so that since its trajectory is essentially unperturbed, the star's final position is $(-800r_s,b,0)$.  The maximum time step of simulation particles is enforced to be $0.2b/V$ so that there are at least $500$ time steps, and we verified that a naive numerical integral with this time stepping scheme accurately reproduces the analytic impulse approximation.

At the end of this simulation, we remove the star and subsequently continue the simulation for the duration $10t_\mathrm{dyn}$ with the same parameters as in Sec.~\ref{sec:sims}.  The time step is no longer artificially small.  We executed this procedure for several different encounter velocities $V$, and the resulting microhalo density profiles are shown in Fig.~\ref{fig:impulse}.  The validity of the impulse approximation can be conditioned on the comparison between the encounter timescale $b/V$ and the microhalo's internal dynamical timescale $t_\mathrm{dyn}$ given by Eq.~(\ref{tdyn}), and we find that deviations from the impulse approximation become significant only when $t_\mathrm{dyn}\lesssim 5 b/V$.  The heightened efficacy of slow encounters at altering a microhalo's structure, apparent in Fig.~\ref{fig:impulse}, can be understood in light of the tendency for postencounter relaxation to make a halo more susceptible to future encounters (see Sec.~\ref{sec:successive}).  If $t_\mathrm{dyn}\lesssim b/V$, the halo begins to relax even as the encounter progresses, increasing its susceptibility to that same encounter.

\section{Encounters in close succession}\label{sec:multiple}

\begin{figure}[t]
	\includegraphics[width=\columnwidth]{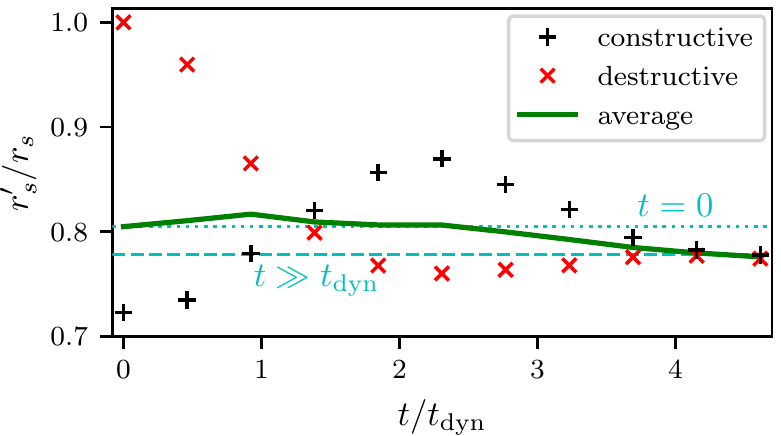}
	\caption{\label{fig:dyn} The impact of two encounters separated by a short time $t$.  We plot the change in $r_s$ for two encounter geometries, one in which the velocity injections add constructively and one in which they add destructively.  The solid line indicates the average between the two cases, computed as described in the text.  The dashed line marks the change in $r_s$ in the $t\gg t_\mathrm{dyn}$ scenario while the dotted line represents the expected impact of simultaneous encounters.  Evidently, encounters should be treated as simultaneous if they are separated by less than a few dynamical time intervals.}
\end{figure}

In this appendix, we test a microhalo's response to multiple encounters occurring within its dynamical timescale.  Recall that on average, encounters separated by a time interval $t\gg t_\mathrm{dyn}$ are more efficient than simultaneous encounters at altering the microhalo's structure.  To test the intermediate $t\lesssim t_\mathrm{dyn}$ regime, we carry out a series of simulations of two identical encounters separated by varying time $t$.  The impact of two closely spaced encounters is highly sensitive to the geometry between them, so we consider two extreme cases.  In the first case, the encounters are collinear, so their velocity injections add constructively.  In the second case, the velocity injections add destructively; this is attained by making them parallel but in perpendicular directions from the halo [e.g., swapping the $y$ and $z$ coordinates in the scenario of Eq.~(\ref{dv})].

The encounters we study here have relative energy injection parameter $q=1/120$, and we plot in Fig.~\ref{fig:dyn} the changes in the microhalo's scale radius $r_s$ that result from these scenarios.  At each time separation $t$, we also compute the average between the constructive and destructive scenarios in the following way.  We compute an effective parameter $q_\mathrm{eff}$ for each double-encounter scenario by inverting Eq.~(\ref{rs}).  Next, we average the effective energy injections $q_\mathrm{eff}$ for the constructive and destructive scenarios, and we use Eq.~(\ref{rs}) to convert the resulting average $q_\mathrm{eff}$ into a change in $r_s$.  This procedure automatically yields the correct expected impact of two encounters of arbitrary geometry in the $t=0$ case, but for $t>0$ it can be considered only a guide.  The result is also plotted in Fig.~\ref{fig:dyn}, and we find that, roughly speaking, two encounters can be treated as simultaneous if they are separated by less than a few dynamical time intervals.

\section{Impact of encounters on the phase-space distribution}\label{sec:phase}

\begin{figure}[t]
	\includegraphics[width=\columnwidth]{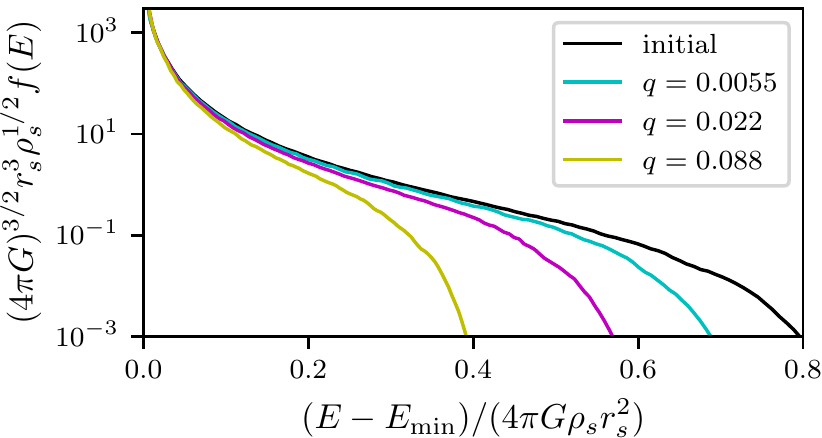}
	\caption{\label{fig:phase} Phase-space density $f(E)=\mathrm{d}^6 m/(\mathrm{d}^3\vec x \mathrm{d}^3\vec v)$, as a function of energy $E$ of halo particles, for a microhalo before and after different stellar encounters; the relative energy parameter $q$ associated with each encounter is listed.  $E$ and $f$ are normalized to properties of the initial density profile, and $E$ is additionally specified relative to the floor $E_\mathrm{min}$ of the potential well.  We choose an initial halo that has already experienced a stellar encounter, so it has a density profile described by Eq.~(\ref{profile}) with $\alpha=0.78$ instead of an NFW profile.  As discussed in Appendix~\ref{sec:phase}, simple interpretations fail to accurately describe how $f(E)$ changes in response to an encounter.}
\end{figure}

Figures \ref{fig:x_re} and~\ref{fig:rhovsr_re} depict a remarkably precise relationship between the relative energy injection parameter $q$ and the change in a microhalo's density profile.  This precision suggests that it should be possible to derive from first principles the parameters $\zeta=0.63$ and $\eta=0.72$ that set these relationships.  However, the change in the density profile is determined in a complicated way by both the initial heating of halo material and the subsequent relaxation of the halo, making such a derivation challenging.  As a demonstration, we plot in Fig.~\ref{fig:phase} the phase-space density $f(E)$ of a microhalo before and after various stellar encounters.  The function $f(E)$ after a stellar encounter is not related in an obvious way to $f(E)$ beforehand; for instance, it is not related, even approximately, by either truncation beyond some energy $E_t$ or a combination of that truncation with the lowering of $f(E)$ to maintain continuity at $E_t$, procedures suggested by Ref.~\cite{drakos2017phase}.

The key challenge to such a simple interpretation is that particles throughout the halo are heated by the stellar encounter, even those that remain tightly bound.  The resulting elevation of halo particles alters the gravitational potential even at small radii, which in turn further changes the energies $E$ of halo particles.  For this reason, Fig.~\ref{fig:phase} shows that $f(E)$ decreases significantly in response to an encounter even at small $E$, an effect these simple interpretations do not capture.  Due to these challenges, we leave to future work a first-principles derivation of Eqs. (\ref{rs}) and~(\ref{rhos}) and their parameters $\zeta$ and $\eta$.

\clearpage

\bibliography{references}

 \newcommand{\noop}[1]{}
\begin{thebibliography}{114}%
\makeatletter
\providecommand \@ifxundefined [1]{%
 \@ifx{#1\undefined}
}%
\providecommand \@ifnum [1]{%
 \ifnum #1\expandafter \@firstoftwo
 \else \expandafter \@secondoftwo
 \fi
}%
\providecommand \@ifx [1]{%
 \ifx #1\expandafter \@firstoftwo
 \else \expandafter \@secondoftwo
 \fi
}%
\providecommand \natexlab [1]{#1}%
\providecommand \enquote  [1]{``#1''}%
\providecommand \bibnamefont  [1]{#1}%
\providecommand \bibfnamefont [1]{#1}%
\providecommand \citenamefont [1]{#1}%
\providecommand \href@noop [0]{\@secondoftwo}%
\providecommand \href [0]{\begingroup \@sanitize@url \@href}%
\providecommand \@href[1]{\@@startlink{#1}\@@href}%
\providecommand \@@href[1]{\endgroup#1\@@endlink}%
\providecommand \@sanitize@url [0]{\catcode `\\12\catcode `\$12\catcode
  `\&12\catcode `\#12\catcode `\^12\catcode `\_12\catcode `\%12\relax}%
\providecommand \@@startlink[1]{}%
\providecommand \@@endlink[0]{}%
\providecommand \url  [0]{\begingroup\@sanitize@url \@url }%
\providecommand \@url [1]{\endgroup\@href {#1}{\urlprefix }}%
\providecommand \urlprefix  [0]{URL }%
\providecommand \Eprint [0]{\href }%
\providecommand \doibase [0]{http://dx.doi.org/}%
\providecommand \selectlanguage [0]{\@gobble}%
\providecommand \bibinfo  [0]{\@secondoftwo}%
\providecommand \bibfield  [0]{\@secondoftwo}%
\providecommand \translation [1]{[#1]}%
\providecommand \BibitemOpen [0]{}%
\providecommand \bibitemStop [0]{}%
\providecommand \bibitemNoStop [0]{.\EOS\space}%
\providecommand \EOS [0]{\spacefactor3000\relax}%
\providecommand \BibitemShut  [1]{\csname bibitem#1\endcsname}%
\let\auto@bib@innerbib\@empty
\bibitem [{\citenamefont {Ghigna}\ \emph {et~al.}(1998)\citenamefont {Ghigna},
  \citenamefont {Moore}, \citenamefont {Governato}, \citenamefont {Lake},
  \citenamefont {Quinn},\ and\ \citenamefont {Stadel}}]{ghigna1998dark}%
  \BibitemOpen
  \bibfield  {author} {\bibinfo {author} {\bibfnamefont {S.}~\bibnamefont
  {Ghigna}}, \bibinfo {author} {\bibfnamefont {B.}~\bibnamefont {Moore}},
  \bibinfo {author} {\bibfnamefont {F.}~\bibnamefont {Governato}}, \bibinfo
  {author} {\bibfnamefont {G.}~\bibnamefont {Lake}}, \bibinfo {author}
  {\bibfnamefont {T.}~\bibnamefont {Quinn}}, \ and\ \bibinfo {author}
  {\bibfnamefont {J.}~\bibnamefont {Stadel}},\ }\href {\doibase
  10.1046/j.1365-8711.1998.01918.x} {\bibfield  {journal} {\bibinfo  {journal}
  {Mon. Not. R. Astron. Soc.}\ }\textbf {\bibinfo {volume} {300}},\ \bibinfo
  {pages} {146} (\bibinfo {year} {1998})},\ \Eprint
  {http://arxiv.org/abs/astro-ph/9801192} {arXiv:astro-ph/9801192} \BibitemShut
  {NoStop}%
\bibitem [{\citenamefont {Tormen}\ \emph {et~al.}(1998)\citenamefont {Tormen},
  \citenamefont {Diaferio},\ and\ \citenamefont {Syer}}]{tormen1998survival}%
  \BibitemOpen
  \bibfield  {author} {\bibinfo {author} {\bibfnamefont {G.}~\bibnamefont
  {Tormen}}, \bibinfo {author} {\bibfnamefont {A.}~\bibnamefont {Diaferio}}, \
  and\ \bibinfo {author} {\bibfnamefont {D.}~\bibnamefont {Syer}},\ }\href
  {\doibase 10.1046/j.1365-8711.1998.01775.x} {\bibfield  {journal} {\bibinfo
  {journal} {Mon. Not. R. Astron. Soc.}\ }\textbf {\bibinfo {volume} {299}},\
  \bibinfo {pages} {728} (\bibinfo {year} {1998})},\ \Eprint
  {http://arxiv.org/abs/astro-ph/9712222} {arXiv:astro-ph/9712222} \BibitemShut
  {NoStop}%
\bibitem [{\citenamefont {Gao}\ \emph {et~al.}(2004)\citenamefont {Gao},
  \citenamefont {White}, \citenamefont {Jenkins}, \citenamefont {Stoehr},\ and\
  \citenamefont {Springel}}]{gao2004subhalo}%
  \BibitemOpen
  \bibfield  {author} {\bibinfo {author} {\bibfnamefont {L.}~\bibnamefont
  {Gao}}, \bibinfo {author} {\bibfnamefont {S.~D.}\ \bibnamefont {White}},
  \bibinfo {author} {\bibfnamefont {A.}~\bibnamefont {Jenkins}}, \bibinfo
  {author} {\bibfnamefont {F.}~\bibnamefont {Stoehr}}, \ and\ \bibinfo {author}
  {\bibfnamefont {V.}~\bibnamefont {Springel}},\ }\href {\doibase
  10.1111/j.1365-2966.2004.08360.x} {\bibfield  {journal} {\bibinfo  {journal}
  {Mon. Not. R. Astron. Soc.}\ }\textbf {\bibinfo {volume} {355}},\ \bibinfo
  {pages} {819} (\bibinfo {year} {2004})},\ \Eprint
  {http://arxiv.org/abs/astro-ph/0404589} {arXiv:astro-ph/0404589} \BibitemShut
  {NoStop}%
\bibitem [{\citenamefont {Kravtsov}\ \emph {et~al.}(2004)\citenamefont
  {Kravtsov}, \citenamefont {Berlind}, \citenamefont {Wechsler}, \citenamefont
  {Klypin}, \citenamefont {Gottl{\"o}ber}, \citenamefont {Allgood},\ and\
  \citenamefont {Primack}}]{kravtsov2004dark}%
  \BibitemOpen
  \bibfield  {author} {\bibinfo {author} {\bibfnamefont {A.~V.}\ \bibnamefont
  {Kravtsov}}, \bibinfo {author} {\bibfnamefont {A.~A.}\ \bibnamefont
  {Berlind}}, \bibinfo {author} {\bibfnamefont {R.~H.}\ \bibnamefont
  {Wechsler}}, \bibinfo {author} {\bibfnamefont {A.~A.}\ \bibnamefont
  {Klypin}}, \bibinfo {author} {\bibfnamefont {S.}~\bibnamefont
  {Gottl{\"o}ber}}, \bibinfo {author} {\bibfnamefont {B.}~\bibnamefont
  {Allgood}}, \ and\ \bibinfo {author} {\bibfnamefont {J.~R.}\ \bibnamefont
  {Primack}},\ }\href {\doibase 10.1086/420959} {\bibfield  {journal} {\bibinfo
   {journal} {Astrophys. J.}\ }\textbf {\bibinfo {volume} {609}},\ \bibinfo
  {pages} {35} (\bibinfo {year} {2004})},\ \Eprint
  {http://arxiv.org/abs/astro-ph/0308519} {arXiv:astro-ph/0308519} \BibitemShut
  {NoStop}%
\bibitem [{\citenamefont {Giocoli}\ \emph {et~al.}(2008)\citenamefont
  {Giocoli}, \citenamefont {Tormen},\ and\ \citenamefont {Van
  Den~Bosch}}]{giocoli2008population}%
  \BibitemOpen
  \bibfield  {author} {\bibinfo {author} {\bibfnamefont {C.}~\bibnamefont
  {Giocoli}}, \bibinfo {author} {\bibfnamefont {G.}~\bibnamefont {Tormen}}, \
  and\ \bibinfo {author} {\bibfnamefont {F.~C.}\ \bibnamefont {Van
  Den~Bosch}},\ }\href {\doibase 10.1111/j.1365-2966.2008.13182.x} {\bibfield
  {journal} {\bibinfo  {journal} {Mon. Not. R. Astron. Soc.}\ }\textbf
  {\bibinfo {volume} {386}},\ \bibinfo {pages} {2135} (\bibinfo {year}
  {2008})},\ \Eprint {http://arxiv.org/abs/0712.1563} {arXiv:0712.1563}
  \BibitemShut {NoStop}%
\bibitem [{\citenamefont {Giocoli}\ \emph {et~al.}(2010)\citenamefont
  {Giocoli}, \citenamefont {Tormen}, \citenamefont {Sheth},\ and\ \citenamefont
  {van~den Bosch}}]{giocoli2010substructure}%
  \BibitemOpen
  \bibfield  {author} {\bibinfo {author} {\bibfnamefont {C.}~\bibnamefont
  {Giocoli}}, \bibinfo {author} {\bibfnamefont {G.}~\bibnamefont {Tormen}},
  \bibinfo {author} {\bibfnamefont {R.~K.}\ \bibnamefont {Sheth}}, \ and\
  \bibinfo {author} {\bibfnamefont {F.~C.}\ \bibnamefont {van~den Bosch}},\
  }\href {\doibase 10.1111/j.1365-2966.2010.16311.x} {\bibfield  {journal}
  {\bibinfo  {journal} {Mon. Not. R. Astron. Soc.}\ }\textbf {\bibinfo {volume}
  {404}},\ \bibinfo {pages} {502} (\bibinfo {year} {2010})},\ \Eprint
  {http://arxiv.org/abs/0911.0436} {arXiv:0911.0436} \BibitemShut {NoStop}%
\bibitem [{\citenamefont {Hofmann}\ \emph {et~al.}(2001)\citenamefont
  {Hofmann}, \citenamefont {Schwarz},\ and\ \citenamefont
  {St{\"o}cker}}]{hofmann2001damping}%
  \BibitemOpen
  \bibfield  {author} {\bibinfo {author} {\bibfnamefont {S.}~\bibnamefont
  {Hofmann}}, \bibinfo {author} {\bibfnamefont {D.~J.}\ \bibnamefont
  {Schwarz}}, \ and\ \bibinfo {author} {\bibfnamefont {H.}~\bibnamefont
  {St{\"o}cker}},\ }\href {\doibase 10.1103/PhysRevD.64.083507} {\bibfield
  {journal} {\bibinfo  {journal} {Phys. Rev. D}\ }\textbf {\bibinfo {volume}
  {64}},\ \bibinfo {pages} {083507} (\bibinfo {year} {2001})},\ \Eprint
  {http://arxiv.org/abs/astro-ph/0104173} {arXiv:astro-ph/0104173} \BibitemShut
  {NoStop}%
\bibitem [{\citenamefont {Green}\ \emph {et~al.}(2004)\citenamefont {Green},
  \citenamefont {Hofmann},\ and\ \citenamefont {Schwarz}}]{green2004power}%
  \BibitemOpen
  \bibfield  {author} {\bibinfo {author} {\bibfnamefont {A.~M.}\ \bibnamefont
  {Green}}, \bibinfo {author} {\bibfnamefont {S.}~\bibnamefont {Hofmann}}, \
  and\ \bibinfo {author} {\bibfnamefont {D.~J.}\ \bibnamefont {Schwarz}},\
  }\href {\doibase 10.1111/j.1365-2966.2004.08232.x} {\bibfield  {journal}
  {\bibinfo  {journal} {Mon. Not. R. Astron. Soc.}\ }\textbf {\bibinfo {volume}
  {353}},\ \bibinfo {pages} {L23} (\bibinfo {year} {2004})},\ \Eprint
  {http://arxiv.org/abs/astro-ph/0309621} {arXiv:astro-ph/0309621} \BibitemShut
  {NoStop}%
\bibitem [{\citenamefont {Diemand}\ \emph {et~al.}(2005)\citenamefont
  {Diemand}, \citenamefont {Moore},\ and\ \citenamefont
  {Stadel}}]{diemand2005earth}%
  \BibitemOpen
  \bibfield  {author} {\bibinfo {author} {\bibfnamefont {J.}~\bibnamefont
  {Diemand}}, \bibinfo {author} {\bibfnamefont {B.}~\bibnamefont {Moore}}, \
  and\ \bibinfo {author} {\bibfnamefont {J.}~\bibnamefont {Stadel}},\ }\href
  {\doibase 10.1038/nature03270} {\bibfield  {journal} {\bibinfo  {journal}
  {Nature (London)}\ }\textbf {\bibinfo {volume} {433}},\ \bibinfo {pages}
  {389} (\bibinfo {year} {2005})},\ \Eprint
  {http://arxiv.org/abs/astro-ph/0501589} {arXiv:astro-ph/0501589} \BibitemShut
  {NoStop}%
\bibitem [{\citenamefont {Goerdt}\ \emph {et~al.}(2007)\citenamefont {Goerdt},
  \citenamefont {Gnedin}, \citenamefont {Moore}, \citenamefont {Diemand},\ and\
  \citenamefont {Stadel}}]{goerdt2007survival}%
  \BibitemOpen
  \bibfield  {author} {\bibinfo {author} {\bibfnamefont {T.}~\bibnamefont
  {Goerdt}}, \bibinfo {author} {\bibfnamefont {O.~Y.}\ \bibnamefont {Gnedin}},
  \bibinfo {author} {\bibfnamefont {B.}~\bibnamefont {Moore}}, \bibinfo
  {author} {\bibfnamefont {J.}~\bibnamefont {Diemand}}, \ and\ \bibinfo
  {author} {\bibfnamefont {J.}~\bibnamefont {Stadel}},\ }\href {\doibase
  10.1111/j.1365-2966.2006.11281.x} {\bibfield  {journal} {\bibinfo  {journal}
  {Mon. Not. R. Astron. Soc.}\ }\textbf {\bibinfo {volume} {375}},\ \bibinfo
  {pages} {191} (\bibinfo {year} {2007})},\ \Eprint
  {http://arxiv.org/abs/astro-ph/0608495} {arXiv:astro-ph/0608495} \BibitemShut
  {NoStop}%
\bibitem [{\citenamefont {Schneider}\ \emph {et~al.}(2010)\citenamefont
  {Schneider}, \citenamefont {Krauss},\ and\ \citenamefont
  {Moore}}]{schneider2010impact}%
  \BibitemOpen
  \bibfield  {author} {\bibinfo {author} {\bibfnamefont {A.}~\bibnamefont
  {Schneider}}, \bibinfo {author} {\bibfnamefont {L.}~\bibnamefont {Krauss}}, \
  and\ \bibinfo {author} {\bibfnamefont {B.}~\bibnamefont {Moore}},\ }\href
  {\doibase 10.1103/PhysRevD.82.063525} {\bibfield  {journal} {\bibinfo
  {journal} {Phys. Rev. D}\ }\textbf {\bibinfo {volume} {82}},\ \bibinfo
  {pages} {063525} (\bibinfo {year} {2010})},\ \Eprint
  {http://arxiv.org/abs/1004.5432} {arXiv:1004.5432} \BibitemShut {NoStop}%
\bibitem [{\citenamefont {Ishiyama}\ \emph {et~al.}(2010)\citenamefont
  {Ishiyama}, \citenamefont {Makino},\ and\ \citenamefont
  {Ebisuzaki}}]{ishiyama2010gamma}%
  \BibitemOpen
  \bibfield  {author} {\bibinfo {author} {\bibfnamefont {T.}~\bibnamefont
  {Ishiyama}}, \bibinfo {author} {\bibfnamefont {J.}~\bibnamefont {Makino}}, \
  and\ \bibinfo {author} {\bibfnamefont {T.}~\bibnamefont {Ebisuzaki}},\ }\href
  {\doibase 10.1088/2041-8205/723/2/l195} {\bibfield  {journal} {\bibinfo
  {journal} {Astrophys. J. Lett.}\ }\textbf {\bibinfo {volume} {723}},\
  \bibinfo {pages} {L195} (\bibinfo {year} {2010})},\ \Eprint
  {http://arxiv.org/abs/1006.3392} {arXiv:1006.3392} \BibitemShut {NoStop}%
\bibitem [{\citenamefont {Anderhalden}\ and\ \citenamefont
  {Diemand}(2013{\natexlab{a}})}]{anderhalden2013density}%
  \BibitemOpen
  \bibfield  {author} {\bibinfo {author} {\bibfnamefont {D.}~\bibnamefont
  {Anderhalden}}\ and\ \bibinfo {author} {\bibfnamefont {J.}~\bibnamefont
  {Diemand}},\ }\href {\doibase 10.1088/1475-7516/2013/04/009} {\bibfield
  {journal} {\bibinfo  {journal} {J. Cosmol. Astropart. Phys.}\ }\textbf
  {\bibinfo {volume} {04}},\ \bibinfo {pages} {009} (\bibinfo {year}
  {2013}{\natexlab{a}})},\ \Eprint {http://arxiv.org/abs/1302.0003}
  {arXiv:1302.0003} \BibitemShut {NoStop}%
\bibitem [{\citenamefont {Anderhalden}\ and\ \citenamefont
  {Diemand}(2013{\natexlab{b}})}]{anderhalden2013erratum}%
  \BibitemOpen
  \bibfield  {author} {\bibinfo {author} {\bibfnamefont {D.}~\bibnamefont
  {Anderhalden}}\ and\ \bibinfo {author} {\bibfnamefont {J.}~\bibnamefont
  {Diemand}},\ }\href {\doibase 10.1088/1475-7516/2013/08/e02} {\ \textbf
  {\bibinfo {volume} {08}},\ \bibinfo {pages} {E02} (\bibinfo {year}
  {2013}{\natexlab{b}})}\BibitemShut {NoStop}%
\bibitem [{\citenamefont {Ishiyama}(2014)}]{ishiyama2014hierarchical}%
  \BibitemOpen
  \bibfield  {author} {\bibinfo {author} {\bibfnamefont {T.}~\bibnamefont
  {Ishiyama}},\ }\href {\doibase 10.1088/0004-637x/788/1/27} {\bibfield
  {journal} {\bibinfo  {journal} {Astrophys. J.}\ }\textbf {\bibinfo {volume}
  {788}},\ \bibinfo {pages} {27} (\bibinfo {year} {2014})},\ \Eprint
  {http://arxiv.org/abs/1404.1650} {arXiv:1404.1650} \BibitemShut {NoStop}%
\bibitem [{\citenamefont {Ishiyama}\ and\ \citenamefont
  {Ando}()}]{ishiyama2019abundance}%
  \BibitemOpen
  \bibfield  {author} {\bibinfo {author} {\bibfnamefont {T.}~\bibnamefont
  {Ishiyama}}\ and\ \bibinfo {author} {\bibfnamefont {S.}~\bibnamefont
  {Ando}},\ }\href@noop {} {\ }\Eprint {http://arxiv.org/abs/1907.03642}
  {arXiv:1907.03642} \BibitemShut {NoStop}%
\bibitem [{\citenamefont {Berezinsky}\ \emph {et~al.}(2003)\citenamefont
  {Berezinsky}, \citenamefont {Dokuchaev},\ and\ \citenamefont
  {Eroshenko}}]{berezinsky2003small}%
  \BibitemOpen
  \bibfield  {author} {\bibinfo {author} {\bibfnamefont {V.}~\bibnamefont
  {Berezinsky}}, \bibinfo {author} {\bibfnamefont {V.}~\bibnamefont
  {Dokuchaev}}, \ and\ \bibinfo {author} {\bibfnamefont {Y.}~\bibnamefont
  {Eroshenko}},\ }\href {\doibase 10.1103/PhysRevD.68.103003} {\bibfield
  {journal} {\bibinfo  {journal} {Phys. Rev. D}\ }\textbf {\bibinfo {volume}
  {68}},\ \bibinfo {pages} {103003} (\bibinfo {year} {2003})},\ \Eprint
  {http://arxiv.org/abs/astro-ph/0301551} {arXiv:astro-ph/0301551} \BibitemShut
  {NoStop}%
\bibitem [{\citenamefont {Pieri}\ \emph {et~al.}(2008)\citenamefont {Pieri},
  \citenamefont {Bertone},\ and\ \citenamefont {Branchini}}]{pieri2008dark}%
  \BibitemOpen
  \bibfield  {author} {\bibinfo {author} {\bibfnamefont {L.}~\bibnamefont
  {Pieri}}, \bibinfo {author} {\bibfnamefont {G.}~\bibnamefont {Bertone}}, \
  and\ \bibinfo {author} {\bibfnamefont {E.}~\bibnamefont {Branchini}},\ }\href
  {\doibase 10.1111/j.1365-2966.2007.12828.x} {\bibfield  {journal} {\bibinfo
  {journal} {Mon. Not. R. Astron. Soc.}\ }\textbf {\bibinfo {volume} {384}},\
  \bibinfo {pages} {1627} (\bibinfo {year} {2008})},\ \Eprint
  {http://arxiv.org/abs/0706.2101} {arXiv:0706.2101} \BibitemShut {NoStop}%
\bibitem [{\citenamefont {Springel}\ \emph {et~al.}(2008)\citenamefont
  {Springel}, \citenamefont {White}, \citenamefont {Frenk}, \citenamefont
  {Navarro}, \citenamefont {Jenkins}, \citenamefont {Vogelsberger},
  \citenamefont {Wang}, \citenamefont {Ludlow},\ and\ \citenamefont
  {Helmi}}]{springel2008prospects}%
  \BibitemOpen
  \bibfield  {author} {\bibinfo {author} {\bibfnamefont {V.}~\bibnamefont
  {Springel}}, \bibinfo {author} {\bibfnamefont {S.~D.}\ \bibnamefont {White}},
  \bibinfo {author} {\bibfnamefont {C.~S.}\ \bibnamefont {Frenk}}, \bibinfo
  {author} {\bibfnamefont {J.~F.}\ \bibnamefont {Navarro}}, \bibinfo {author}
  {\bibfnamefont {A.}~\bibnamefont {Jenkins}}, \bibinfo {author} {\bibfnamefont
  {M.}~\bibnamefont {Vogelsberger}}, \bibinfo {author} {\bibfnamefont
  {J.}~\bibnamefont {Wang}}, \bibinfo {author} {\bibfnamefont {A.}~\bibnamefont
  {Ludlow}}, \ and\ \bibinfo {author} {\bibfnamefont {A.}~\bibnamefont
  {Helmi}},\ }\href {\doibase 10.1038/nature07411} {\bibfield  {journal}
  {\bibinfo  {journal} {Nature (London)}\ }\textbf {\bibinfo {volume} {456}},\
  \bibinfo {pages} {73} (\bibinfo {year} {2008})},\ \Eprint
  {http://arxiv.org/abs/0809.0894} {arXiv:0809.0894} \BibitemShut {NoStop}%
\bibitem [{\citenamefont {Berezinsky}\ \emph {et~al.}(2008)\citenamefont
  {Berezinsky}, \citenamefont {Dokuchaev},\ and\ \citenamefont
  {Eroshenko}}]{berezinsky2008remnants}%
  \BibitemOpen
  \bibfield  {author} {\bibinfo {author} {\bibfnamefont {V.}~\bibnamefont
  {Berezinsky}}, \bibinfo {author} {\bibfnamefont {V.}~\bibnamefont
  {Dokuchaev}}, \ and\ \bibinfo {author} {\bibfnamefont {Y.}~\bibnamefont
  {Eroshenko}},\ }\href {\doibase 10.1103/physrevd.77.083519} {\bibfield
  {journal} {\bibinfo  {journal} {Phys. Rev. D}\ }\textbf {\bibinfo {volume}
  {77}},\ \bibinfo {pages} {083519} (\bibinfo {year} {2008})},\ \Eprint
  {http://arxiv.org/abs/0712.3499} {arXiv:0712.3499} \BibitemShut {NoStop}%
\bibitem [{\citenamefont {Gao}\ \emph {et~al.}(2012)\citenamefont {Gao},
  \citenamefont {Frenk}, \citenamefont {Jenkins}, \citenamefont {Springel},\
  and\ \citenamefont {White}}]{gao2011will}%
  \BibitemOpen
  \bibfield  {author} {\bibinfo {author} {\bibfnamefont {L.}~\bibnamefont
  {Gao}}, \bibinfo {author} {\bibfnamefont {C.}~\bibnamefont {Frenk}}, \bibinfo
  {author} {\bibfnamefont {A.}~\bibnamefont {Jenkins}}, \bibinfo {author}
  {\bibfnamefont {V.}~\bibnamefont {Springel}}, \ and\ \bibinfo {author}
  {\bibfnamefont {S.}~\bibnamefont {White}},\ }\href {\doibase
  10.1111/j.1365-2966.2011.19836.x} {\bibfield  {journal} {\bibinfo  {journal}
  {Mon. Not. R. Astron. Soc.}\ }\textbf {\bibinfo {volume} {419}},\ \bibinfo
  {pages} {1721} (\bibinfo {year} {2012})},\ \Eprint
  {http://arxiv.org/abs/1107.1916} {arXiv:1107.1916} \BibitemShut {NoStop}%
\bibitem [{\citenamefont {Belotsky}\ \emph {et~al.}(2014)\citenamefont
  {Belotsky}, \citenamefont {Kirillov},\ and\ \citenamefont
  {Khlopov}}]{belotsky2014gamma}%
  \BibitemOpen
  \bibfield  {author} {\bibinfo {author} {\bibfnamefont {K.}~\bibnamefont
  {Belotsky}}, \bibinfo {author} {\bibfnamefont {A.}~\bibnamefont {Kirillov}},
  \ and\ \bibinfo {author} {\bibfnamefont {M.}~\bibnamefont {Khlopov}},\ }\href
  {\doibase 10.1134/S0202289314010022} {\bibfield  {journal} {\bibinfo
  {journal} {Gravitation Cosmol.}\ }\textbf {\bibinfo {volume} {20}},\ \bibinfo
  {pages} {47} (\bibinfo {year} {2014})},\ \Eprint
  {http://arxiv.org/abs/1212.6087} {arXiv:1212.6087} \BibitemShut {NoStop}%
\bibitem [{\citenamefont {S{\'a}nchez-Conde}\ and\ \citenamefont
  {Prada}(2014)}]{sanchez2014flattening}%
  \BibitemOpen
  \bibfield  {author} {\bibinfo {author} {\bibfnamefont {M.~A.}\ \bibnamefont
  {S{\'a}nchez-Conde}}\ and\ \bibinfo {author} {\bibfnamefont {F.}~\bibnamefont
  {Prada}},\ }\href {\doibase 10.1093/mnras/stu1014} {\bibfield  {journal}
  {\bibinfo  {journal} {Mon. Not. R. Astron. Soc.}\ }\textbf {\bibinfo {volume}
  {442}},\ \bibinfo {pages} {2271} (\bibinfo {year} {2014})},\ \Eprint
  {http://arxiv.org/abs/1312.1729} {arXiv:1312.1729} \BibitemShut {NoStop}%
\bibitem [{\citenamefont {Bartels}\ and\ \citenamefont
  {Ando}(2015)}]{bartels2015boosting}%
  \BibitemOpen
  \bibfield  {author} {\bibinfo {author} {\bibfnamefont {R.}~\bibnamefont
  {Bartels}}\ and\ \bibinfo {author} {\bibfnamefont {S.}~\bibnamefont {Ando}},\
  }\href {\doibase 10.1103/physrevd.92.123508} {\bibfield  {journal} {\bibinfo
  {journal} {Phys. Rev. D}\ }\textbf {\bibinfo {volume} {92}},\ \bibinfo
  {pages} {123508} (\bibinfo {year} {2015})},\ \Eprint
  {http://arxiv.org/abs/1507.08656} {arXiv:1507.08656} \BibitemShut {NoStop}%
\bibitem [{\citenamefont {Anderson}\ \emph {et~al.}(2016)\citenamefont
  {Anderson}, \citenamefont {Zimmer}, \citenamefont {Conrad}, \citenamefont
  {Gustafsson}, \citenamefont {S{\'a}nchez-Conde},\ and\ \citenamefont
  {Caputo}}]{anderson2016search}%
  \BibitemOpen
  \bibfield  {author} {\bibinfo {author} {\bibfnamefont {B.}~\bibnamefont
  {Anderson}}, \bibinfo {author} {\bibfnamefont {S.}~\bibnamefont {Zimmer}},
  \bibinfo {author} {\bibfnamefont {J.}~\bibnamefont {Conrad}}, \bibinfo
  {author} {\bibfnamefont {M.}~\bibnamefont {Gustafsson}}, \bibinfo {author}
  {\bibfnamefont {M.}~\bibnamefont {S{\'a}nchez-Conde}}, \ and\ \bibinfo
  {author} {\bibfnamefont {R.}~\bibnamefont {Caputo}},\ }\href {\doibase
  10.1088/1475-7516/2016/02/026} {\bibfield  {journal} {\bibinfo  {journal} {J.
  Cosmol. Astropart. Phys.}\ }\textbf {\bibinfo {volume} {02}},\ \bibinfo
  {pages} {026} (\bibinfo {year} {2016})},\ \Eprint
  {http://arxiv.org/abs/1511.00014} {arXiv:1511.00014} \BibitemShut {NoStop}%
\bibitem [{\citenamefont {Stref}\ and\ \citenamefont
  {Lavalle}(2017)}]{stref2017modeling}%
  \BibitemOpen
  \bibfield  {author} {\bibinfo {author} {\bibfnamefont {M.}~\bibnamefont
  {Stref}}\ and\ \bibinfo {author} {\bibfnamefont {J.}~\bibnamefont
  {Lavalle}},\ }\href {\doibase 10.1103/physrevd.95.063003} {\bibfield
  {journal} {\bibinfo  {journal} {Phys. Rev. D}\ }\textbf {\bibinfo {volume}
  {95}},\ \bibinfo {pages} {063003} (\bibinfo {year} {2017})},\ \Eprint
  {http://arxiv.org/abs/1610.02233} {arXiv:1610.02233} \BibitemShut {NoStop}%
\bibitem [{\citenamefont {Hiroshima}\ \emph {et~al.}(2018)\citenamefont
  {Hiroshima}, \citenamefont {Ando},\ and\ \citenamefont
  {Ishiyama}}]{hiroshima2018modeling}%
  \BibitemOpen
  \bibfield  {author} {\bibinfo {author} {\bibfnamefont {N.}~\bibnamefont
  {Hiroshima}}, \bibinfo {author} {\bibfnamefont {S.}~\bibnamefont {Ando}}, \
  and\ \bibinfo {author} {\bibfnamefont {T.}~\bibnamefont {Ishiyama}},\ }\href
  {\doibase 10.1103/physrevd.97.123002} {\bibfield  {journal} {\bibinfo
  {journal} {Phys. Rev. D}\ }\textbf {\bibinfo {volume} {97}},\ \bibinfo
  {pages} {123002} (\bibinfo {year} {2018})},\ \Eprint
  {http://arxiv.org/abs/1803.07691} {arXiv:1803.07691} \BibitemShut {NoStop}%
\bibitem [{\citenamefont {Stref}\ \emph {et~al.}(2019)\citenamefont {Stref},
  \citenamefont {Lacroix},\ and\ \citenamefont {Lavalle}}]{stref2019remnants}%
  \BibitemOpen
  \bibfield  {author} {\bibinfo {author} {\bibfnamefont {M.}~\bibnamefont
  {Stref}}, \bibinfo {author} {\bibfnamefont {T.}~\bibnamefont {Lacroix}}, \
  and\ \bibinfo {author} {\bibfnamefont {J.}~\bibnamefont {Lavalle}},\ }\href
  {\doibase 10.3390/galaxies7020065} {\bibfield  {journal} {\bibinfo  {journal}
  {Galaxies}\ }\textbf {\bibinfo {volume} {7}},\ \bibinfo {pages} {65}
  (\bibinfo {year} {2019})},\ \Eprint {http://arxiv.org/abs/1905.02008}
  {arXiv:1905.02008} \BibitemShut {NoStop}%
\bibitem [{\citenamefont {Ando}\ \emph {et~al.}(2019)\citenamefont {Ando},
  \citenamefont {Ishiyama},\ and\ \citenamefont {Hiroshima}}]{ando2019halo}%
  \BibitemOpen
  \bibfield  {author} {\bibinfo {author} {\bibfnamefont {S.}~\bibnamefont
  {Ando}}, \bibinfo {author} {\bibfnamefont {T.}~\bibnamefont {Ishiyama}}, \
  and\ \bibinfo {author} {\bibfnamefont {N.}~\bibnamefont {Hiroshima}},\ }\href
  {\doibase 10.3390/galaxies7030068} {\bibfield  {journal} {\bibinfo  {journal}
  {Galaxies}\ }\textbf {\bibinfo {volume} {7}},\ \bibinfo {pages} {68}
  (\bibinfo {year} {2019})},\ \Eprint {http://arxiv.org/abs/1903.11427}
  {arXiv:1903.11427} \BibitemShut {NoStop}%
\bibitem [{\citenamefont {Boden}\ \emph {et~al.}(1998)\citenamefont {Boden},
  \citenamefont {Shao},\ and\ \citenamefont
  {Van~Buren}}]{boden1998astrometric}%
  \BibitemOpen
  \bibfield  {author} {\bibinfo {author} {\bibfnamefont {A.}~\bibnamefont
  {Boden}}, \bibinfo {author} {\bibfnamefont {M.}~\bibnamefont {Shao}}, \ and\
  \bibinfo {author} {\bibfnamefont {D.}~\bibnamefont {Van~Buren}},\ }\href
  {\doibase 10.1086/305913} {\bibfield  {journal} {\bibinfo  {journal}
  {Astrophys. J.}\ }\textbf {\bibinfo {volume} {502}},\ \bibinfo {pages} {538}
  (\bibinfo {year} {1998})},\ \Eprint {http://arxiv.org/abs/astro-ph/9802179}
  {arXiv:astro-ph/9802179} \BibitemShut {NoStop}%
\bibitem [{\citenamefont {Chen}\ and\ \citenamefont
  {Koushiappas}(2010)}]{chen2010gravitational}%
  \BibitemOpen
  \bibfield  {author} {\bibinfo {author} {\bibfnamefont {J.}~\bibnamefont
  {Chen}}\ and\ \bibinfo {author} {\bibfnamefont {S.~M.}\ \bibnamefont
  {Koushiappas}},\ }\href {\doibase 10.1088/0004-637x/724/1/400} {\bibfield
  {journal} {\bibinfo  {journal} {Astrophys. J.}\ }\textbf {\bibinfo {volume}
  {724}},\ \bibinfo {pages} {400} (\bibinfo {year} {2010})},\ \Eprint
  {http://arxiv.org/abs/1008.2385} {arXiv:1008.2385} \BibitemShut {NoStop}%
\bibitem [{\citenamefont {Erickcek}\ and\ \citenamefont
  {Law}(2011)}]{erickcek2011astrometric}%
  \BibitemOpen
  \bibfield  {author} {\bibinfo {author} {\bibfnamefont {A.~L.}\ \bibnamefont
  {Erickcek}}\ and\ \bibinfo {author} {\bibfnamefont {N.~M.}\ \bibnamefont
  {Law}},\ }\href {\doibase 10.1088/0004-637x/729/1/49} {\bibfield  {journal}
  {\bibinfo  {journal} {Astrophys. J.}\ }\textbf {\bibinfo {volume} {729}},\
  \bibinfo {pages} {49} (\bibinfo {year} {2011})},\ \Eprint
  {http://arxiv.org/abs/1007.4228} {arXiv:1007.4228} \BibitemShut {NoStop}%
\bibitem [{\citenamefont {Van~Tilburg}\ \emph {et~al.}(2018)\citenamefont
  {Van~Tilburg}, \citenamefont {Taki},\ and\ \citenamefont
  {Weiner}}]{van2018halometry}%
  \BibitemOpen
  \bibfield  {author} {\bibinfo {author} {\bibfnamefont {K.}~\bibnamefont
  {Van~Tilburg}}, \bibinfo {author} {\bibfnamefont {A.-M.}\ \bibnamefont
  {Taki}}, \ and\ \bibinfo {author} {\bibfnamefont {N.}~\bibnamefont
  {Weiner}},\ }\href {\doibase 10.1088/1475-7516/2018/07/041} {\bibfield
  {journal} {\bibinfo  {journal} {J. Cosmol. Astropart. Phys.}\ }\textbf
  {\bibinfo {volume} {07}},\ \bibinfo {pages} {041} (\bibinfo {year} {2018})},\
  \Eprint {http://arxiv.org/abs/1804.01991} {arXiv:1804.01991} \BibitemShut
  {NoStop}%
\bibitem [{\citenamefont {Siegel}\ \emph {et~al.}(2007)\citenamefont {Siegel},
  \citenamefont {Hertzberg},\ and\ \citenamefont {Fry}}]{siegel2007probing}%
  \BibitemOpen
  \bibfield  {author} {\bibinfo {author} {\bibfnamefont {E.~R.}\ \bibnamefont
  {Siegel}}, \bibinfo {author} {\bibfnamefont {M.}~\bibnamefont {Hertzberg}}, \
  and\ \bibinfo {author} {\bibfnamefont {J.}~\bibnamefont {Fry}},\ }\href
  {\doibase 10.1111/j.1365-2966.2007.12435.x} {\bibfield  {journal} {\bibinfo
  {journal} {Mon. Not. R. Astron. Soc.}\ }\textbf {\bibinfo {volume} {382}},\
  \bibinfo {pages} {879} (\bibinfo {year} {2007})},\ \Eprint
  {http://arxiv.org/abs/astro-ph/0702546} {arXiv:astro-ph/0702546} \BibitemShut
  {NoStop}%
\bibitem [{\citenamefont {Baghram}\ \emph {et~al.}(2011)\citenamefont
  {Baghram}, \citenamefont {Afshordi},\ and\ \citenamefont
  {Zurek}}]{baghram2011prospects}%
  \BibitemOpen
  \bibfield  {author} {\bibinfo {author} {\bibfnamefont {S.}~\bibnamefont
  {Baghram}}, \bibinfo {author} {\bibfnamefont {N.}~\bibnamefont {Afshordi}}, \
  and\ \bibinfo {author} {\bibfnamefont {K.~M.}\ \bibnamefont {Zurek}},\ }\href
  {\doibase 10.1103/physrevd.84.043511} {\bibfield  {journal} {\bibinfo
  {journal} {Phys. Rev. D}\ }\textbf {\bibinfo {volume} {84}},\ \bibinfo
  {pages} {043511} (\bibinfo {year} {2011})},\ \Eprint
  {http://arxiv.org/abs/1101.5487} {arXiv:1101.5487} \BibitemShut {NoStop}%
\bibitem [{\citenamefont {Kashiyama}\ and\ \citenamefont
  {Oguri}()}]{kashiyama2018detectability}%
  \BibitemOpen
  \bibfield  {author} {\bibinfo {author} {\bibfnamefont {K.}~\bibnamefont
  {Kashiyama}}\ and\ \bibinfo {author} {\bibfnamefont {M.}~\bibnamefont
  {Oguri}},\ }\href@noop {} {\ }\Eprint {http://arxiv.org/abs/1801.07847}
  {arXiv:1801.07847} \BibitemShut {NoStop}%
\bibitem [{\citenamefont {Gonzalez-Morales}\ \emph {et~al.}(2013)\citenamefont
  {Gonzalez-Morales}, \citenamefont {Valenzuela},\ and\ \citenamefont
  {Aguilar}}]{gonzalez2013constraining}%
  \BibitemOpen
  \bibfield  {author} {\bibinfo {author} {\bibfnamefont {A.~X.}\ \bibnamefont
  {Gonzalez-Morales}}, \bibinfo {author} {\bibfnamefont {O.}~\bibnamefont
  {Valenzuela}}, \ and\ \bibinfo {author} {\bibfnamefont {L.~A.}\ \bibnamefont
  {Aguilar}},\ }\href {\doibase 10.1088/1475-7516/2013/03/001} {\bibfield
  {journal} {\bibinfo  {journal} {J. Cosmol. Astropart. Phys.}\ }\textbf
  {\bibinfo {volume} {03}},\ \bibinfo {pages} {001} (\bibinfo {year} {2013})},\
  \Eprint {http://arxiv.org/abs/1211.6745} {arXiv:1211.6745} \BibitemShut
  {NoStop}%
\bibitem [{\citenamefont {Erkal}\ and\ \citenamefont
  {Belokurov}(2015)}]{erkal2015forensics}%
  \BibitemOpen
  \bibfield  {author} {\bibinfo {author} {\bibfnamefont {D.}~\bibnamefont
  {Erkal}}\ and\ \bibinfo {author} {\bibfnamefont {V.}~\bibnamefont
  {Belokurov}},\ }\href {\doibase 10.1093/mnras/stv655} {\bibfield  {journal}
  {\bibinfo  {journal} {Mon. Not. R. Astron. Soc.}\ }\textbf {\bibinfo {volume}
  {450}},\ \bibinfo {pages} {1136} (\bibinfo {year} {2015})},\ \Eprint
  {http://arxiv.org/abs/1412.6035} {arXiv:1412.6035} \BibitemShut {NoStop}%
\bibitem [{\citenamefont {Erkal}\ \emph {et~al.}(2016)\citenamefont {Erkal},
  \citenamefont {Belokurov}, \citenamefont {Bovy},\ and\ \citenamefont
  {Sanders}}]{erkal2016number}%
  \BibitemOpen
  \bibfield  {author} {\bibinfo {author} {\bibfnamefont {D.}~\bibnamefont
  {Erkal}}, \bibinfo {author} {\bibfnamefont {V.}~\bibnamefont {Belokurov}},
  \bibinfo {author} {\bibfnamefont {J.}~\bibnamefont {Bovy}}, \ and\ \bibinfo
  {author} {\bibfnamefont {J.~L.}\ \bibnamefont {Sanders}},\ }\href {\doibase
  10.1093/mnras/stw1957} {\bibfield  {journal} {\bibinfo  {journal} {Mon. Not.
  R. Astron. Soc.}\ }\textbf {\bibinfo {volume} {463}},\ \bibinfo {pages} {102}
  (\bibinfo {year} {2016})},\ \Eprint {http://arxiv.org/abs/1606.04946}
  {arXiv:1606.04946} \BibitemShut {NoStop}%
\bibitem [{\citenamefont {Buschmann}\ \emph {et~al.}(2018)\citenamefont
  {Buschmann}, \citenamefont {Kopp}, \citenamefont {Safdi},\ and\ \citenamefont
  {Wu}}]{buschmann2018stellar}%
  \BibitemOpen
  \bibfield  {author} {\bibinfo {author} {\bibfnamefont {M.}~\bibnamefont
  {Buschmann}}, \bibinfo {author} {\bibfnamefont {J.}~\bibnamefont {Kopp}},
  \bibinfo {author} {\bibfnamefont {B.~R.}\ \bibnamefont {Safdi}}, \ and\
  \bibinfo {author} {\bibfnamefont {C.-L.}\ \bibnamefont {Wu}},\ }\href
  {\doibase 10.1103/physrevlett.120.211101} {\bibfield  {journal} {\bibinfo
  {journal} {Phys. Rev. Lett.}\ }\textbf {\bibinfo {volume} {120}},\ \bibinfo
  {pages} {211101} (\bibinfo {year} {2018})},\ \Eprint
  {http://arxiv.org/abs/1711.03554} {arXiv:1711.03554} \BibitemShut {NoStop}%
\bibitem [{\citenamefont {Pe{\~n}arrubia}(2018)}]{penarrubia2017fluctuations}%
  \BibitemOpen
  \bibfield  {author} {\bibinfo {author} {\bibfnamefont {J.}~\bibnamefont
  {Pe{\~n}arrubia}},\ }\href {\doibase 10.1093/mnras/stx2773} {\bibfield
  {journal} {\bibinfo  {journal} {Mon. Not. R. Astron. Soc.}\ }\textbf
  {\bibinfo {volume} {474}},\ \bibinfo {pages} {1482} (\bibinfo {year}
  {2018})},\ \Eprint {http://arxiv.org/abs/1710.06443} {arXiv:1710.06443}
  \BibitemShut {NoStop}%
\bibitem [{\citenamefont {Delos}\ \emph {et~al.}(2019)\citenamefont {Delos},
  \citenamefont {Bruff},\ and\ \citenamefont {Erickcek}}]{delos2019predicting}%
  \BibitemOpen
  \bibfield  {author} {\bibinfo {author} {\bibfnamefont {M.~S.}\ \bibnamefont
  {Delos}}, \bibinfo {author} {\bibfnamefont {M.}~\bibnamefont {Bruff}}, \ and\
  \bibinfo {author} {\bibfnamefont {A.~L.}\ \bibnamefont {Erickcek}},\ }\href
  {\doibase 10.1103/physrevd.100.023523} {\bibfield  {journal} {\bibinfo
  {journal} {Phys. Rev. D}\ }\textbf {\bibinfo {volume} {100}},\ \bibinfo
  {pages} {023523} (\bibinfo {year} {2019})},\ \Eprint
  {http://arxiv.org/abs/1905.05766} {arXiv:1905.05766} \BibitemShut {NoStop}%
\bibitem [{\citenamefont {Diemer}\ and\ \citenamefont
  {Kravtsov}(2015)}]{diemer2015universal}%
  \BibitemOpen
  \bibfield  {author} {\bibinfo {author} {\bibfnamefont {B.}~\bibnamefont
  {Diemer}}\ and\ \bibinfo {author} {\bibfnamefont {A.~V.}\ \bibnamefont
  {Kravtsov}},\ }\href {\doibase 10.1088/0004-637x/799/1/108} {\bibfield
  {journal} {\bibinfo  {journal} {Astrophys. J.}\ }\textbf {\bibinfo {volume}
  {799}},\ \bibinfo {pages} {108} (\bibinfo {year} {2015})},\ \Eprint
  {http://arxiv.org/abs/1407.4730} {arXiv:1407.4730} \BibitemShut {NoStop}%
\bibitem [{\citenamefont {Okoli}\ and\ \citenamefont
  {Afshordi}(2016)}]{okoli2015concentration}%
  \BibitemOpen
  \bibfield  {author} {\bibinfo {author} {\bibfnamefont {C.}~\bibnamefont
  {Okoli}}\ and\ \bibinfo {author} {\bibfnamefont {N.}~\bibnamefont
  {Afshordi}},\ }\href {\doibase 10.1093/mnras/stv2905} {\bibfield  {journal}
  {\bibinfo  {journal} {Mon. Not. R. Astron. Soc.}\ }\textbf {\bibinfo {volume}
  {456}},\ \bibinfo {pages} {3068} (\bibinfo {year} {2016})},\ \Eprint
  {http://arxiv.org/abs/1510.03868} {arXiv:1510.03868} \BibitemShut {NoStop}%
\bibitem [{\citenamefont {Ludlow}\ \emph {et~al.}(2016)\citenamefont {Ludlow},
  \citenamefont {Bose}, \citenamefont {Angulo}, \citenamefont {Wang},
  \citenamefont {Hellwing}, \citenamefont {Navarro}, \citenamefont {Cole},\
  and\ \citenamefont {Frenk}}]{ludlow2016mass}%
  \BibitemOpen
  \bibfield  {author} {\bibinfo {author} {\bibfnamefont {A.~D.}\ \bibnamefont
  {Ludlow}}, \bibinfo {author} {\bibfnamefont {S.}~\bibnamefont {Bose}},
  \bibinfo {author} {\bibfnamefont {R.~E.}\ \bibnamefont {Angulo}}, \bibinfo
  {author} {\bibfnamefont {L.}~\bibnamefont {Wang}}, \bibinfo {author}
  {\bibfnamefont {W.~A.}\ \bibnamefont {Hellwing}}, \bibinfo {author}
  {\bibfnamefont {J.~F.}\ \bibnamefont {Navarro}}, \bibinfo {author}
  {\bibfnamefont {S.}~\bibnamefont {Cole}}, \ and\ \bibinfo {author}
  {\bibfnamefont {C.~S.}\ \bibnamefont {Frenk}},\ }\href {\doibase
  10.1093/mnras/stw1046} {\bibfield  {journal} {\bibinfo  {journal} {Mon. Not.
  R. Astron. Soc.}\ }\textbf {\bibinfo {volume} {460}},\ \bibinfo {pages}
  {1214} (\bibinfo {year} {2016})},\ \Eprint {http://arxiv.org/abs/1601.02624}
  {arXiv:1601.02624} \BibitemShut {NoStop}%
\bibitem [{\citenamefont {Diemer}\ and\ \citenamefont
  {Joyce}(2019)}]{diemer2019accurate}%
  \BibitemOpen
  \bibfield  {author} {\bibinfo {author} {\bibfnamefont {B.}~\bibnamefont
  {Diemer}}\ and\ \bibinfo {author} {\bibfnamefont {M.}~\bibnamefont {Joyce}},\
  }\href {\doibase 10.3847/1538-4357/aafad6} {\bibfield  {journal} {\bibinfo
  {journal} {Astrophys. J.}\ }\textbf {\bibinfo {volume} {871}},\ \bibinfo
  {pages} {168} (\bibinfo {year} {2019})},\ \Eprint
  {http://arxiv.org/abs/1809.07326} {arXiv:1809.07326} \BibitemShut {NoStop}%
\bibitem [{\citenamefont {Silk}\ and\ \citenamefont
  {Turner}(1987)}]{silk1987double}%
  \BibitemOpen
  \bibfield  {author} {\bibinfo {author} {\bibfnamefont {J.}~\bibnamefont
  {Silk}}\ and\ \bibinfo {author} {\bibfnamefont {M.~S.}\ \bibnamefont
  {Turner}},\ }\href {\doibase 10.1103/physrevd.35.419} {\bibfield  {journal}
  {\bibinfo  {journal} {Phys. Rev. D}\ }\textbf {\bibinfo {volume} {35}},\
  \bibinfo {pages} {419} (\bibinfo {year} {1987})}\BibitemShut {NoStop}%
\bibitem [{\citenamefont {Salopek}\ \emph {et~al.}(1989)\citenamefont
  {Salopek}, \citenamefont {Bond},\ and\ \citenamefont
  {Bardeen}}]{salopek1989designing}%
  \BibitemOpen
  \bibfield  {author} {\bibinfo {author} {\bibfnamefont {D.~S.}\ \bibnamefont
  {Salopek}}, \bibinfo {author} {\bibfnamefont {J.~R.}\ \bibnamefont {Bond}}, \
  and\ \bibinfo {author} {\bibfnamefont {J.~M.}\ \bibnamefont {Bardeen}},\
  }\href {\doibase 10.1103/physrevd.40.1753} {\bibfield  {journal} {\bibinfo
  {journal} {Phys. Rev. D}\ }\textbf {\bibinfo {volume} {40}},\ \bibinfo
  {pages} {1753} (\bibinfo {year} {1989})}\BibitemShut {NoStop}%
\bibitem [{\citenamefont {Starobinskij}(1992)}]{starobinsky1992}%
  \BibitemOpen
  \bibfield  {author} {\bibinfo {author} {\bibfnamefont {A.~A.}\ \bibnamefont
  {Starobinskij}},\ }\href@noop {} {\bibfield  {journal} {\bibinfo  {journal}
  {JETP Lett.}\ }\textbf {\bibinfo {volume} {55}},\ \bibinfo {pages} {489}
  (\bibinfo {year} {1992})}\BibitemShut {NoStop}%
\bibitem [{\citenamefont {Ivanov}\ \emph {et~al.}(1994)\citenamefont {Ivanov},
  \citenamefont {Naselsky},\ and\ \citenamefont
  {Novikov}}]{ivanov1994inflation}%
  \BibitemOpen
  \bibfield  {author} {\bibinfo {author} {\bibfnamefont {P.}~\bibnamefont
  {Ivanov}}, \bibinfo {author} {\bibfnamefont {P.}~\bibnamefont {Naselsky}}, \
  and\ \bibinfo {author} {\bibfnamefont {I.}~\bibnamefont {Novikov}},\ }\href
  {\doibase 10.1103/physrevd.50.7173} {\bibfield  {journal} {\bibinfo
  {journal} {Phys. Rev. D}\ }\textbf {\bibinfo {volume} {50}},\ \bibinfo
  {pages} {7173} (\bibinfo {year} {1994})}\BibitemShut {NoStop}%
\bibitem [{\citenamefont {Randall}\ \emph {et~al.}(1996)\citenamefont
  {Randall}, \citenamefont {Solja{\v{c}}i{\'c}},\ and\ \citenamefont
  {Guth}}]{randall1996supernatural}%
  \BibitemOpen
  \bibfield  {author} {\bibinfo {author} {\bibfnamefont {L.}~\bibnamefont
  {Randall}}, \bibinfo {author} {\bibfnamefont {M.}~\bibnamefont
  {Solja{\v{c}}i{\'c}}}, \ and\ \bibinfo {author} {\bibfnamefont {A.~H.}\
  \bibnamefont {Guth}},\ }\href {\doibase 10.1016/0550-3213(96)00174-5}
  {\bibfield  {journal} {\bibinfo  {journal} {Nucl. Phys.}\ }\textbf {\bibinfo
  {volume} {B472}},\ \bibinfo {pages} {377} (\bibinfo {year} {1996})},\ \Eprint
  {http://arxiv.org/abs/hep-ph/9512439} {arXiv:hep-ph/9512439} \BibitemShut
  {NoStop}%
\bibitem [{\citenamefont {Stewart}(1997)}]{stewart1997flattening}%
  \BibitemOpen
  \bibfield  {author} {\bibinfo {author} {\bibfnamefont {E.~D.}\ \bibnamefont
  {Stewart}},\ }\href {\doibase 10.1103/physrevd.56.2019} {\bibfield  {journal}
  {\bibinfo  {journal} {Phys. Rev. D}\ }\textbf {\bibinfo {volume} {56}},\
  \bibinfo {pages} {2019} (\bibinfo {year} {1997})},\ \Eprint
  {http://arxiv.org/abs/hep-ph/9703232} {arXiv:hep-ph/9703232} \BibitemShut
  {NoStop}%
\bibitem [{\citenamefont {Adams}\ \emph {et~al.}(1997)\citenamefont {Adams},
  \citenamefont {Ross},\ and\ \citenamefont {Sarkar}}]{adams1997multiple}%
  \BibitemOpen
  \bibfield  {author} {\bibinfo {author} {\bibfnamefont {J.~A.}\ \bibnamefont
  {Adams}}, \bibinfo {author} {\bibfnamefont {G.~G.}\ \bibnamefont {Ross}}, \
  and\ \bibinfo {author} {\bibfnamefont {S.}~\bibnamefont {Sarkar}},\ }\href
  {\doibase 10.1016/s0550-3213(97)00431-8} {\bibfield  {journal} {\bibinfo
  {journal} {Nucl. Phys.}\ }\textbf {\bibinfo {volume} {B503}},\ \bibinfo
  {pages} {405} (\bibinfo {year} {1997})},\ \Eprint
  {http://arxiv.org/abs/hep-ph/9704286} {arXiv:hep-ph/9704286} \BibitemShut
  {NoStop}%
\bibitem [{\citenamefont {Starobinsky}(1998)}]{starobinsky1998beyond}%
  \BibitemOpen
  \bibfield  {author} {\bibinfo {author} {\bibfnamefont {A.~A.}\ \bibnamefont
  {Starobinsky}},\ }\href {\doibase 10.1086/174743} {\bibfield  {journal}
  {\bibinfo  {journal} {Gravitation Cosmol.}\ }\textbf {\bibinfo {volume}
  {4}},\ \bibinfo {pages} {489} (\bibinfo {year} {1998})},\ \Eprint
  {http://arxiv.org/abs/astro-ph/9811360} {arXiv:astro-ph/9811360} \BibitemShut
  {NoStop}%
\bibitem [{\citenamefont {Covi}\ and\ \citenamefont
  {Lyth}(1999)}]{covi1999running}%
  \BibitemOpen
  \bibfield  {author} {\bibinfo {author} {\bibfnamefont {L.}~\bibnamefont
  {Covi}}\ and\ \bibinfo {author} {\bibfnamefont {D.~H.}\ \bibnamefont
  {Lyth}},\ }\href {\doibase 10.1103/physrevd.59.063515} {\bibfield  {journal}
  {\bibinfo  {journal} {Phys. Rev. D}\ }\textbf {\bibinfo {volume} {59}},\
  \bibinfo {pages} {063515} (\bibinfo {year} {1999})},\ \Eprint
  {http://arxiv.org/abs/hep-ph/9809562} {arXiv:hep-ph/9809562} \BibitemShut
  {NoStop}%
\bibitem [{\citenamefont {Martin}\ \emph {et~al.}(2000)\citenamefont {Martin},
  \citenamefont {Riazuelo},\ and\ \citenamefont
  {Sakellariadou}}]{martin2000nonvacuum}%
  \BibitemOpen
  \bibfield  {author} {\bibinfo {author} {\bibfnamefont {J.}~\bibnamefont
  {Martin}}, \bibinfo {author} {\bibfnamefont {A.}~\bibnamefont {Riazuelo}}, \
  and\ \bibinfo {author} {\bibfnamefont {M.}~\bibnamefont {Sakellariadou}},\
  }\href {\doibase 10.1103/physrevd.61.083518} {\bibfield  {journal} {\bibinfo
  {journal} {Phys. Rev. D}\ }\textbf {\bibinfo {volume} {61}},\ \bibinfo
  {pages} {083518} (\bibinfo {year} {2000})},\ \Eprint
  {http://arxiv.org/abs/astro-ph/9904167} {arXiv:astro-ph/9904167} \BibitemShut
  {NoStop}%
\bibitem [{\citenamefont {Chung}\ \emph {et~al.}(2000)\citenamefont {Chung},
  \citenamefont {Kolb}, \citenamefont {Riotto},\ and\ \citenamefont
  {Tkachev}}]{chung2000probing}%
  \BibitemOpen
  \bibfield  {author} {\bibinfo {author} {\bibfnamefont {D.~J.~H.}\
  \bibnamefont {Chung}}, \bibinfo {author} {\bibfnamefont {E.~W.}\ \bibnamefont
  {Kolb}}, \bibinfo {author} {\bibfnamefont {A.}~\bibnamefont {Riotto}}, \ and\
  \bibinfo {author} {\bibfnamefont {I.~I.}\ \bibnamefont {Tkachev}},\ }\href
  {\doibase 10.1103/physrevd.62.043508} {\bibfield  {journal} {\bibinfo
  {journal} {Phys. Rev. D}\ }\textbf {\bibinfo {volume} {62}},\ \bibinfo
  {pages} {043508} (\bibinfo {year} {2000})},\ \Eprint
  {http://arxiv.org/abs/hep-ph/9910437} {arXiv:hep-ph/9910437} \BibitemShut
  {NoStop}%
\bibitem [{\citenamefont {Martin}\ and\ \citenamefont
  {Brandenberger}(2001)}]{martin2001trans}%
  \BibitemOpen
  \bibfield  {author} {\bibinfo {author} {\bibfnamefont {J.}~\bibnamefont
  {Martin}}\ and\ \bibinfo {author} {\bibfnamefont {R.~H.}\ \bibnamefont
  {Brandenberger}},\ }\href {\doibase 10.1103/physrevd.63.123501} {\bibfield
  {journal} {\bibinfo  {journal} {Phys. Rev. D}\ }\textbf {\bibinfo {volume}
  {63}},\ \bibinfo {pages} {123501} (\bibinfo {year} {2001})},\ \Eprint
  {http://arxiv.org/abs/hep-th/0005209} {arXiv:hep-th/0005209} \BibitemShut
  {NoStop}%
\bibitem [{\citenamefont {Joy}\ \emph {et~al.}(2008)\citenamefont {Joy},
  \citenamefont {Sahni},\ and\ \citenamefont {Starobinsky}}]{joy2008new}%
  \BibitemOpen
  \bibfield  {author} {\bibinfo {author} {\bibfnamefont {M.}~\bibnamefont
  {Joy}}, \bibinfo {author} {\bibfnamefont {V.}~\bibnamefont {Sahni}}, \ and\
  \bibinfo {author} {\bibfnamefont {A.~A.}\ \bibnamefont {Starobinsky}},\
  }\href {\doibase 10.1103/physrevd.77.023514} {\bibfield  {journal} {\bibinfo
  {journal} {Phys. Rev. D}\ }\textbf {\bibinfo {volume} {77}},\ \bibinfo
  {pages} {023514} (\bibinfo {year} {2008})},\ \Eprint
  {http://arxiv.org/abs/0711.1585} {arXiv:0711.1585} \BibitemShut {NoStop}%
\bibitem [{\citenamefont {Barnaby}\ and\ \citenamefont
  {Huang}(2009)}]{barnaby2009particle}%
  \BibitemOpen
  \bibfield  {author} {\bibinfo {author} {\bibfnamefont {N.}~\bibnamefont
  {Barnaby}}\ and\ \bibinfo {author} {\bibfnamefont {Z.}~\bibnamefont
  {Huang}},\ }\href {\doibase 10.1103/physrevd.80.126018} {\bibfield  {journal}
  {\bibinfo  {journal} {Phys. Rev. D}\ }\textbf {\bibinfo {volume} {80}},\
  \bibinfo {pages} {126018} (\bibinfo {year} {2009})},\ \Eprint
  {http://arxiv.org/abs/0909.0751} {arXiv:0909.0751} \BibitemShut {NoStop}%
\bibitem [{\citenamefont {Barnaby}(2010)}]{barnaby2010features}%
  \BibitemOpen
  \bibfield  {author} {\bibinfo {author} {\bibfnamefont {N.}~\bibnamefont
  {Barnaby}},\ }\href {\doibase 10.1103/physrevd.82.106009} {\bibfield
  {journal} {\bibinfo  {journal} {Phys. Rev. D}\ }\textbf {\bibinfo {volume}
  {82}},\ \bibinfo {pages} {106009} (\bibinfo {year} {2010})},\ \Eprint
  {http://arxiv.org/abs/1006.4615} {arXiv:1006.4615} \BibitemShut {NoStop}%
\bibitem [{\citenamefont {Ben-Dayan}\ and\ \citenamefont
  {Brustein}(2010)}]{ben2010cosmic}%
  \BibitemOpen
  \bibfield  {author} {\bibinfo {author} {\bibfnamefont {I.}~\bibnamefont
  {Ben-Dayan}}\ and\ \bibinfo {author} {\bibfnamefont {R.}~\bibnamefont
  {Brustein}},\ }\href {\doibase 10.1088/1475-7516/2010/09/007} {\bibfield
  {journal} {\bibinfo  {journal} {J. Cosmol. Astropart. Phys.}\ }\textbf
  {\bibinfo {volume} {09}},\ \bibinfo {pages} {007} (\bibinfo {year} {2010})},\
  \Eprint {http://arxiv.org/abs/0907.2384} {arXiv:0907.2384} \BibitemShut
  {NoStop}%
\bibitem [{\citenamefont {Gong}\ and\ \citenamefont
  {Sasaki}(2011)}]{gong2011waterfall}%
  \BibitemOpen
  \bibfield  {author} {\bibinfo {author} {\bibfnamefont {J.-O.}\ \bibnamefont
  {Gong}}\ and\ \bibinfo {author} {\bibfnamefont {M.}~\bibnamefont {Sasaki}},\
  }\href {\doibase 10.1088/1475-7516/2011/03/028} {\bibfield  {journal}
  {\bibinfo  {journal} {J. Cosmol. Astropart. Phys.}\ }\textbf {\bibinfo
  {volume} {03}},\ \bibinfo {pages} {028} (\bibinfo {year} {2011})},\ \Eprint
  {http://arxiv.org/abs/1010.3405} {arXiv:1010.3405} \BibitemShut {NoStop}%
\bibitem [{\citenamefont {Lyth}(2011)}]{lyth2011contribution}%
  \BibitemOpen
  \bibfield  {author} {\bibinfo {author} {\bibfnamefont {D.~H.}\ \bibnamefont
  {Lyth}},\ }\href {\doibase 10.1088/1475-7516/2011/07/035} {\bibfield
  {journal} {\bibinfo  {journal} {J. Cosmol. Astropart. Phys.}\ }\textbf
  {\bibinfo {volume} {07}},\ \bibinfo {pages} {035} (\bibinfo {year} {2011})},\
  \Eprint {http://arxiv.org/abs/1012.4617} {arXiv:1012.4617} \BibitemShut
  {NoStop}%
\bibitem [{\citenamefont {Bugaev}\ and\ \citenamefont
  {Klimai}(2011)}]{bugaev2011curvature}%
  \BibitemOpen
  \bibfield  {author} {\bibinfo {author} {\bibfnamefont {E.}~\bibnamefont
  {Bugaev}}\ and\ \bibinfo {author} {\bibfnamefont {P.}~\bibnamefont
  {Klimai}},\ }\href {\doibase 10.1088/1475-7516/2011/11/028} {\bibfield
  {journal} {\bibinfo  {journal} {J. Cosmol. Astropart. Phys.}\ }\textbf
  {\bibinfo {volume} {11}},\ \bibinfo {pages} {028} (\bibinfo {year} {2011})},\
  \Eprint {http://arxiv.org/abs/1107.3754} {arXiv:1107.3754} \BibitemShut
  {NoStop}%
\bibitem [{\citenamefont {Barnaby}\ and\ \citenamefont
  {Peloso}(2011)}]{barnaby2011large}%
  \BibitemOpen
  \bibfield  {author} {\bibinfo {author} {\bibfnamefont {N.}~\bibnamefont
  {Barnaby}}\ and\ \bibinfo {author} {\bibfnamefont {M.}~\bibnamefont
  {Peloso}},\ }\href {\doibase 10.1103/physrevlett.106.181301} {\bibfield
  {journal} {\bibinfo  {journal} {Phys. Rev. Lett.}\ }\textbf {\bibinfo
  {volume} {106}},\ \bibinfo {pages} {181301} (\bibinfo {year} {2011})},\
  \Eprint {http://arxiv.org/abs/1011.1500} {arXiv:1011.1500} \BibitemShut
  {NoStop}%
\bibitem [{\citenamefont {Ach{\'u}carro}\ \emph {et~al.}(2011)\citenamefont
  {Ach{\'u}carro}, \citenamefont {Gong}, \citenamefont {Hardeman},
  \citenamefont {Palma},\ and\ \citenamefont {Patil}}]{achucarro2011features}%
  \BibitemOpen
  \bibfield  {author} {\bibinfo {author} {\bibfnamefont {A.}~\bibnamefont
  {Ach{\'u}carro}}, \bibinfo {author} {\bibfnamefont {J.-O.}\ \bibnamefont
  {Gong}}, \bibinfo {author} {\bibfnamefont {S.}~\bibnamefont {Hardeman}},
  \bibinfo {author} {\bibfnamefont {G.~A.}\ \bibnamefont {Palma}}, \ and\
  \bibinfo {author} {\bibfnamefont {S.~P.}\ \bibnamefont {Patil}},\ }\href
  {\doibase 10.1088/1475-7516/2011/01/030} {\bibfield  {journal} {\bibinfo
  {journal} {J. Cosmol. Astropart. Phys.}\ }\textbf {\bibinfo {volume} {01}},\
  \bibinfo {pages} {030} (\bibinfo {year} {2011})},\ \Eprint
  {http://arxiv.org/abs/1010.3693} {arXiv:1010.3693} \BibitemShut {NoStop}%
\bibitem [{\citenamefont {Cespedes}\ \emph {et~al.}(2012)\citenamefont
  {Cespedes}, \citenamefont {Atal},\ and\ \citenamefont
  {Palma}}]{cespedes2012importance}%
  \BibitemOpen
  \bibfield  {author} {\bibinfo {author} {\bibfnamefont {S.}~\bibnamefont
  {Cespedes}}, \bibinfo {author} {\bibfnamefont {V.}~\bibnamefont {Atal}}, \
  and\ \bibinfo {author} {\bibfnamefont {G.~A.}\ \bibnamefont {Palma}},\ }\href
  {\doibase 10.1088/1475-7516/2012/05/008} {\bibfield  {journal} {\bibinfo
  {journal} {J. Cosmol. Astropart. Phys.}\ }\textbf {\bibinfo {volume} {05}},\
  \bibinfo {pages} {008} (\bibinfo {year} {2012})},\ \Eprint
  {http://arxiv.org/abs/1201.4848} {arXiv:1201.4848} \BibitemShut {NoStop}%
\bibitem [{\citenamefont {Barnaby}\ \emph {et~al.}(2012)\citenamefont
  {Barnaby}, \citenamefont {Pajer},\ and\ \citenamefont
  {Peloso}}]{barnaby2012gauge}%
  \BibitemOpen
  \bibfield  {author} {\bibinfo {author} {\bibfnamefont {N.}~\bibnamefont
  {Barnaby}}, \bibinfo {author} {\bibfnamefont {E.}~\bibnamefont {Pajer}}, \
  and\ \bibinfo {author} {\bibfnamefont {M.}~\bibnamefont {Peloso}},\ }\href
  {\doibase 10.1103/physrevd.85.023525} {\bibfield  {journal} {\bibinfo
  {journal} {Phys. Rev. D}\ }\textbf {\bibinfo {volume} {85}},\ \bibinfo
  {pages} {023525} (\bibinfo {year} {2012})},\ \Eprint
  {http://arxiv.org/abs/1110.3327} {arXiv:1110.3327} \BibitemShut {NoStop}%
\bibitem [{\citenamefont {Erickcek}\ and\ \citenamefont
  {Sigurdson}(2011)}]{erickcek2011reheating}%
  \BibitemOpen
  \bibfield  {author} {\bibinfo {author} {\bibfnamefont {A.~L.}\ \bibnamefont
  {Erickcek}}\ and\ \bibinfo {author} {\bibfnamefont {K.}~\bibnamefont
  {Sigurdson}},\ }\href {\doibase 10.1103/physrevd.84.083503} {\bibfield
  {journal} {\bibinfo  {journal} {Phys. Rev. D}\ }\textbf {\bibinfo {volume}
  {84}},\ \bibinfo {pages} {083503} (\bibinfo {year} {2011})},\ \Eprint
  {http://arxiv.org/abs/1106.0536} {arXiv:1106.0536} \BibitemShut {NoStop}%
\bibitem [{\citenamefont {Barenboim}\ and\ \citenamefont
  {Rasero}(2014)}]{barenboim2014structure}%
  \BibitemOpen
  \bibfield  {author} {\bibinfo {author} {\bibfnamefont {G.}~\bibnamefont
  {Barenboim}}\ and\ \bibinfo {author} {\bibfnamefont {J.}~\bibnamefont
  {Rasero}},\ }\href {\doibase 10.1007/jhep04(2014)138} {\bibfield  {journal}
  {\bibinfo  {journal} {J. High Energy Phys.}\ }\textbf {\bibinfo {volume}
  {4}},\ \bibinfo {pages} {138} (\bibinfo {year} {2014})},\ \Eprint
  {http://arxiv.org/abs/1311.4034} {arXiv:1311.4034} \BibitemShut {NoStop}%
\bibitem [{\citenamefont {Fan}\ \emph {et~al.}(2014)\citenamefont {Fan},
  \citenamefont {{\"O}zsoy},\ and\ \citenamefont {Watson}}]{fan2014nonthermal}%
  \BibitemOpen
  \bibfield  {author} {\bibinfo {author} {\bibfnamefont {J.~J.}\ \bibnamefont
  {Fan}}, \bibinfo {author} {\bibfnamefont {O.}~\bibnamefont {{\"O}zsoy}}, \
  and\ \bibinfo {author} {\bibfnamefont {S.}~\bibnamefont {Watson}},\ }\href
  {\doibase 10.1103/physrevd.90.043536} {\bibfield  {journal} {\bibinfo
  {journal} {Phys. Rev. D}\ }\textbf {\bibinfo {volume} {90}},\ \bibinfo
  {pages} {043536} (\bibinfo {year} {2014})},\ \Eprint
  {http://arxiv.org/abs/1405.7373} {arXiv:1405.7373} \BibitemShut {NoStop}%
\bibitem [{\citenamefont {Erickcek}(2015)}]{erickcek2015dark}%
  \BibitemOpen
  \bibfield  {author} {\bibinfo {author} {\bibfnamefont {A.~L.}\ \bibnamefont
  {Erickcek}},\ }\href {\doibase 10.1103/physrevd.92.103505} {\bibfield
  {journal} {\bibinfo  {journal} {Phys. Rev. D}\ }\textbf {\bibinfo {volume}
  {92}},\ \bibinfo {pages} {103505} (\bibinfo {year} {2015})},\ \Eprint
  {http://arxiv.org/abs/1504.03335} {arXiv:1504.03335} \BibitemShut {NoStop}%
\bibitem [{\citenamefont {Redmond}\ \emph {et~al.}(2018)\citenamefont
  {Redmond}, \citenamefont {Trezza},\ and\ \citenamefont
  {Erickcek}}]{redmond2018growth}%
  \BibitemOpen
  \bibfield  {author} {\bibinfo {author} {\bibfnamefont {K.}~\bibnamefont
  {Redmond}}, \bibinfo {author} {\bibfnamefont {A.}~\bibnamefont {Trezza}}, \
  and\ \bibinfo {author} {\bibfnamefont {A.~L.}\ \bibnamefont {Erickcek}},\
  }\href {\doibase 10.1103/physrevd.98.063504} {\bibfield  {journal} {\bibinfo
  {journal} {Phys. Rev. D}\ }\textbf {\bibinfo {volume} {98}},\ \bibinfo
  {pages} {063504} (\bibinfo {year} {2018})},\ \Eprint
  {http://arxiv.org/abs/1807.01327} {arXiv:1807.01327} \BibitemShut {NoStop}%
\bibitem [{\citenamefont {Blanco}\ \emph {et~al.}()\citenamefont {Blanco},
  \citenamefont {Delos}, \citenamefont {Erickcek},\ and\ \citenamefont
  {Hooper}}]{blanco2019annihilation}%
  \BibitemOpen
  \bibfield  {author} {\bibinfo {author} {\bibfnamefont {C.}~\bibnamefont
  {Blanco}}, \bibinfo {author} {\bibfnamefont {M.~S.}\ \bibnamefont {Delos}},
  \bibinfo {author} {\bibfnamefont {A.~L.}\ \bibnamefont {Erickcek}}, \ and\
  \bibinfo {author} {\bibfnamefont {D.}~\bibnamefont {Hooper}},\ }\href@noop {}
  {\ }\Eprint {http://arxiv.org/abs/1906.00010} {arXiv:1906.00010} \BibitemShut
  {NoStop}%
\bibitem [{\citenamefont {Taylor}\ and\ \citenamefont
  {Babul}(2001)}]{taylor2001dynamics}%
  \BibitemOpen
  \bibfield  {author} {\bibinfo {author} {\bibfnamefont {J.~E.}\ \bibnamefont
  {Taylor}}\ and\ \bibinfo {author} {\bibfnamefont {A.}~\bibnamefont {Babul}},\
  }\href {\doibase 10.1086/322276} {\bibfield  {journal} {\bibinfo  {journal}
  {Astrophys. J.}\ }\textbf {\bibinfo {volume} {559}},\ \bibinfo {pages} {716}
  (\bibinfo {year} {2001})},\ \Eprint {http://arxiv.org/abs/astro-ph/0012305}
  {arXiv:astro-ph/0012305} \BibitemShut {NoStop}%
\bibitem [{\citenamefont {Hayashi}\ \emph {et~al.}(2003)\citenamefont
  {Hayashi}, \citenamefont {Navarro}, \citenamefont {Taylor}, \citenamefont
  {Stadel},\ and\ \citenamefont {Quinn}}]{hayashi2003structural}%
  \BibitemOpen
  \bibfield  {author} {\bibinfo {author} {\bibfnamefont {E.}~\bibnamefont
  {Hayashi}}, \bibinfo {author} {\bibfnamefont {J.~F.}\ \bibnamefont
  {Navarro}}, \bibinfo {author} {\bibfnamefont {J.~E.}\ \bibnamefont {Taylor}},
  \bibinfo {author} {\bibfnamefont {J.}~\bibnamefont {Stadel}}, \ and\ \bibinfo
  {author} {\bibfnamefont {T.}~\bibnamefont {Quinn}},\ }\href {\doibase
  10.1086/345788} {\bibfield  {journal} {\bibinfo  {journal} {Astrophys. J.}\
  }\textbf {\bibinfo {volume} {584}},\ \bibinfo {pages} {541} (\bibinfo {year}
  {2003})},\ \Eprint {http://arxiv.org/abs/astro-ph/0203004}
  {arXiv:astro-ph/0203004} \BibitemShut {NoStop}%
\bibitem [{\citenamefont {Pe{\~n}arrubia}\ and\ \citenamefont
  {Benson}(2005)}]{penarrubia2005effects}%
  \BibitemOpen
  \bibfield  {author} {\bibinfo {author} {\bibfnamefont {J.}~\bibnamefont
  {Pe{\~n}arrubia}}\ and\ \bibinfo {author} {\bibfnamefont {A.~J.}\
  \bibnamefont {Benson}},\ }\href {\doibase 10.1111/j.1365-2966.2005.09633.x}
  {\bibfield  {journal} {\bibinfo  {journal} {Mon. Not. R. Astron. Soc.}\
  }\textbf {\bibinfo {volume} {364}},\ \bibinfo {pages} {977} (\bibinfo {year}
  {2005})},\ \Eprint {http://arxiv.org/abs/astro-ph/0412370}
  {arXiv:astro-ph/0412370} \BibitemShut {NoStop}%
\bibitem [{\citenamefont {van~den Bosch}\ \emph {et~al.}(2005)\citenamefont
  {van~den Bosch}, \citenamefont {Tormen},\ and\ \citenamefont
  {Giocoli}}]{van2005mass}%
  \BibitemOpen
  \bibfield  {author} {\bibinfo {author} {\bibfnamefont {F.~C.}\ \bibnamefont
  {van~den Bosch}}, \bibinfo {author} {\bibfnamefont {G.}~\bibnamefont
  {Tormen}}, \ and\ \bibinfo {author} {\bibfnamefont {C.}~\bibnamefont
  {Giocoli}},\ }\href {\doibase 10.1111/j.1365-2966.2005.08964.x} {\bibfield
  {journal} {\bibinfo  {journal} {Mon. Not. R. Astron. Soc.}\ }\textbf
  {\bibinfo {volume} {359}},\ \bibinfo {pages} {1029} (\bibinfo {year}
  {2005})},\ \Eprint {http://arxiv.org/abs/astro-ph/0409201}
  {arXiv:astro-ph/0409201} \BibitemShut {NoStop}%
\bibitem [{\citenamefont {Zentner}\ \emph {et~al.}(2005)\citenamefont
  {Zentner}, \citenamefont {Berlind}, \citenamefont {Bullock}, \citenamefont
  {Kravtsov},\ and\ \citenamefont {Wechsler}}]{zentner2005physics}%
  \BibitemOpen
  \bibfield  {author} {\bibinfo {author} {\bibfnamefont {A.~R.}\ \bibnamefont
  {Zentner}}, \bibinfo {author} {\bibfnamefont {A.~A.}\ \bibnamefont
  {Berlind}}, \bibinfo {author} {\bibfnamefont {J.~S.}\ \bibnamefont
  {Bullock}}, \bibinfo {author} {\bibfnamefont {A.~V.}\ \bibnamefont
  {Kravtsov}}, \ and\ \bibinfo {author} {\bibfnamefont {R.~H.}\ \bibnamefont
  {Wechsler}},\ }\href {\doibase 10.1086/428898} {\bibfield  {journal}
  {\bibinfo  {journal} {Astrophys. J.}\ }\textbf {\bibinfo {volume} {624}},\
  \bibinfo {pages} {505} (\bibinfo {year} {2005})},\ \Eprint
  {http://arxiv.org/abs/astro-ph/0411586} {arXiv:astro-ph/0411586} \BibitemShut
  {NoStop}%
\bibitem [{\citenamefont {Kampakoglou}\ and\ \citenamefont
  {Benson}(2007)}]{kampakoglou2006tidal}%
  \BibitemOpen
  \bibfield  {author} {\bibinfo {author} {\bibfnamefont {M.}~\bibnamefont
  {Kampakoglou}}\ and\ \bibinfo {author} {\bibfnamefont {A.~J.}\ \bibnamefont
  {Benson}},\ }\href {\doibase 10.1111/j.1365-2966.2006.11223.x} {\bibfield
  {journal} {\bibinfo  {journal} {Mon. Not. R. Astron. Soc.}\ }\textbf
  {\bibinfo {volume} {374}},\ \bibinfo {pages} {775} (\bibinfo {year}
  {2007})},\ \Eprint {http://arxiv.org/abs/astro-ph/0607024}
  {arXiv:astro-ph/0607024} \BibitemShut {NoStop}%
\bibitem [{\citenamefont {Gan}\ \emph {et~al.}(2010)\citenamefont {Gan},
  \citenamefont {Kang}, \citenamefont {van~den Bosch},\ and\ \citenamefont
  {Hou}}]{gan2010improved}%
  \BibitemOpen
  \bibfield  {author} {\bibinfo {author} {\bibfnamefont {J.}~\bibnamefont
  {Gan}}, \bibinfo {author} {\bibfnamefont {X.}~\bibnamefont {Kang}}, \bibinfo
  {author} {\bibfnamefont {F.~C.}\ \bibnamefont {van~den Bosch}}, \ and\
  \bibinfo {author} {\bibfnamefont {J.}~\bibnamefont {Hou}},\ }\href {\doibase
  10.1111/j.1365-2966.2010.17266.x} {\bibfield  {journal} {\bibinfo  {journal}
  {Mon. Not. R. Astron. Soc.}\ }\textbf {\bibinfo {volume} {408}},\ \bibinfo
  {pages} {2201} (\bibinfo {year} {2010})},\ \Eprint
  {http://arxiv.org/abs/1007.0023} {arXiv:1007.0023} \BibitemShut {NoStop}%
\bibitem [{\citenamefont {Penarrubia}\ \emph {et~al.}(2010)\citenamefont
  {Penarrubia}, \citenamefont {Benson}, \citenamefont {Walker}, \citenamefont
  {Gilmore}, \citenamefont {McConnachie},\ and\ \citenamefont
  {Mayer}}]{penarrubia2010impact}%
  \BibitemOpen
  \bibfield  {author} {\bibinfo {author} {\bibfnamefont {J.}~\bibnamefont
  {Penarrubia}}, \bibinfo {author} {\bibfnamefont {A.~J.}\ \bibnamefont
  {Benson}}, \bibinfo {author} {\bibfnamefont {M.~G.}\ \bibnamefont {Walker}},
  \bibinfo {author} {\bibfnamefont {G.}~\bibnamefont {Gilmore}}, \bibinfo
  {author} {\bibfnamefont {A.~W.}\ \bibnamefont {McConnachie}}, \ and\ \bibinfo
  {author} {\bibfnamefont {L.}~\bibnamefont {Mayer}},\ }\href {\doibase
  10.1111/j.1365-2966.2010.16762.x} {\bibfield  {journal} {\bibinfo  {journal}
  {Mon. Not. R. Astron. Soc.}\ }\textbf {\bibinfo {volume} {406}},\ \bibinfo
  {pages} {1290} (\bibinfo {year} {2010})},\ \Eprint
  {http://arxiv.org/abs/1002.3376} {arXiv:1002.3376} \BibitemShut {NoStop}%
\bibitem [{\citenamefont {Pullen}\ \emph {et~al.}(2014)\citenamefont {Pullen},
  \citenamefont {Benson},\ and\ \citenamefont
  {Moustakas}}]{pullen2014nonlinear}%
  \BibitemOpen
  \bibfield  {author} {\bibinfo {author} {\bibfnamefont {A.~R.}\ \bibnamefont
  {Pullen}}, \bibinfo {author} {\bibfnamefont {A.~J.}\ \bibnamefont {Benson}},
  \ and\ \bibinfo {author} {\bibfnamefont {L.~A.}\ \bibnamefont {Moustakas}},\
  }\href {\doibase 10.1088/0004-637x/792/1/24} {\bibfield  {journal} {\bibinfo
  {journal} {Astrophys. J.}\ }\textbf {\bibinfo {volume} {792}},\ \bibinfo
  {pages} {24} (\bibinfo {year} {2014})},\ \Eprint
  {http://arxiv.org/abs/1407.8189} {arXiv:1407.8189} \BibitemShut {NoStop}%
\bibitem [{\citenamefont {Jiang}\ and\ \citenamefont {van~den
  Bosch}(2016)}]{jiang2016statistics}%
  \BibitemOpen
  \bibfield  {author} {\bibinfo {author} {\bibfnamefont {F.}~\bibnamefont
  {Jiang}}\ and\ \bibinfo {author} {\bibfnamefont {F.~C.}\ \bibnamefont
  {van~den Bosch}},\ }\href {\doibase 10.1093/mnras/stw439} {\bibfield
  {journal} {\bibinfo  {journal} {Mon. Not. R. Astron. Soc.}\ }\textbf
  {\bibinfo {volume} {458}},\ \bibinfo {pages} {2848} (\bibinfo {year}
  {2016})},\ \Eprint {http://arxiv.org/abs/1403.6827} {arXiv:1403.6827}
  \BibitemShut {NoStop}%
\bibitem [{\citenamefont {van~den Bosch}(2017)}]{van2017dissecting}%
  \BibitemOpen
  \bibfield  {author} {\bibinfo {author} {\bibfnamefont {F.~C.}\ \bibnamefont
  {van~den Bosch}},\ }\href {\doibase 10.1093/mnras/stx520} {\bibfield
  {journal} {\bibinfo  {journal} {Mon. Not. R. Astron. Soc.}\ }\textbf
  {\bibinfo {volume} {468}},\ \bibinfo {pages} {885} (\bibinfo {year}
  {2017})},\ \Eprint {http://arxiv.org/abs/1611.02657} {arXiv:1611.02657}
  \BibitemShut {NoStop}%
\bibitem [{\citenamefont {van~den Bosch}\ \emph {et~al.}(2018)\citenamefont
  {van~den Bosch}, \citenamefont {Ogiya}, \citenamefont {Hahn},\ and\
  \citenamefont {Burkert}}]{van2017disruption}%
  \BibitemOpen
  \bibfield  {author} {\bibinfo {author} {\bibfnamefont {F.~C.}\ \bibnamefont
  {van~den Bosch}}, \bibinfo {author} {\bibfnamefont {G.}~\bibnamefont
  {Ogiya}}, \bibinfo {author} {\bibfnamefont {O.}~\bibnamefont {Hahn}}, \ and\
  \bibinfo {author} {\bibfnamefont {A.}~\bibnamefont {Burkert}},\ }\href
  {\doibase 10.1093/mnras/stx2956} {\bibfield  {journal} {\bibinfo  {journal}
  {Mon. Not. R. Astron. Soc.}\ }\textbf {\bibinfo {volume} {474}},\ \bibinfo
  {pages} {3043} (\bibinfo {year} {2018})},\ \Eprint
  {http://arxiv.org/abs/1711.05276} {arXiv:1711.05276} \BibitemShut {NoStop}%
\bibitem [{\citenamefont {van~den Bosch}\ and\ \citenamefont
  {Ogiya}(2018)}]{van2018dark}%
  \BibitemOpen
  \bibfield  {author} {\bibinfo {author} {\bibfnamefont {F.~C.}\ \bibnamefont
  {van~den Bosch}}\ and\ \bibinfo {author} {\bibfnamefont {G.}~\bibnamefont
  {Ogiya}},\ }\href {\doibase 10.1093/mnras/sty084} {\bibfield  {journal}
  {\bibinfo  {journal} {Mon. Not. R. Astron. Soc.}\ }\textbf {\bibinfo {volume}
  {475}},\ \bibinfo {pages} {4066} (\bibinfo {year} {2018})},\ \Eprint
  {http://arxiv.org/abs/1801.05427} {arXiv:1801.05427} \BibitemShut {NoStop}%
\bibitem [{\citenamefont {Ogiya}\ \emph {et~al.}(2019)\citenamefont {Ogiya},
  \citenamefont {Van~den Bosch}, \citenamefont {Hahn}, \citenamefont {Green},
  \citenamefont {Miller},\ and\ \citenamefont {Burkert}}]{ogiya2019dash}%
  \BibitemOpen
  \bibfield  {author} {\bibinfo {author} {\bibfnamefont {G.}~\bibnamefont
  {Ogiya}}, \bibinfo {author} {\bibfnamefont {F.~C.}\ \bibnamefont {Van~den
  Bosch}}, \bibinfo {author} {\bibfnamefont {O.}~\bibnamefont {Hahn}}, \bibinfo
  {author} {\bibfnamefont {S.~B.}\ \bibnamefont {Green}}, \bibinfo {author}
  {\bibfnamefont {T.~B.}\ \bibnamefont {Miller}}, \ and\ \bibinfo {author}
  {\bibfnamefont {A.}~\bibnamefont {Burkert}},\ }\href {\doibase
  10.1093/mnras/stz375} {\bibfield  {journal} {\bibinfo  {journal} {Mon. Not.
  R. Astron. Soc.}\ }\textbf {\bibinfo {volume} {485}},\ \bibinfo {pages} {189}
  (\bibinfo {year} {2019})},\ \Eprint {http://arxiv.org/abs/1901.08601}
  {arXiv:1901.08601} \BibitemShut {NoStop}%
\bibitem [{\citenamefont {Errani}\ and\ \citenamefont
  {Pe{\~n}arrubia}()}]{errani2019can}%
  \BibitemOpen
  \bibfield  {author} {\bibinfo {author} {\bibfnamefont {R.}~\bibnamefont
  {Errani}}\ and\ \bibinfo {author} {\bibfnamefont {J.}~\bibnamefont
  {Pe{\~n}arrubia}},\ }\href@noop {} {\ }\Eprint
  {http://arxiv.org/abs/1906.01642} {arXiv:1906.01642} \BibitemShut {NoStop}%
\bibitem [{\citenamefont {Delos}(2019)}]{delos2019tidal}%
  \BibitemOpen
  \bibfield  {author} {\bibinfo {author} {\bibfnamefont {M.~S.}\ \bibnamefont
  {Delos}},\ }\href {\doibase 10.1103/physrevd.100.063505} {\bibfield
  {journal} {\bibinfo  {journal} {Phys. Rev. D}\ }\textbf {\bibinfo {volume}
  {100}},\ \bibinfo {pages} {063505} (\bibinfo {year} {2019})},\ \Eprint
  {http://arxiv.org/abs/1906.10690} {arXiv:1906.10690} \BibitemShut {NoStop}%
\bibitem [{\citenamefont {Berezinsky}\ \emph {et~al.}(2006)\citenamefont
  {Berezinsky}, \citenamefont {Dokuchaev},\ and\ \citenamefont
  {Eroshenko}}]{berezinsky2006destruction}%
  \BibitemOpen
  \bibfield  {author} {\bibinfo {author} {\bibfnamefont {V.}~\bibnamefont
  {Berezinsky}}, \bibinfo {author} {\bibfnamefont {V.}~\bibnamefont
  {Dokuchaev}}, \ and\ \bibinfo {author} {\bibfnamefont {Y.}~\bibnamefont
  {Eroshenko}},\ }\href {\doibase 10.1103/physrevd.73.063504} {\bibfield
  {journal} {\bibinfo  {journal} {Phys. Rev. D}\ }\textbf {\bibinfo {volume}
  {73}},\ \bibinfo {pages} {063504} (\bibinfo {year} {2006})},\ \Eprint
  {http://arxiv.org/abs/astro-ph/0511494} {arXiv:astro-ph/0511494} \BibitemShut
  {NoStop}%
\bibitem [{\citenamefont {Berezinsky}\ \emph {et~al.}(2014)\citenamefont
  {Berezinsky}, \citenamefont {Dokuchaev},\ and\ \citenamefont
  {Eroshenko}}]{berezinsky2014small}%
  \BibitemOpen
  \bibfield  {author} {\bibinfo {author} {\bibfnamefont {V.~S.}\ \bibnamefont
  {Berezinsky}}, \bibinfo {author} {\bibfnamefont {V.~I.}\ \bibnamefont
  {Dokuchaev}}, \ and\ \bibinfo {author} {\bibfnamefont {Y.~N.}\ \bibnamefont
  {Eroshenko}},\ }\href {\doibase 10.3367/UFNe.0184.201401a.0003} {\bibfield
  {journal} {\bibinfo  {journal} {Phys. Usp.}\ }\textbf {\bibinfo {volume}
  {57}},\ \bibinfo {pages} {1} (\bibinfo {year} {2014})},\ \Eprint
  {http://arxiv.org/abs/1405.2204} {arXiv:1405.2204} \BibitemShut {NoStop}%
\bibitem [{\citenamefont {Green}\ and\ \citenamefont
  {Goodwin}(2007)}]{green2007mini}%
  \BibitemOpen
  \bibfield  {author} {\bibinfo {author} {\bibfnamefont {A.~M.}\ \bibnamefont
  {Green}}\ and\ \bibinfo {author} {\bibfnamefont {S.~P.}\ \bibnamefont
  {Goodwin}},\ }\href {\doibase 10.1111/j.1365-2966.2007.11397.x} {\bibfield
  {journal} {\bibinfo  {journal} {Mon. Not. R. Astron. Soc.}\ }\textbf
  {\bibinfo {volume} {375}},\ \bibinfo {pages} {1111} (\bibinfo {year}
  {2007})},\ \Eprint {http://arxiv.org/abs/astro-ph/0604142}
  {arXiv:astro-ph/0604142} \BibitemShut {NoStop}%
\bibitem [{\citenamefont {Zhao}\ \emph {et~al.}(2007)\citenamefont {Zhao},
  \citenamefont {Hooper}, \citenamefont {Angus}, \citenamefont {Taylor},\ and\
  \citenamefont {Silk}}]{zhao2007tidal}%
  \BibitemOpen
  \bibfield  {author} {\bibinfo {author} {\bibfnamefont {H.}~\bibnamefont
  {Zhao}}, \bibinfo {author} {\bibfnamefont {D.}~\bibnamefont {Hooper}},
  \bibinfo {author} {\bibfnamefont {G.~W.}\ \bibnamefont {Angus}}, \bibinfo
  {author} {\bibfnamefont {J.~E.}\ \bibnamefont {Taylor}}, \ and\ \bibinfo
  {author} {\bibfnamefont {J.}~\bibnamefont {Silk}},\ }\href {\doibase
  10.1086/509649} {\bibfield  {journal} {\bibinfo  {journal} {Astrophys. J.}\
  }\textbf {\bibinfo {volume} {654}},\ \bibinfo {pages} {697} (\bibinfo {year}
  {2007})},\ \Eprint {http://arxiv.org/abs/astro-ph/0508215}
  {arXiv:astro-ph/0508215} \BibitemShut {NoStop}%
\bibitem [{\citenamefont {Angus}\ and\ \citenamefont
  {Zhao}(2007)}]{angus2007cold}%
  \BibitemOpen
  \bibfield  {author} {\bibinfo {author} {\bibfnamefont {G.}~\bibnamefont
  {Angus}}\ and\ \bibinfo {author} {\bibfnamefont {H.}~\bibnamefont {Zhao}},\
  }\href {\doibase 10.1111/j.1365-2966.2007.11400.x} {\bibfield  {journal}
  {\bibinfo  {journal} {Mon. Not. R. Astron. Soc.}\ }\textbf {\bibinfo {volume}
  {375}},\ \bibinfo {pages} {1146} (\bibinfo {year} {2007})},\ \Eprint
  {http://arxiv.org/abs/astro-ph/0608580} {arXiv:astro-ph/0608580} \BibitemShut
  {NoStop}%
\bibitem [{\citenamefont {Polisensky}\ and\ \citenamefont
  {Ricotti}(2015)}]{polisensky2015fingerprints}%
  \BibitemOpen
  \bibfield  {author} {\bibinfo {author} {\bibfnamefont {E.}~\bibnamefont
  {Polisensky}}\ and\ \bibinfo {author} {\bibfnamefont {M.}~\bibnamefont
  {Ricotti}},\ }\href {\doibase 10.1093/mnras/stv736} {\bibfield  {journal}
  {\bibinfo  {journal} {Mon. Not. R. Astron. Soc.}\ }\textbf {\bibinfo {volume}
  {450}},\ \bibinfo {pages} {2172} (\bibinfo {year} {2015})},\ \Eprint
  {http://arxiv.org/abs/1504.02126} {arXiv:1504.02126} \BibitemShut {NoStop}%
\bibitem [{\citenamefont {Ogiya}\ and\ \citenamefont
  {Hahn}(2018)}]{ogiya2017sets}%
  \BibitemOpen
  \bibfield  {author} {\bibinfo {author} {\bibfnamefont {G.}~\bibnamefont
  {Ogiya}}\ and\ \bibinfo {author} {\bibfnamefont {O.}~\bibnamefont {Hahn}},\
  }\href {\doibase 10.1093/mnras/stx2639} {\bibfield  {journal} {\bibinfo
  {journal} {Mon. Not. R. Astron. Soc.}\ }\textbf {\bibinfo {volume} {473}},\
  \bibinfo {pages} {4339} (\bibinfo {year} {2018})},\ \Eprint
  {http://arxiv.org/abs/1707.07693} {arXiv:1707.07693} \BibitemShut {NoStop}%
\bibitem [{\citenamefont {Delos}\ \emph
  {et~al.}(2018{\natexlab{a}})\citenamefont {Delos}, \citenamefont {Erickcek},
  \citenamefont {Bailey},\ and\ \citenamefont
  {Alvarez}}]{delos2018ultracompact}%
  \BibitemOpen
  \bibfield  {author} {\bibinfo {author} {\bibfnamefont {M.~S.}\ \bibnamefont
  {Delos}}, \bibinfo {author} {\bibfnamefont {A.~L.}\ \bibnamefont {Erickcek}},
  \bibinfo {author} {\bibfnamefont {A.~P.}\ \bibnamefont {Bailey}}, \ and\
  \bibinfo {author} {\bibfnamefont {M.~A.}\ \bibnamefont {Alvarez}},\ }\href
  {\doibase 10.1103/physrevd.97.041303} {\bibfield  {journal} {\bibinfo
  {journal} {Phys. Rev. D}\ }\textbf {\bibinfo {volume} {97}},\ \bibinfo
  {pages} {041303(R)} (\bibinfo {year} {2018}{\natexlab{a}})},\ \Eprint
  {http://arxiv.org/abs/1712.05421} {arXiv:1712.05421} \BibitemShut {NoStop}%
\bibitem [{\citenamefont {Delos}\ \emph
  {et~al.}(2018{\natexlab{b}})\citenamefont {Delos}, \citenamefont {Erickcek},
  \citenamefont {Bailey},\ and\ \citenamefont {Alvarez}}]{delos2018density}%
  \BibitemOpen
  \bibfield  {author} {\bibinfo {author} {\bibfnamefont {M.~S.}\ \bibnamefont
  {Delos}}, \bibinfo {author} {\bibfnamefont {A.~L.}\ \bibnamefont {Erickcek}},
  \bibinfo {author} {\bibfnamefont {A.~P.}\ \bibnamefont {Bailey}}, \ and\
  \bibinfo {author} {\bibfnamefont {M.~A.}\ \bibnamefont {Alvarez}},\ }\href
  {\doibase 10.1103/physrevd.98.063527} {\bibfield  {journal} {\bibinfo
  {journal} {Phys. Rev. D}\ }\textbf {\bibinfo {volume} {98}},\ \bibinfo
  {pages} {063527} (\bibinfo {year} {2018}{\natexlab{b}})},\ \Eprint
  {http://arxiv.org/abs/1806.07389} {arXiv:1806.07389} \BibitemShut {NoStop}%
\bibitem [{\citenamefont {Angulo}\ \emph {et~al.}(2017)\citenamefont {Angulo},
  \citenamefont {Hahn}, \citenamefont {Ludlow},\ and\ \citenamefont
  {Bonoli}}]{angulo2017earth}%
  \BibitemOpen
  \bibfield  {author} {\bibinfo {author} {\bibfnamefont {R.~E.}\ \bibnamefont
  {Angulo}}, \bibinfo {author} {\bibfnamefont {O.}~\bibnamefont {Hahn}},
  \bibinfo {author} {\bibfnamefont {A.~D.}\ \bibnamefont {Ludlow}}, \ and\
  \bibinfo {author} {\bibfnamefont {S.}~\bibnamefont {Bonoli}},\ }\href
  {\doibase 10.1093/mnras/stx1658} {\bibfield  {journal} {\bibinfo  {journal}
  {Mon. Not. R. Astron. Soc.}\ }\textbf {\bibinfo {volume} {471}},\ \bibinfo
  {pages} {4687} (\bibinfo {year} {2017})},\ \Eprint
  {http://arxiv.org/abs/1604.03131} {arXiv:1604.03131} \BibitemShut {NoStop}%
\bibitem [{\citenamefont {Ogiya}\ \emph {et~al.}(2016)\citenamefont {Ogiya},
  \citenamefont {Nagai},\ and\ \citenamefont {Ishiyama}}]{ogiya2016dynamical}%
  \BibitemOpen
  \bibfield  {author} {\bibinfo {author} {\bibfnamefont {G.}~\bibnamefont
  {Ogiya}}, \bibinfo {author} {\bibfnamefont {D.}~\bibnamefont {Nagai}}, \ and\
  \bibinfo {author} {\bibfnamefont {T.}~\bibnamefont {Ishiyama}},\ }\href
  {\doibase 10.1093/mnras/stw1551} {\bibfield  {journal} {\bibinfo  {journal}
  {Mon. Not. R. Astron. Soc.}\ }\textbf {\bibinfo {volume} {461}},\ \bibinfo
  {pages} {3385} (\bibinfo {year} {2016})},\ \Eprint
  {http://arxiv.org/abs/1604.02866} {arXiv:1604.02866} \BibitemShut {NoStop}%
\bibitem [{\citenamefont {Navarro}\ \emph {et~al.}(1996)\citenamefont
  {Navarro}, \citenamefont {Frenk},\ and\ \citenamefont
  {White}}]{navarro1996structure}%
  \BibitemOpen
  \bibfield  {author} {\bibinfo {author} {\bibfnamefont {J.~F.}\ \bibnamefont
  {Navarro}}, \bibinfo {author} {\bibfnamefont {C.~S.}\ \bibnamefont {Frenk}},
  \ and\ \bibinfo {author} {\bibfnamefont {S.~D.~M.}\ \bibnamefont {White}},\
  }\href {\doibase 10.1086/177173} {\bibfield  {journal} {\bibinfo  {journal}
  {Astrophys. J.}\ }\textbf {\bibinfo {volume} {462}},\ \bibinfo {pages} {563}
  (\bibinfo {year} {1996})},\ \Eprint {http://arxiv.org/abs/astro-ph/9508025}
  {arXiv:astro-ph/9508025} \BibitemShut {NoStop}%
\bibitem [{\citenamefont {Navarro}\ \emph {et~al.}(1997)\citenamefont
  {Navarro}, \citenamefont {Frenk},\ and\ \citenamefont
  {White}}]{navarro1997universal}%
  \BibitemOpen
  \bibfield  {author} {\bibinfo {author} {\bibfnamefont {J.~F.}\ \bibnamefont
  {Navarro}}, \bibinfo {author} {\bibfnamefont {C.~S.}\ \bibnamefont {Frenk}},
  \ and\ \bibinfo {author} {\bibfnamefont {S.~D.~M.}\ \bibnamefont {White}},\
  }\href {\doibase 10.1086/304888} {\bibfield  {journal} {\bibinfo  {journal}
  {Astrophys. J.}\ }\textbf {\bibinfo {volume} {490}},\ \bibinfo {pages} {493}
  (\bibinfo {year} {1997})},\ \Eprint {http://arxiv.org/abs/astro-ph/9611107}
  {arXiv:astro-ph/9611107} \BibitemShut {NoStop}%
\bibitem [{\citenamefont {Widrow}(2000)}]{widrow2000distribution}%
  \BibitemOpen
  \bibfield  {author} {\bibinfo {author} {\bibfnamefont {L.~M.}\ \bibnamefont
  {Widrow}},\ }\href {\doibase 10.1086/317367} {\bibfield  {journal} {\bibinfo
  {journal} {Astrophys. J. Suppl. Ser.}\ }\textbf {\bibinfo {volume} {131}},\
  \bibinfo {pages} {39} (\bibinfo {year} {2000})}\BibitemShut {NoStop}%
\bibitem [{\citenamefont {Binney}\ and\ \citenamefont
  {Knebe}(2002)}]{binney2002two}%
  \BibitemOpen
  \bibfield  {author} {\bibinfo {author} {\bibfnamefont {J.}~\bibnamefont
  {Binney}}\ and\ \bibinfo {author} {\bibfnamefont {A.}~\bibnamefont {Knebe}},\
  }\href {\doibase 10.1046/j.1365-8711.2002.05400.x} {\bibfield  {journal}
  {\bibinfo  {journal} {Mon. Not. R. Astron. Soc.}\ }\textbf {\bibinfo {volume}
  {333}},\ \bibinfo {pages} {378} (\bibinfo {year} {2002})},\ \Eprint
  {http://arxiv.org/abs/astro-ph/0105183} {arXiv:astro-ph/0105183} \BibitemShut
  {NoStop}%
\bibitem [{\citenamefont {Binney}\ and\ \citenamefont
  {Tremaine}(1987)}]{binney1987galactic}%
  \BibitemOpen
  \bibfield  {author} {\bibinfo {author} {\bibfnamefont {J.}~\bibnamefont
  {Binney}}\ and\ \bibinfo {author} {\bibfnamefont {S.}~\bibnamefont
  {Tremaine}},\ }\href {\doibase 10.2307/j.ctvc778ff} {\emph {\bibinfo {title}
  {Galactic Dynamics}}}\ (\bibinfo  {publisher} {Princeton University Press},\
  \bibinfo {address} {Princeton, NJ},\ \bibinfo {year} {1987})\BibitemShut
  {NoStop}%
\bibitem [{\citenamefont {Green}\ and\ \citenamefont {van~den
  Bosch}()}]{green2019tidal}%
  \BibitemOpen
  \bibfield  {author} {\bibinfo {author} {\bibfnamefont {S.~B.}\ \bibnamefont
  {Green}}\ and\ \bibinfo {author} {\bibfnamefont {F.~C.}\ \bibnamefont
  {van~den Bosch}},\ }\href@noop {} {\ }\Eprint
  {http://arxiv.org/abs/1908.08537} {arXiv:1908.08537} \BibitemShut {NoStop}%
\bibitem [{\citenamefont {Einasto}(1965)}]{einasto2965}%
  \BibitemOpen
  \bibfield  {author} {\bibinfo {author} {\bibfnamefont {J.}~\bibnamefont
  {Einasto}},\ }\href@noop {} {\bibfield  {journal} {\bibinfo  {journal} {Tr.
  Astrofiz. Inst. Alma-Ata}\ }\textbf {\bibinfo {volume} {5}},\ \bibinfo
  {pages} {87} (\bibinfo {year} {1965})}\BibitemShut {NoStop}%
\bibitem [{\citenamefont {Moore}(1993)}]{moore1993upper}%
  \BibitemOpen
  \bibfield  {author} {\bibinfo {author} {\bibfnamefont {B.}~\bibnamefont
  {Moore}},\ }\href {\doibase 10.1086/186967} {\bibfield  {journal} {\bibinfo
  {journal} {Astrophys. J.}\ }\textbf {\bibinfo {volume} {413}},\ \bibinfo
  {pages} {L93} (\bibinfo {year} {1993})},\ \Eprint
  {http://arxiv.org/abs/astro-ph/9306004} {arXiv:astro-ph/9306004} \BibitemShut
  {NoStop}%
\bibitem [{\citenamefont {King}(1966)}]{king1966structure}%
  \BibitemOpen
  \bibfield  {author} {\bibinfo {author} {\bibfnamefont {I.~R.}\ \bibnamefont
  {King}},\ }\href {\doibase 10.1086/109857} {\bibfield  {journal} {\bibinfo
  {journal} {Astron. J.}\ }\textbf {\bibinfo {volume} {71}},\ \bibinfo {pages}
  {64} (\bibinfo {year} {1966})}\BibitemShut {NoStop}%
\bibitem [{\citenamefont {Delos}\ \emph {et~al.}()\citenamefont {Delos},
  \citenamefont {Linden},\ and\ \citenamefont {Erickcek}}]{delos2019gamma}%
  \BibitemOpen
  \bibfield  {author} {\bibinfo {author} {\bibfnamefont {M.~S.}\ \bibnamefont
  {Delos}}, \bibinfo {author} {\bibfnamefont {T.}~\bibnamefont {Linden}}, \
  and\ \bibinfo {author} {\bibfnamefont {A.~L.}\ \bibnamefont {Erickcek}},\
  }\href@noop {} {\ }\Eprint {http://arxiv.org/abs/1910.08553}
  {arXiv:1910.08553} \BibitemShut {NoStop}%
\bibitem [{\citenamefont {Green}(2011)}]{green2011colour}%
  \BibitemOpen
  \bibfield  {author} {\bibinfo {author} {\bibfnamefont {D.~A.}\ \bibnamefont
  {Green}},\ }\href@noop {} {\bibfield  {journal} {\bibinfo  {journal} {Bull.
  Astron. Soc. India}\ }\textbf {\bibinfo {volume} {39}},\ \bibinfo {pages}
  {289} (\bibinfo {year} {2011})},\ \Eprint {http://arxiv.org/abs/1108.5083}
  {arXiv:1108.5083} \BibitemShut {NoStop}%
\bibitem [{\citenamefont {Drakos}\ \emph {et~al.}(2017)\citenamefont {Drakos},
  \citenamefont {Taylor},\ and\ \citenamefont {Benson}}]{drakos2017phase}%
  \BibitemOpen
  \bibfield  {author} {\bibinfo {author} {\bibfnamefont {N.~E.}\ \bibnamefont
  {Drakos}}, \bibinfo {author} {\bibfnamefont {J.~E.}\ \bibnamefont {Taylor}},
  \ and\ \bibinfo {author} {\bibfnamefont {A.~J.}\ \bibnamefont {Benson}},\
  }\href {\doibase 10.1093/mnras/stx652} {\bibfield  {journal} {\bibinfo
  {journal} {Mon. Not. R. Astron. Soc.}\ }\textbf {\bibinfo {volume} {468}},\
  \bibinfo {pages} {2345} (\bibinfo {year} {2017})},\ \Eprint
  {http://arxiv.org/abs/1703.07836} {arXiv:1703.07836} \BibitemShut {NoStop}%
\end{thebibliography}%

\end{document}